\documentclass[conference]{IEEEtran}
\IEEEoverridecommandlockouts
\usepackage{cite}
\usepackage{amsmath,amssymb,amsfonts}
\usepackage{algorithmic}
\usepackage{graphicx}
\usepackage{textcomp}
\usepackage{xcolor}
\usepackage{pifont}
\usepackage{hyperref}
\usepackage{cleveref}
\usepackage{flushend}

\def\BibTeX{{\rm B\kern-.05em{\sc i\kern-.025em b}\kern-.08em
    T\kern-.1667em\lower.7ex\hbox{E}\kern-.125emX}}
\begin{document}

\title{Physiological Closed-Loop Control (PCLC) Systems: Review of a Modern Frontier in Automation\\
}

\author{Mohammad Javad Khodaei$^{1,*}$, Nicholas Candelino$^{1,*}$, Amin Mehrvarz$^{1}$, and Nader Jalili$^{1,2}$ \\
$^{1}$Department of Mechanical and Industrial Engineering, Northeastern University, Boston, MA, USA \\
$^{2}$Professor and Head, Department of Mechanical Engineering, University of Alabama,
Tuscaloosa, AL, USA\\
$^{*}$ Co-first author}

\maketitle

\begin{abstract}
Over the past decade, there has been an unprecedented international focus on improved quality and availability of medical care, which has reignited interest in clinical automation and drawn researchers toward novel solutions in the field of physiological closed-loop control systems (PCLCs). Today, multidisciplinary groups of expert scientists, engineers, clinicians, mathematicians, and policy-makers are combining their knowledge and experience to develop both the next generation of PCLC-based medical equipment and a collaborative commercial/academic infrastructure to support this rapidly expanding frontier. In the following article, we provide a robust introduction to the various aspects of this growing field motivated by the recent and ongoing work supporting two leading technologies: the artificial pancreas (AP) and automated anesthesia. Following a brief high-level overview of the main concepts in automated therapy and some relevant tools from systems and control theory, we explore -- separately -- the developments, challenges, state-of-the-art, and probable directions for AP and automated anesthesia systems. We then close the review with a consideration of the common lessons gleaned from these ventures and the implications they present for future investigations and adjacent research.
\end{abstract}


\section{Introduction}
\label{sec:introduction}
The modern advance of automated and intelligent systems toward more-and-more visible roles in society (e.g., commercially available automated vehicles, humanoid and biomimetic robots, household vacuums and thermostats, etc.) has accelerated the development of many new technologies and restored interest in several challenging fields. Principal among these recent works, are the broad ongoing efforts to deploy collaborative human-machine systems, such as those designed to provide advanced support in medical diagnostics and therapy. In this review, we focus on a special class of these systems that apply concepts from feedback control to attain automatic regulation of physiological variables -- such as blood pressure, depth of anesthesia, blood glucose concentrations, and others (see examples in \textcolor{blue}{{Table }\ref{table1}}) -- known as physiological closed-loop control systems (PCLCs). While the pioneering works toward many kinds of PCLCs were performed over a half-century ago, we are now only nascent in an era of suitable computational/technological availability and physiological/pharmaceutical scientific understanding to begin realizing these devices on a broad commercial scale.

From the outset, it is important to understand that these devices do not and are not intended to replace expert and supporting clinical staff. In fact, automated devices rely on the knowledge of experts and simply put, promise to do a much better job at following clinical instruction than present technology. By introducing sensors capable of continuous monitoring (i.e., generating measurements every few seconds or minutes depending on the treatment) these devices stand to recognize and eliminate unintentional variations in therapies, including missed doses and faults like sensor detachment or blocked fluid pathways. Furthermore, post analysis of the measurements obtained by these devices and the treatment profiles they generate can provide important insights into the effectiveness of treatment strategies (e.g., the clinical setpoint is maintained to no effect), possible heterogeneity in disease pathologies, and goals for future pharmaceuticals (e.g., short or longer acting doses). The development and adoption of these new systems is currently providing research directions in control engineering, artificial intelligence, human-machine interactions, social science, medical practice, therapeutic technology, pharmaceutical development, and biological/physiological systems modeling. Further, since poor controller performance in physiological systems can lead to immediate adverse, potentially irreversible, and even fatal physiological responses, PCLCs are a rapidly developing field of safety critical control systems and additional research on safe PCLC operation will be of particular value in the transition to common medical practice (esp. as networked and ambulatory systems are realized).

As a primer for researchers in these and other disciplines, we have sought in this review to introduce key information from various aspects of the ongoing research to provide both a background and up-to-date perception of the field for unfamiliar readers. This includes a brief introduction to modeling and control, within the scope of those methods common to physiological systems, in \textcolor{blue}{Section \ref{Preliminaries}}; the use of case studies in \textcolor{blue}{Sections \ref{diabetes}-\ref{automated anesthesia}} to frame and discuss important topics in current research; and a minimalist presentation of the detailed mathematical models, which are all very well documented in the cited references. Following the case studies, in \textcolor{blue}{Section \ref{discussion}} we will discuss several aspects from these investigations that are of great importance in the broader scope of physiological control as well as possible directions for the future.
\begin{table*}[h!]
\centering
\caption{Examples of research in physiological closed-loop systems (PCLCs).}
\label{table1}
\begin{tabular}{cccc}
\hline \hline
Treatment Target & Treatment Actuator & Treatment Sensor & Control Method \\ \hline \hline
Blood Glucose (e.g.\textcolor{blue}{\cite{palerm2011}}) & Insulin infusion/bolus & \begin{tabular}[c]{@{}c@{}}Blood Glucose \\ Concentration\end{tabular} & PID \\
Anesthesia (e.g.,\textcolor{blue}{\cite{doctor2016}}) & Propofol/Remifentanil & \begin{tabular}[c]{@{}c@{}}Electroencephalogram\\ /Bispectral Index\end{tabular} & Fuzzy-Logic \\
Blood Pressure (e.g.,\textcolor{blue}{\cite{luspay2015}}) & \begin{tabular}[c]{@{}c@{}}Vasoactive drugs such\\  as phenylephrine\end{tabular} & \begin{tabular}[c]{@{}c@{}}Mean Arterial \\ Pressure\end{tabular} & \begin{tabular}[c]{@{}c@{}}Linear Parameter\\  Varying Control\end{tabular} \\
\begin{tabular}[c]{@{}c@{}}Anemia in End-stage-\\ Renal-Disease (e.g.,\textcolor{blue}{\cite{gaweda2008}})\end{tabular} & \begin{tabular}[c]{@{}c@{}}Erythropoiesis-\\ stimulating agents\end{tabular} & Hemoglobin Level & \begin{tabular}[c]{@{}c@{}}Model Predictive \\ Control\end{tabular} \\
Tumor Growth (e.g.,\textcolor{blue}{\cite{cunningham2018}}) & Abiraterone & Tumor Composition & \begin{tabular}[c]{@{}c@{}}Evolutionary\\  Game Theory\end{tabular} \\Heart Failure (e.g.,\textcolor{blue}{\cite{alomari2012, petrou2018}}) & \begin{tabular}[c]{@{}c@{}}Implantable Rotary \\ Blood Pump\end{tabular} & \begin{tabular}[c]{@{}c@{}}Left Ventricle Pressure\\ /Flow Pulsatility\end{tabular} & \begin{tabular}[c]{@{}c@{}}Sliding Mode\\  Control\end{tabular} \\
Fluid Resuscitation (e.g.,\textcolor{blue}{\cite{rinehart2013}}) & \begin{tabular}[c]{@{}c@{}}Intravenous fluid\\  administration (IV injection \\ of replacement fluids/blood)\end{tabular} & \begin{tabular}[c]{@{}c@{}}Stroke Volume; Heart Rate;\\  Mean Arterial Pressure;\\  Pulse-Pressure Variation\end{tabular} & \begin{tabular}[c]{@{}c@{}}Predictive Rule-\\ Based Control\end{tabular} \\ \hline
\end{tabular}
\end{table*}

\begin{figure}
\label{fig1}
\centering
\includegraphics[width=80mm,scale=0.5]{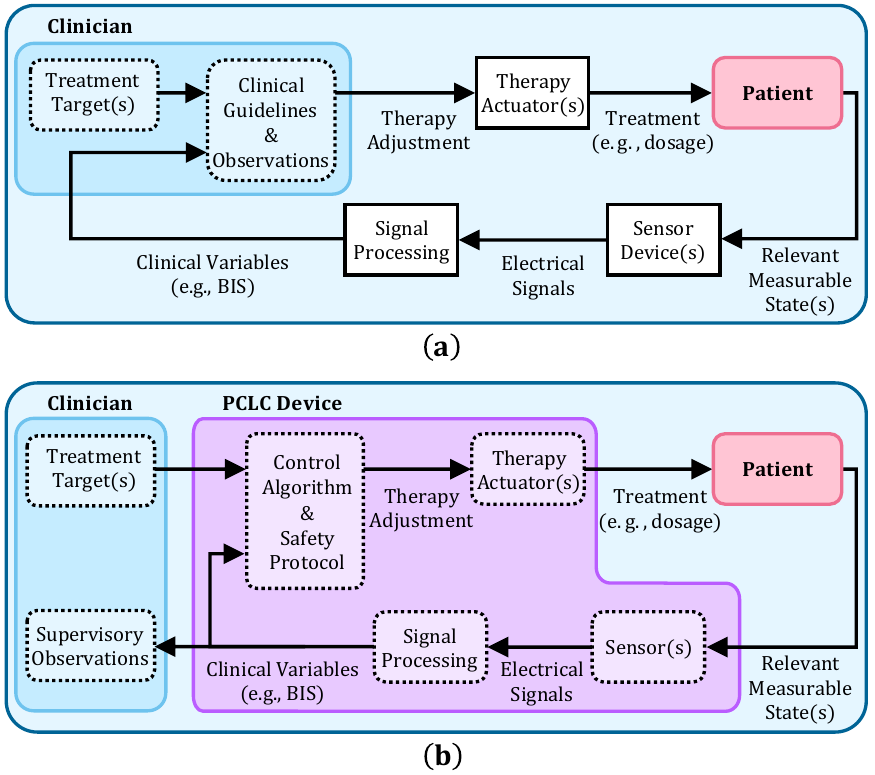}
\caption{Comparison of (a) manual therapy applied by a clinician according to observations and set guidelines for making therapy adjustments and (b) automated therapy applied by a physiological closed-loop control (PCLC) device according to a control algorithm with the possibility of a clinician acting in a supervisory capacity (e.g. monitoring Bispectral Index (BIS) for additional safety in anesthesia).}
\end{figure}

\section{Preliminaries}\label{Preliminaries}
Regardless of the controlled variables and outputs, the general process implemented in PCLCs can be represented conceptually by a diagram resembling \textcolor{blue}{{Fig. }\ref{fig1}}. Here, a control algorithm -- normally implemented on a computer/processor -- provides an interface between (i) target values for physiological variables that have been specified by clinicians, (ii) feedback signals containing measurements and estimates of system variables, and (iii) the infusion pumps or other actuators that apply treatment to the patient. Notice, from a high-level perspective, there is little difference between the "automated" approach and the classical manual approach aside from the simplification of tasks required from the clinician/patient, who no longer has to interact with the infusion pump (or other actuator) directly and instead can focus on specifying treatment goals and assessing therapy performance.

Recognizing the generality of this concept, many agencies and organizations are in the process of developing more rigorous guidelines for commercial PCLCs that will help to establish the scope of this up-and-coming class of devices. As noted in a recent paper from researchers at the Center for Devices and Radiological Health (CDRH) at the US Food and Drug Administration (FDA), the CDRH maintains a working definition of PCLC medical devices (or PCLC devices) as, "a medical device that incorporates physiological sensor(s) for automatic manipulation of a physiological variable through actuation of therapy that is conventionally made by a clinician"\textcolor{blue}{\cite{parvinian2018}}. This definition -- demonstrated in \textcolor{blue}{{Fig. }\ref{fig1}b} -- recognizes the role of automation in PCLC while allowing a great deal of flexibility in the disorder being treated, the types of sensors and therapies used (many are certainly yet to be discovered), and the extent to which the PCLC operates independent of clinician intervention -- known as the level of automation (LOA)\textcolor{blue}{\cite{parvinian2018}}. To help guide the development of commercial PCLC devices, the United States\textcolor{blue}{\cite{food2014food}}, Canada\textcolor{blue}{\cite{health2019}}, and the European Union\textcolor{blue}{\cite{European2017}} have formally recognized versions of the recently developed IEC 60601-1-10 consensus standard\textcolor{blue}{\cite{medical2007}}, which is the first standard to provide, "requirements for the development (analysis, design, verification and validation) of a physiologic closed-loop controller (PCLC) as part of a physiologic closed-loop control system (PCLCS) in medical electrical equipment and medical electrical systems to control a physiologic variable"\textcolor{blue}{\cite{medical2007}}. Beyond regulatory concerns, such guidance may help the broader research and development communities establish a common language (possibly even one that is shared among medical and engineering professionals); a set of safe design practices, testing procedures, and quality management objectives; and a platform to qualify and address issues raised by designers, clinicians, and patients.

Treating this feedback system similar to those in more traditional engineering disciplines, the design and implementation of PCLCs requires accurate models of the patient, sensor, and actuator dynamics, including any fundamental limitations and safety constraints. While it may be obvious that the sensor and actuator dynamics can be described using traditional differential equations, similar mathematical modeling techniques can also be applied to develop equations representing patient physiology and pharmaceutical action. In fact, the human body is generally modeled using compartmental techniques (see \textcolor{blue}{Section \ref{compartmental}}) from mathematical systems physiology, which can describe the behavior of healthy and perturbed organ/tissue interactions as well as the distribution/transportation of pharmaceuticals in the body (pharmacokinetics) and the action of these pharmaceuticals on certain organ systems (pharmacodynamics) in terms of nonlinear differential equations.

\subsection{Compartmental Models}\label{compartmental}
\begin{figure}
\centering
\includegraphics[width=80mm,scale=0.5]{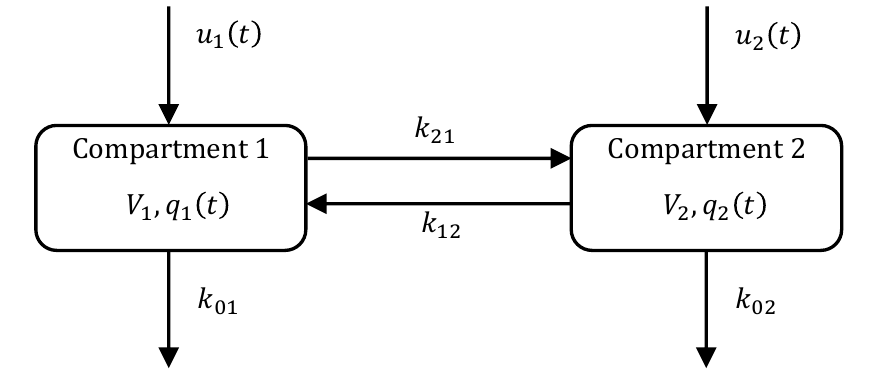}
\caption{General compartmental model for 2 compartment systems}
\label{fig2}
\end{figure}

When working with physiological or biological systems, the dynamic interactions of system states/variables are often depicted in the form of compartmental models. This representation is slightly different from the standard block diagram models of systems theory but conveys similar information. Under this modeling paradigm, a system is comprised of one or more compartments, each of which has a (normally) fixed volume $V_i$ and contains a variable state quantity $q_i(t)$. The representation of this quantity $q_i(t)$ as a fraction of the compartment volume gives the concentration $c_i(t)=q_i (t)/V_i$ . When the change in these quantities or concentrations takes the form of ordinary differential equations, a compartmental modeling diagram can be drawn to depict the contributions of each state quantity to the dynamics (similar to state-variable representation). In the case of a linear time-invariant system, every compartment is associated with a linear first order differential equation described by the signals entering and leaving the respective compartment. Considering the general two-compartment model shown in \textcolor{blue}{{Fig. }\ref{fig2}}, the change in $q_1(t)$ is given by
\begin{equation}
    \dot{q}_1(t)=u_1(t)-k_{21}q_1(t)+k_{12}q_2(t)-k_{01}q_1(t)\label{eq1}
\end{equation}
where $u_1(t)$ is the input to compartment 1; $k_{01}q_1(t)$ is the leakage from this compartment to the environment; $k_{21} q_1(t)$ is the exchange from compartment 1 to compartment 2; and $k_{12}q_2(t)$ is the exchange from compartment 2 to compartment 1. Similarly, based on the arrows pointing toward and away from compartment 2, the change in $q_2(t)$ can be written as
\begin{equation}
    \dot{q}_2(t)=u_2(t)+k_{21}q_1(t)-k_{12}q_2(t)-k_{02}q_2(t)\label{eq2}
\end{equation}
where $u_2(t)$ is the input to compartment 2; $k_{02}q_2(t)$ is the leakage/loss from compartment 2 to the environment; and all other terms are the same as described previously.

Together, these equations describe the exchange between compartments and the environment. In general, there may be $n$ compartments, each of which may have a constant or changing volume $V_i$ and can be linked to any or all other compartments by linear or nonlinear interactions. For more detailed information on compartmental modeling, readers are referred to the extensive coverage in\textcolor{blue}{\cite{distefano2015}}.

\subsection{Control Methodologies}
In this section, we briefly review the fundamentals of the 3 most common control methods currently used in physiological systems: (i) proportional-integral-derivative (PID) control, (ii) model predictive control (MPC), and (iii) fuzzy-logic (F-L) control. The application of these control schemes to specific physiological systems, including the advantages, variations, and recent developments to each strategy, are described in later sections. Of course, many other control methods are also being pursued in the literature and will be discussed as needed.

\subsubsection{Proportional-Integral-Derivative (PID) Control}
PID control is one of the most frequently used control approaches in industrial systems due to its simple mathematics, well-known gain tuning methods, and broad application to both linear and nonlinear systems. The Laplace domain transfer function of the ideal PID controller can be given by
\begin{equation}
    \frac{U(s)}{E(s)}={{k}_{p}}\left( 1+\frac{1}{s{{T}_{i}}}+{{T}_{d}}s \right)\label{eq3}
\end{equation}
where $s$ is the complex-valued Laplace variable; $E(s)$ and $U(s)$ are, respectively, the Laplace domain input and output of the PID controller; $k_p$ is the proportional gain; $T_i$ is the integral time-constant; and $T_d$ is the derivative time-constant. In most cases, the coefficients $k_p$, $T_i$, and $T_d$ are free-parameters that take constant values and are designed to obtain desirable system properties. The values of these parameters may be obtained by either theory-based design strategies or numerical/experimental tuning methods.

\subsubsection{Model Predictive Control (MPC)}
Model predictive control is a robust and optimal control technique used to handle systems with constraints while optimizing the controller action and predicting system outputs for robustness against noise and disturbances. The basic formulation of this controller can be given by
\begin{equation}
	U_{t}^{*}\left( x\left( t \right) \right):=\arg \min \sum\limits_{k=0}^{N-1}{q\left( {{x}_{t+k}},{{u}_{t+k}} \right)}\label{eq4}
\end{equation}
subject to\\
	$x_t=x(t)$\hspace{85pt}(Measurement or estimation)\\
	$x_{(t+k+1)}=Ax_{(t+k)}+Bu_{(t+k)}$	\hspace{50pt}(System model)\\
	$x_{(t+k)}\in\chi$\hspace{126pt}(State constraints)\\
	$u_{(t+k)}\in{u}$\hspace{125pt}(Input constraints)\\
	$U_t^* (x(t))=\{{u_t^*,u_{(t+1)}^*,…,u_{(t+N-1)}^*}\}$(Optimization variables)\\
where $U_t^*(x(t))$ is the optimal input sequence for the entire planning window $N$; $q$ is the cost function; and $\chi$ and $u$ are, respectively, sets of state and input constraints. At any given time $t$, this controller finds the set of optimal control actions $U_t^*$ based on the predicted dynamics over the entire window $N$ and implements only the first control action $u_t^*$ in this sequence.

\subsubsection{Fuzzy-Logic (F-L) Control}
\begin{figure}
\centering
\includegraphics[width=80mm,scale=0.5]{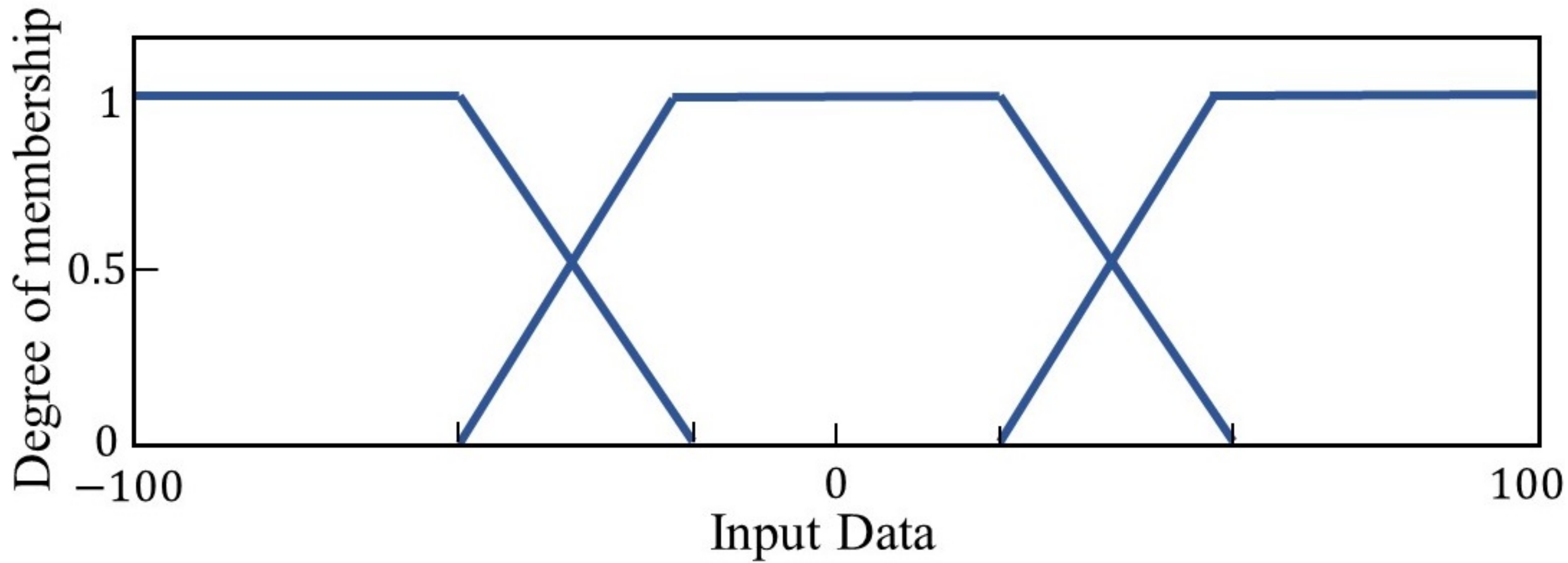}
\caption{Fuzzy logic sets}
\label{fig3}
\end{figure}

Fuzzy systems are knowledge-based or rule-based systems consisting of fuzzy "if-then" rules that represent heuristic verbal conditions as continuous membership functions \textcolor{blue}{{Fig. }\ref{fig3}}\textcolor{blue}{\cite{soliman2011}}. This method can be used to model and control complex dynamics, nonlinear systems, and systems that are difficult to express mathematically. In general, fuzzy-logic systems can be classified as Mamdani and Sugeno systems\textcolor{blue}{\cite{mamdani1975, takagi1985, sugeno1988}}. Fuzzy-logic systems have been successfully implemented in many industries.
\subsection{Case Studies}
In order to demonstrate and contextualize many important principles in PCLCs, we will proceed over \textcolor{blue}{Sections \ref{diabetes}} and \textcolor{blue}{\ref{automated anesthesia}} with detailed case studies of the two most active areas in physiological control: (i) automated anesthesia and (ii) automated insulin therapy for diabetes. For each system, a detailed background on therapeutic goals, physiology, and mathematical representations is given prior to discussions on current trends in the control methodologies and specific challenges encountered in each area.

\section{Diabetes and the Artificial Pancreas}\label{diabetes}
To understand the problem of blood glucose regulation, let us begin with a brief review of the essential physiology of a properly functioning glucoregulatory system and build towards a fundamental understanding of the diabetic disorders (pathophysiology) as well as their underlying causes (etiology) and associated health risks (complications and comorbidities). By way of this introduction, we hope to provide clarity for the coming discussions on system models and insights when broaching the subjects of treatment and closed-loop therapy.

\subsection{Physiology of the Healthy Blood Glucose Regulatory System}
The natural regulation of blood glucose in a healthy individual is the result of a variety of interacting physiological processes involving hormones, neurotransmitters, and our own consumption of food. Historically, the accepted metric for judging the quality of this regulation is the amount of glucose present in the blood stream, measured as a molar (mmol/L) or mass (mg/dL) concentration. This blood glucose concentration is commonly referred to as an individual's blood glucose level and is generally well-maintained between about 4-6 mmol/L (72-108 mg/dL) during healthy glucose homeostasis. However, even in healthy subjects, there are occasional excursions above this range (e.g., after recent meals) or slightly below it (e.g., during or after periods of high activity). The interactions that govern glucose concentrations are highly diverse and most, if not all, of these interactions are reciprocal -- both the glucose concentration and the acting processes are mutually affected by the regulatory procedure. The systems involved in this regulation include the brain, liver, kidney, gastrointestinal tract, adipose and muscle tissues, and pancreas. Of these systems, the pancreas is arguably the most critical for maintaining proper blood glucose and will be our focus in the sequel.

The pancreas, located behind the stomach, is a glandular organ primarily responsible for secreting digestive enzymes through ducts to the small intestines and releasing pancreatic hormones into the blood stream. While both enzymes and hormones play important roles in proper metabolic homeostasis, it is the latter (i.e., hormones) that are principally responsible for regulating blood glucose levels\textcolor{blue}{\cite{roder2016pancreatic}}.

These hormones, known as the pancreatic hormones, are produced at small clusters of endocrine cells (i.e., cells that secrete directly into the blood stream), known as islets, that are distributed sparsely throughout the pancreatic tissue (comprising $\textless2\%$). These pancreatic islets (or islets of Langerhans) are constructed from a combination of 5 kinds of endocrine cells called $\alpha$-cells, $\beta$-cells, $\gamma$-cells, $\delta$-cells, and $\epsilon$-cells. Each of these cell types secrete a specific set of hormones that are known to be involved in at least some aspects of normal glucoregulatory function. Of these cells, the $\alpha$ and $\beta$ types make-up $\textgreater80\%$ of all pancreatic islet cells and are considered the 2 most essential endocrine cells for blood glucose regulation\textcolor{blue}{\cite{roder2016pancreatic}}. The first of these, $\alpha$-cells, produce a single hormone called glucagon, which drives the production of endogenous glucose by catabolizing (breaking-down) stored glycogen in the liver (known as hepatic glycogenolysis) and to a lesser extent by promoting the generation of glucose from non-carbohydrate compounds in the liver and kidneys through processes termed, respectively, hepatic and renal gluconeogenesis. The latter type of cells, $\beta$-cells, produce 3 hormones called $C$-peptide, amylin, and insulin, all of which have important roles in glucose regulation and are believed to correspond with various symptoms of blood glucose pathophysiology. Of these 3 hormones, insulin is the most well-known, and is responsible for lowering blood glucose concentrations through promotion of glucose utilization by adipose and muscle tissues, as well as by prompting the creation of glycogen and fatty acid energy stores from glucose through glycogenesis and lipogenesis, respectively. Thus, in contrast to glucagon, which raises blood glucose levels, insulin provides the primary means to reduce this concentration\textcolor{blue}{\cite{roder2016pancreatic}}.

As one might expect, based on their roles in the blood glucose regulatory function, glucagon is primarily secreted during periods of low blood glucose such as sleep or fasting, while insulin secretion is stimulated by elevated blood glucose concentrations and the incretion effect promoted by ingestion of glucose, fructose, amino acids, and long-chain free fatty acids\textcolor{blue}{\cite{roder2016pancreatic}}. Of course, insulin secretion is highest following meals (i.e., when blood glucose levels are at their highest). Together, glucagon and insulin provide the basis for many simple and intuitive bi-hormonal models of the pancreatic endocrine system. While many of the physio-chemical mechanisms for these secretions and uptakes are quite well known, it is important to understand that these hormones are released directly into blood stream and interact with many different organ systems. Thus, the specific mechanisms and outcomes of these interactions remain highly active research topics. 

Of course, the 2 other $\beta$-cell hormones and those secreted from the $\gamma$, $\delta$, and $\epsilon$-cells are also intrinsic to proper physiology. Unfortunately, the functions of these hormones are not yet as well understood as those of insulin and glucagon and are often (perhaps necessarily) disregarded in the control-oriented literature. However, there is reason to believe study of these pancreatic hormones, and their roles in both physiology and pathophysiology, will likely lead to better multi-hormonal models and treatment approaches in the coming years\textcolor{blue}{\cite{roder2016pancreatic, aronoff2004}}. For example, amylin -- produced by $\beta$-cells along with insulin -- is believed to perform complementary functions to the insulin-based removal of glucose, instead inhibiting the appearance of absorbed (exogenous) and glycogenolytic glucose by, respectively, reducing the flow of food from the stomach to the small intestines (gastric emptying) and blocking the release of postprandial glucagon (i.e., inhibiting unnecessary endogenous glucose secretion following meals)\textcolor{blue}{\cite{aronoff2004}}. While an FDA-approved amylin analog (i.e., pramlintide) is available, the cost of treatment with such complimentary hormone analogs may be prohibitively expensive and perhaps even detrimental if not properly (e.g., physiologically or pseudo-physiologically) controlled.

\subsubsection{Pathophysiology and Etiology}

Type 1 Diabetes Mellitus (T1DM):
The term type 1 diabetes mellitus describes blood glucose regulatory disorders resulting from the complete (or near complete) destruction of all pancreatic $\beta$-cells, leading to hyperglycemia and life-threatening complications. T1DM is generally caused by an auto-immune response, but does occur infrequently without any known or identified cause (idiopathic T1DM)\textcolor{blue}{\cite{2association2018, punthakee2018}}. The characteristic $\beta$-cell destruction of T1DM leads to a dependence on external sources of insulin (i.e., insulin therapy) for survival; hence, this form (esp. the auto-immune form) was previously identified by the term insulin-dependent diabetes mellitus (IDDM)\textcolor{blue}{\cite{2association2018,ozougwu2013}}. Type 1 diabetes accounts for 5-10\% of all cases\textcolor{blue}{\cite{2association2018}}.

Considering the preceding discussions on healthy glucoregulatory function, it is clear that destruction of $\beta$-cells in T1DM suppresses the production of the insulin, amylin, and $C$-peptide hormones. Without these hormones (particularly insulin), cells throughout the body cannot remove glucose from the blood stream for use in metabolic processes (e.g., storage as glycogen). As a result, the body turns toward the metabolism of fats, which leads to increased levels of free fatty acids and ultimately ketones in the blood stream\textcolor{blue}{\cite{ozougwu2013}}. Overproduction of the latter leads to life-threatening diabetic ketoacidosis. Furthermore, complications due to elevated blood glucose levels (hyperglycemia) appear across numerous organ systems and include microvascular diseases such as neuropathy, retinopathy, and nephropathy\textcolor{blue}{\cite{ozougwu2013, dimeglio2018}}. In T1DM, the absence of interactions between pancreatic $\alpha$-cells and $\beta$-cells removes an important pathway for natural glucagon regulation, leading to continued glucagon production during periods of elevated blood glucose, resulting in more severe hyperglycemia and even greater risk of complications\textcolor{blue}{\cite{ozougwu2013}}. Unfortunately, while many of the risks associated with T1DM can be reduced by tight regulation of blood glucose levels\textcolor{blue}{\cite{ozougwu2013}}, the destruction of $\beta$-cells affects multiple hormones and many systems, some of which are not fully restored by external blood glucose regulation. For instance, increased risk of cardiovascular disease -- the leading cause of reduced life-expectancy in type 1 diabetics -- remains prevalent in type 1 diabetics with good blood glucose control\textcolor{blue}{\cite{ozougwu2013}}.\\

Type 2 Diabetes Mellitus (T2DM):
The term type 2 diabetes mellitus describes a family of blood glucose regulatory disorders that are marked by relative insulin deficiency, due to a combination of poor insulin production and poor insulin utilization, leading to episodes of hyperglycemia\textcolor{blue}{\cite{2association2018, punthakee2018}}. Previously identified as non-insulin-dependent diabetes mellitus (NIDDM), current opinion is trending towards a more graduated classification system that may improve differentiation of T2DM by severity and pathophysiology of impaired insulin-secretion (production) and insulin-sensitivity (utilization)\textcolor{blue}{\cite{2association2018, stidsen2018, ahlqvist2018}}. Type 2 diabetes accounts for about 85\% of all cases\textcolor{blue}{\cite{2association2018, ozougwu2013, ahlqvist2018}}.

Unlike T1DM, T2DM is not an autoimmune disorder and has neither a single underlying cause (etiology) nor a unique physical origin. Instead, the term T2DM applies to various non-specific disorders (i.e., not specifically meeting the criteria for definition as any distinct disorder such as monogenic, gestational, or type 1 diabetes mellitus) that result in chronic excess blood glucose\textcolor{blue}{\cite{2association2018, punthakee2018}}. While in some cases, T2DM is marked by a reduced pancreatic insulin secretion that results in a relative insulin deficiency, in others, insulin production is unaffected, but cells exhibit an inability to utilize available insulin (i.e., insulin resistance). However, most of the type 2 diabetic population exhibit both insulin deficiency and insulin resistance\textcolor{blue}{\cite{2association2018, punthakee2018, stidsen2018, ahlqvist2018}}. Thus, it is becoming more common to characterize each instance of T2DM along gradients of insulin-sensitivity and insulin secretion (among other variables including coexisting conditions), to better identify root-causes and optimal treatment regimens\textcolor{blue}{\cite{ahlqvist2018, stidsen2018}}. While T2DM is not normally life-threatening by itself, it is associated with reduced life-expectancy due to macrovascular complications, as well as numerous microvascular complications that are believed to be linked with episodes of hyperglycemia. Like T2DM itself, the accompanying complications are diverse and, some research suggests, potentially well-correlated with pathophysiologically distinct forms of the disorder\textcolor{blue}{\cite{stidsen2018, ahlqvist2018, gedebjerg2018}}.\\

Gestational Diabetes Mellitus (GDM):
Glucose intolerance that develops in women during pregnancy, indicated by the emergence of persistent hyperglycemia, is known as gestational diabetes mellitus\textcolor{blue}{\cite{2association2018, punthakee2018, wexler2018}}. Like T2DM, GDM is comprised of a highly heterogeneous set of disorders, which may present predominantly insulin resistant or insulin deficient characteristics\textcolor{blue}{\cite{wexler2018}}. While standards for diagnosing GDM vary, the critical feature of GDM is that it is developed during pregnancy and was not present prior\textcolor{blue}{\cite{2association2018}}. The pathophysiology of GDM is not yet well-understood\textcolor{blue}{\cite{wexler2018}}, but ongoing research may benefit from more accurate and detailed sensing technologies. GDM is a disorder of international interest and can result in complications for both mother and child and can progress into a form of T2DM after pregnancy\textcolor{blue}{\cite{2association2018}}.\\

Other Specific Types of Diabetes:
In addition to T1DM, T2DM, and GDM, there are numerous, less common diabetic disorders (together comprising <5\% of cases) with more specific association with distinct etiological factors such as genetics, drug or chemical exposure, and pancreatic diseases\textcolor{blue}{\cite{2association2018, punthakee2018}}. Differentiation of these specific forms from T1DM and T2DM is critical, as the treatment requirements for many of these disorders differ from general guidelines and mischaracterization can lead to poor or detrimental outcomes and unnecessary expenses\textcolor{blue}{\cite{2association2018, stidsen2018}}.

\subsection{Traditional Monitoring and Treatment}
Considering the variety of disorders that may exist, or coexist, in diabetic individuals, it is not surprising that there are a plethora of diabetic diagnostic tools and interventions, some of which have been shown to be most effective in monitoring and treating specific disorders, while simultaneously potentially detrimental for treating others. Likewise, treatment guidelines are often adjusted for improved effectiveness in specific demographics often based on age, gender, and overall health, with the final recommendation for individuals tailored by their physician\textcolor{blue}{\cite{6association2018, 8association2018}}.

\subsubsection{Laboratory Screening}
For the past half century, blood glucose regulatory function and concentration targets for diabetics and non-diabetics have primarily been specified according to 2 standard measurements: (i) static blood glucose concentrations and (ii) percent glycated hemoglobin (HbA1c, or less formally A1C). While a static measurement of blood glucose provides insight into the current glycemic state (i.e., the current mmol/L concentration of glucose in the blood stream), it does not provide any insight into recent concentrations or an indication of whether this value is increasing or decreasing\textcolor{blue}{\cite{adofsson2018}}. Conversely, A1C provides an approximate indication of the average blood glucose concentration over a period of about the preceding 2-3 months\textcolor{blue}{\cite{6association2018, adofsson2018}}. While A1C is less sensitive to daily blood glucose variations and a primary predictor of diabetes related complications, acceptable A1C values have been shown to vary significantly across populations and with various environmental factors, and further do not provide any specific information regarding occurrences of hypoglycemia, hyperglycemia, or their severity\textcolor{blue}{\cite{2association2018, 6association2018}}.

In clinical settings, diabetic screening is often performed using plasma blood glucose concentrations that are measured by accredited laboratories using standardized tests\textcolor{blue}{\cite{sacks2011}} to determine levels under the moderately controlled circumstances either following a period of fasting to obtain the so-called fasting plasma glucose (FPG) level or 1-2 hours into an oral glucose tolerance test (OGTT) administered after a period of fasting. If the FPG or OGTT results fall below certain thresholds (see characteristic thresholds from the ADA\textcolor{blue}{\cite{2association2018}} in \textcolor{blue}{{Table }\ref{table2}}), these tests results, respectively, indicate impaired fasting glucose (IFG) or impaired glucose tolerance (IGT) associated with metabolic disfunction. The accuracy of these tests depends heavily on patient adherence to the fasting guidelines and clinical sample handling. Thus, for clinical A1C (also performed in laboratories\textcolor{blue}{\cite{sacks2011}}), FPG, and OGTT testing, confirmation of an initial diagnosis of diabetes often requires that 2 of these values fall outside the acceptable thresholds\textcolor{blue}{\cite{2association2018}}.
\begin{table*}
\centering
\caption{FPG, OGTT, and A1C thresholds for initial diagnosis of diabetes.}
\label{table2}
\begin{tabular}{ccc}
\hline \hline
Test & Prediabetes Thresholds & Diabetes Threshold \\ \hline \hline
FPG & 100-125 mg/dL (5.6-6.9 mmol/L) & $\geq$126 mg/dL (7.0 mmol/L) \\
75-g OGTT (2 hour) & 140-199 mg/dL (7.8-11.0 mmol/L) & $\geq$200 mg/dL (11.1 mmol/L) \\
A1C & 5.7-6.4\% (39-47 mmol/mol) & $\geq$6.5\% (48 mmol/mol) \\ \hline
\end{tabular}
\end{table*}

\subsubsection{Self-Monitoring of Blood Glucose (i.e., "Finger-Stick" Monitoring)}
Once diagnosed, most diabetics (especially those with T1DM or T2DM requiring insulin therapy) will require portable diagnostic equipment so that blood glucose concentrations can be monitored according to the severity of their disorders or in the event of symptoms that indicate excursions from normal glycemic control. For many years, the predominant method for this daily testing was self-monitoring of blood glucose (SMBG) by so-called "finger-stick" devices\textcolor{blue}{\cite{adofsson2018, leelarathna2018}}, which require drawing a small sample of blood during every test. Initially accepted for their marked improvement over urine testing equipment, use of these finger-stick devices causes pain, and measurements can be uncomfortable or unsafe (due to blood spill) to perform in public\textcolor{blue}{\cite{adofsson2018}}. While significantly less accurate than accredited laboratory testing\textcolor{blue}{\cite{sacks2011}}, these portable diagnostic tools have enabled diabetics to maintain tighter glycemic control by regular assessment of their blood glucose levels, especially prior to meals, sleep, and insulin dosing. However, like static laboratory measurements, SMBG does not provide any historical or predictive information by itself and must be combined with knowledge of recent past/future carbohydrate intake, exercise, and medication dosing to estimate appropriate corrective actions. Coupled with the inconvenience and limitations of regular blood sampling, the static nature of SMBG has prompted the modern quest for safe, accurate, and continuous blood glucose monitors (CGMs).

\subsubsection{Continuous and Flash Glucose Monitors}
Starting in the late 1990s, diagnostic tools for continuous home monitoring of blood glucose began to emerge as a potential remedy to the shortcomings of traditional SMBG\textcolor{blue}{\cite{adofsson2018}}. Originally designed to record, but not display, blood glucose measurements over several days -- to be analyzed retrospectively by a physician\textcolor{blue}{\cite{leelarathna2018, sacks2011, wood2018}}-- more recent real-time CGM (rtCGM) devices that log data and provide immediate access to measurements and analytical trends are now available on a broad commercial scale (\textcolor{blue}{{Table }\ref{table3}})\textcolor{blue}{\cite{adofsson2018}}. Unlike SMBG, which requires taking blood samples at each measurement time, most rtCGM devices are minimally invasive and measure glucose concentrations subcutaneously through the interstitial fluid rather than whole blood or plasma\textcolor{blue}{\cite{sacks2011, wood2018}}. These sensors can be worn continuously for up to a week, and provide updated measurements every 5 minutes, requiring finger-stick measurements for calibration (up to twice per day)\textcolor{blue}{\cite{sacks2011}}. While only one such device is FDA approved for non-adjunctive use in insulin therapy (i.e., for use in determining insulin dosage without the need to confirm blood glucose levels by traditional SMBG), this is a recent development and more devices may be expected in the future\textcolor{blue}{\cite{6association2018}}.

\begin{table*}[]
\centering
\caption{Examples of commercially available CGM components and device\textcolor{blue}{\cite{adofsson2018, medtronic2014, medtronic2017,medtronic2018, medtronic20183, dexcom2017, 5dexcom2018, 6dexcom2018}}.}
\label{table3}
\begin{tabular}{ccccc}
\hline \hline
\multicolumn{1}{c}{Manufacturer} & Device & Accuracy (MARD \%) & Duration & \multicolumn{1}{c}{\begin{tabular}[c]{@{}c@{}}Required\\ Calibration\end{tabular}} \\ \hline \hline
\multicolumn{1}{c}{Medtronic} & Enlite Sensor* & - & 6 days & \multicolumn{1}{c}{-} \\
\multicolumn{1}{c}{} & \begin{tabular}[c]{@{}c@{}}Enlite with Gaurdian\\ 2 link transmitter\end{tabular} & 11 & 6 days & \multicolumn{1}{c}{2/day} \\
\multicolumn{1}{c}{} & \begin{tabular}[c]{@{}c@{}}Enlite with MiniLink\\  transmitter\end{tabular} & 13.6 & 6 days & \multicolumn{1}{c}{2/day} \\
\multicolumn{1}{c}{} & \begin{tabular}[c]{@{}c@{}}Enlite with Guardian \\ connect transmitter\end{tabular} & 11 & 6 days & \multicolumn{1}{c}{2/day} \\
\multicolumn{1}{c}{} & Guardian Sensor 3* & - & 7 days & \multicolumn{1}{c}{-} \\
\multicolumn{1}{c}{} & \begin{tabular}[c]{@{}c@{}}Guardian Sensor 3 \\ with Guardian link 3\\ transmitter\end{tabular} & 10.6 & 7 days & \multicolumn{1}{c}{2/day} \\
\multicolumn{1}{c}{} & \begin{tabular}[c]{@{}c@{}}Guardian connect \\ system\end{tabular} & \begin{tabular}[c]{@{}c@{}}10.6 (abdomen)\\ 9.0 (arm)\end{tabular} & 7 days & \multicolumn{1}{c}{2/day} \\
\multicolumn{1}{c}{Dexcom} & G4 & 13 & 7 days & \multicolumn{1}{c}{2/day} \\
\multicolumn{1}{c}{} & G5 & 9 & 7 days & \multicolumn{1}{c}{2/day} \\
\multicolumn{1}{c}{} & G6 & 9.8 & 10 days & \multicolumn{1}{c}{none} \\
\multicolumn{1}{c}{Abbot} & FreeStyle Libre & 11.4 & 14 days & \multicolumn{1}{c}{none} \\ \hline
\multicolumn{5}{l}{\begin{tabular}[c]{@{}l@{}}*Sensing elements may limit the duration CGM systems and affect their accuracy but do not alone give \\ the mean absolute relative difference (MARD) or calibration schedule.\end{tabular}}
\end{tabular}
\end{table*}

This consistent availability of current data has led to several important features that are impossible with traditional SMBG. First, many rtCGM systems allow users to set alarms that indicate or even predict the onset of hypoglycemia, allowing diabetics to increase their carbohydrate/glucose intake and avoid hypoglycemia, which is especially useful for those suffering from hypoglycemia unawareness\textcolor{blue}{\cite{adofsson2018}}. Second, many rtCGM systems allow users to automatically share data/trends with caregivers and physicians, having important implications for young children and their parents\textcolor{blue}{\cite{adofsson2018}}, as well as elderly and disabled individuals who are living alone. Finally, rtCGM systems not only provide improved predictive measures for users through real-time measurements, measurement histories, and predicted glycemic trajectories, many of these systems can be interfaced with portable insulin infusion systems to automatically suspend insulin dosing when near or approaching hypoglycemia\textcolor{blue}{\cite{adofsson2018, wood2018}}. Combinations of rtCGM and insulin infusion systems, known as sensor augmented pump (SAP) therapy, are further enabling improved overnight glycemic control, and represent a step toward wide-spread acceptance of closed-loop insulin therapy systems.

More recently, a new class of glucose monitor known as the "flash" or "intermittently sensed" glucose monitor (isCGM) have become available that provide immediate, on demand glucose measurements (lagging true blood glucose concentrations by about 5 minutes) in a minimally invasive package worn continuously for 14 days without calibration (the sensing element is replaced after 14 days without requiring any SMBG calibrations)\textcolor{blue}{\cite{leelarathna2018,wood2018}}. While this device may not be suitable for those with hypoglycemic unawareness (developed by repeated hypoglycemic excursions) as it does not currently provide an autonomous alert, various studies have demonstrated improved time-in-range and greatly reduced time in hypoglycemia in both type 1 and type 2 hypoglycemic aware populations previously using finger-stick SMBG. Small observational studies indicate that these results may follow from the rapid and pain-free nature of blood glucose self-monitoring with these devices, which allows users to follow even stringent (>10 times per day) testing guidelines without experiencing discomfort\textcolor{blue}{\cite{leelarathna2018}}. This research has also reported behavioral changes in patients, including better adherence to monitoring timelines, which may produce additional societal benefits, as unchecked low blood sugar is associated with otherwise avoidable injuries and a increased risk of car accidents\textcolor{blue}{\cite{houlden2018}}. Due to time lag and measurement inaccuracy -- possibly exceeding 15\% mean absolute relative difference (MARD) from matched reference measurements -- users of these devices must still perform traditional SMBG tests when hypoglycemia is expected, measurements vary rapidly or do not corroborate symptoms, and when required by law\textcolor{blue}{\cite{leelarathna2018}}. While these flash monitors are available and gaining significant interest, there remain significant concerns in areas including sensor placement, failure modes, user insulin stacking due to time lags, and measurement variations over the duration of the 14-day sensor life-span.

Studies on CGM devices have consistently indicated that, compared to control groups using SMBG, insulin treated diabetics using CGM systems show improved HbA1c as well as significantly decreased frequency and severity of hypoglycemia and hyperglycemia, with rtCGM demonstrating superior outcomes to isCGM\textcolor{blue}{\cite{6association2018,adofsson2018, leelarathna2018, wood2018}}.  Despite their benefits, modern CGM sensors suffer from delays due to glucose transport from plasma to the interstitial fluid, may be subject to sensor interference from mechanical\textcolor{blue}{\cite{castle2010}} and pharmaceutical (e.g., acetaminophen\textcolor{blue}{\cite{basu2017}}) sources, and are known to demonstrate reduced accuracy during the first day of use and at low glucose concentrations (i.e., near hypoglycemia)\textcolor{blue}{\cite{adofsson2018, castle2010}}. With further research, these issues may be resolved in the near future.

\subsubsection{Non-Invasive, Wearable, and Implantable Monitoring}
During the past decade, there has been increased academic and commercial interest in the development of non-invasive and continuous monitoring of blood glucose levels through wearable technologies. Unlike traditional monitors, many of these sensors do not measure blood glucose directly, but attempt to leverage alternative fluids such as urine, sweat, saliva, breath, ocular fluid, and interstitial fluid\textcolor{blue}{\cite{bruen2017}}. While some of these devices are promising, with many designed to be discrete and comfortable, measurements taken from these alternative physiological fluids tend to be less reliable than traditional methods, and any such technology will surely require rigorous testing and clinical studies before any statements can be made regarding safety and accuracy\textcolor{blue}{\cite{kim2018}}.

\subsubsection{Non-Insulin Therapy}
While not generally suitable for T1DM, non-insulin oral antihyperglycemic agents are common in treating T2DM. In fact, the first-line pharmaceutical therapy for T2DM is a non-insulin antihyperglycemic called metformin, which reduces hepatic glucose production (i.e., in the liver) and is often highly effective at reducing HbA1c and associated cardiovascular risks, without raising significant concerns of hypoglycemia\textcolor{blue}{\cite{8association2018}}. According to ADA guidelines, metformin is frequently prescribed as part of a dual or triple pharmaceutical regimen (combination therapy) when metformin alone does not provide sufficient reductions in HbA1c -- used in this case as a primary indicator of glycemic control. These additional treatments may be oral agents such as sulfonylureas (increases insulin secretion), DPP-4 inhibitors (increases glucose dependent insulin secretion and decreases glucose dependent glucagon secretion), and Thiazolidinediones (increases insulin sensitivity), but may also include subcutaneous injections of insulin or GLP-1 receptor agonists, which act on multiple levels\textcolor{blue}{\cite{8association2018}}.

Considering the underlying lifestyle factors (e.g., sedentary behavior, obesity, and diets that are high in fats and sugars\textcolor{blue}{\cite{chatterjee2017}}) that are known to promote the development of T2DM, prevention and treatment through lifestyle modification is a significant area of study\textcolor{blue}{\cite{chatterjee2017, tomaro2017, 4associaion2018}}. Intensive lifestyle modifications have been shown to improve HbA1c, and are believed to be associated with reduced microcardiovascular and macrocardiovascular complications as well as lower rates of diabetes related mortality in T2DM. Similarly, there is evidence to support the benefits of nutritional balance in T1DM; however, the direct evidence for reduction of diabetes related ailments (beyond general health and fitness) associated with physical activity and obesity are much better supported for T2DM. In fact, due to the increased glycemic variability of individuals with T1DM, the ADA recommends physicians take care in prescribing specific exercises and durations for individuals when using physical activity as a means of glycemic control\textcolor{blue}{\cite{4associaion2018}}.

While these non-insulin therapies are often effective at slowing the progression of T2DM, in some cases leading to remission of the disorder, most individuals with T2DM will often require insulin therapy at later stages of treatment, even with careful adherence to therapy guidelines and healthy lifestyle choices\textcolor{blue}{\cite{8association2018}}.

\subsubsection{Multiple Daily Insulin Injections}
Recalling that severe insulin deficiency is the hallmark of T1DM related $\beta$-cell destruction, it is not surprising that primary treatment for T1DM is insulin therapy\textcolor{blue}{\cite{8association2018}}. The most common method of insulin therapy consists of the regular periodic injection of basal (baseline) insulin multiple times a day -- known as multiple daily insulin injections (MDI)-- supported by additional insulin doses (boluses) and oral glucose or glucagon as needed to maintain normoglycemic conditions (e.g., at meal times). While several different practical guidelines exist for calculating the required basal and bolus insulin doses, dosage is always described in terms of units of insulin or units/day (based on weight as units/kg/day). 

Recent pharmaceutical advances have produced a variety of injectable insulins that may be categorized as either synthetic human insulin or modified insulin analogs. Referring to \textcolor{blue}{{Table }\ref{table4}}\textcolor{blue}{\cite{8association2018, moore2018}}, variations in these insulin formulations have led to markedly different properties, and so it is common to classify insulins according to their concentration in units/mL (e.g., an insulin specified as U-\# contains \# units/mL), the time it takes for the insulin to begin reducing blood glucose levels (onset time), the time it takes for the insulin action to reach its peak effectiveness (peak time), and the length of time that the formulation will remain effective after injection (duration)\textcolor{blue}{\cite{moore2018}}. When prescribing or taking insulin, each of these factors comes into play and (in an MDI scheme) make some forms better suited as basal insulin and others as mealtime boluses\textcolor{blue}{\cite{8association2018}}. Thus, even within insulin therapy, there is now an array of options that allow for personalization and situational evaluation of treatments.

\begin{table*}[]
\centering
\caption{Characteristics of various commercially available insulin formulations.}
\label{table4}
\begin{tabular}{ccccc}
\hline \hline
Classification & Formulation & Onset Time & Peak Time & Duration \\ \hline \hline
Rapid acting & Aspart & \textless{}15 min & 30-90 min & 3-5 hours \\
 & Lispro & \textless{}15 min & 30-90 min & 3-5 hours \\
 & Glulisine & \textless{}15 min & 30-90 min & 3-5 hours \\
Short acting & Regular Human Insulin & \textless{}1 hour & 2-4 hours & 4-8 hours \\
Intermediate acting & Human NPH & 1-2 hours & 4-10 hours & 10-18 hours \\
Long acting (basal) & Glargine & 1-2 hours & - & 24 hours \\
 & Detemir & 1-2 hours & - & 24 hours \\
Ultra-long acting (basal) & Degludec & 30-90 min & - & 42 hours \\ \hline
\end{tabular}
\end{table*}

Note, the use of insulin is not limited to T1DM, and there are, of course, many factors affecting the selection, efficacy, and safety of different insulin formulations and treatment regimens. In addition to single insulin formulations, some injections are developed as a mix of rapid acting (i.e., quick onset and peak times with short duration) and long acting (i.e., delayed onset time, low or no peak, and long duration) insulins to provide both basal and bolus action from a single injection, thereby reducing the number of injections required per day\textcolor{blue}{\cite{8association2018, moore2018}}.

In some cases, an analog of the pancreatic $\beta$-cell hormone amylin may be used to augment insulin therapy\textcolor{blue}{\cite{8association2018}}. However, addition of such agents may increase treatment costs significantly. Further, in certain cases, it may be beneficial to perform pancreas or islet transplantation, in place of insulin therapy.

\subsubsection{Continuous Subcutaneous Insulin Infusion}
Initially developed over 40 years ago\textcolor{blue}{\cite{pickup1978}}, continuous subcutaneous insulin infusion (CSII) systems are portable pump therapy devices that are generally constructed as a combination of an onboard insulin reservoir, an electromechanical infusion pump, and an infusion apparatus (tubing and cannula)\textcolor{blue}{\cite{pickup1978, pozzilli2016}}. According to various studies, these systems can be operated using synthetic human insulin or rapid-acting insulin analogs (RAIA), with RAIA providing superior performance to synthetic human insulin\textcolor{blue}{\cite{pozzilli2016}}. In most cases, CSII uses the same basal dosage as MDI, with the basal insulin dosage applied more consistently over the day in CSII\textcolor{blue}{\cite{pozzilli2016}}.

While pragmatic evidence supporting substantial improvements to patient outcomes achieved by switch from MDI to CSII is minimal or frequently contradicted\textcolor{blue}{\cite{8association2018, pozzilli2016}}, there is evidence that switching to CSII is beneficial for diabetics with poor glycemic control under MDI\textcolor{blue}{\cite{pozzilli2016}}, and CSII is beginning to see adoption as a recommended first-line therapy for some type 1 populations (e.g., adolescents)\textcolor{blue}{\cite{8association2018, pozzilli2016}}. However, when CSII is used in conjunction with rtCGM for sensor augmented pump therapy, the benefits are more pronounced. Connected systems of monitoring and treatment devices, like SAP, provide therapeutic opportunities that cannot be obtained using MDI, such as overnight glycemic regulation. As an extension of this concept, the ability to continuously monitor and control insulin concentrations (whether by CSII or other novel future technology) is paramount to the development of closed-loop insulin therapies. Like SAP, closed-loop therapy systems are designed to utilize the features of continuous infusion that cannot be matched by MDI, with or without CGM, to achieve tighter glycemic control than is possible by any current treatment.

As with all therapy devices, CSII must be appropriately used and maintained, as device performance depends on proper operation (e.g., timely replacement of consumables) to avoid failure modes such as clogged or impeded infusion pathways, which can lead to insulin deficiency and hyperglycemia\textcolor{blue}{\cite{pozzilli2016}}.

\subsection{Modeling for Analysis and Control of Diabetes}
Building on nearly 2 centuries of research and modeling studies focused on the pancreatic endocrine system, the recent worldwide interest in diabetes has instigated a surge in the development of glucoregulatory models and simulation platforms, primarily aimed at T1DM\textcolor{blue}{\cite{oviedo2017, fritzen2018, colmegna2014, cobelli2014, ajmera2013}} and, to a lesser extent, T2DM\textcolor{blue}{\cite{fritzen2018, cobelli2014, ajmera2013}}. In fact, according to several reviews\textcolor{blue}{\cite{oviedo2017, ajmera2013}}, there are well over 100 different mathematical models describing, reproducing, and/or predicting various aspects of the blood glucose regulatory dynamics from cellular mechanisms and neurological signaling to long-term outcomes, disease progression, and risks of complications\textcolor{blue}{\cite{ajmera2013}}. The mathematical descriptions used in these dynamic models are diverse and include ordinary, partial, and stochastic differential equations, which often include time-delays, and are frequently treated with a degree of modularity -- as new models may be established as composites of subsystems from multiple predecessors\textcolor{blue}{\cite{ajmera2013}}.

Among these models, some, such as the mechanistic models that describe the $Ca^{2+}$ and $K^+$ ion channel mediated release of insulin from pancreatic $\beta$-cells, have solid footing in the first principles of electrochemistry and physics. However, the inclusion off such models tends to add unwarranted complexity as the pathophysiological and etiological relevance of these states are not well-understood, while producing negligible variations in the observable input-output behaviors that are presently more relevant to healthcare outcomes. Thus, in the development and analysis of systems for the closed-loop treatment of diabetes, models are generally categorized according to their level of detail and may be considered as either (i) reduced complexity control-oriented models -- for controller synthesis methods\textcolor{blue}{\cite{oviedo2017}}, or (ii) high-fidelity models for analysis and validation\textcolor{blue}{\cite{colmegna2014, cobelli2014}}. In either case, these models attempt to replicate the glucoregulatory dynamics of diabetic patients. However, while the former is concerned with representing accurate input-output relationships, in the latter category it is also important that models are structured such that variables and parameters maintain physiological significance and produce physiologically viable results at internal states. Thus, models of this second kind, can provide additional insights into substance and organ interactions, and have become known as virtual patient models -- as they are generally used as simulated analogs of real patients.

While early models of virtual patients utilized nominal or average parameter values, obtained from population studies, their more recent use in developing closed-loop therapy systems has led to a need for individualized model behaviors that closely match the severity and mode of dysfunction in specific patients. Accordingly, while some parameters may still be estimated by nominal values from population data\textcolor{blue}{\cite{wilinska2008}}, others have been identified from measurements on specific patients. For several models, this method has been used to generate arrays of virtual patients -- called cohorts -- that can be used to represent the interpatient-variability that may exist within a particular pathophysiology. When used for internal model type control schemes, the ability or inability to identify these individualized parameters from readily available data frequently dictates whether a simulation model is useful for real-time control. The availability of such data may also dictate the environments in which it is possible to implement certain control schemes. This is of particular concern when adaptation/tuning or online identification is required, as some diagnostic tools may only be accessible, or practical, in a clinical setting.

While many physiological models have been studied, several, such as the Bergman minimal model\textcolor{blue}{\cite{bergman1979, bergman2005}}, UVa/Padova simulator\textcolor{blue}{\cite{man2014}}, and Cambridge model\textcolor{blue}{\cite{wilinska2010}} represent the overwhelming majority of simulation studies using virtual patients.

\subsubsection{Minimal Glucoregulatory Models}
While any glucoregulatory or subsystem model that depicts the essential dynamics while using the fewest possible states may be considered minimal\textcolor{blue}{\cite{cobelli2014, cobelli2009}}, the Bergman minimal model (BMM) is by-far the most well-known and frequently adapted. Despite simplifications and neglected dynamics, the parameters and variables in minimal models must both convey clear physiological significance and be readily identified from measurable clinical data\textcolor{blue}{\cite{cobelli2009}}. Notice, this requisite simplicity and identifiability are often shared with the kinds of models that are sought in controller synthesis. Hence, in addition to its primary analytical purpose, the BMM has also been adapted for use in the implementation of closed-loop blood glucose control strategies.

\begin{figure}
\centering
\includegraphics[width=80mm,scale=0.5]{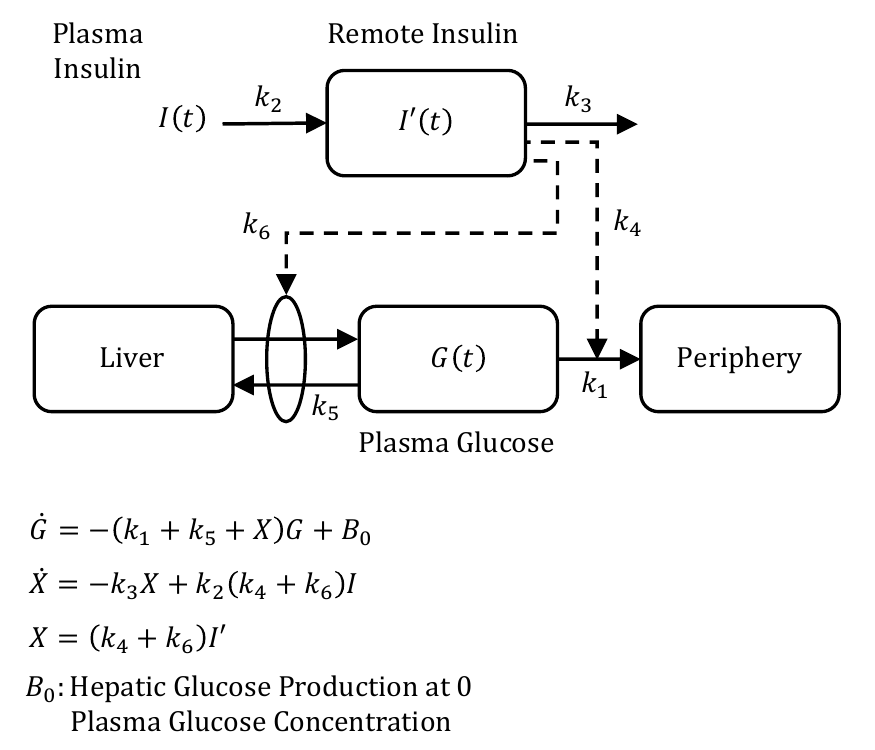}
\caption{Original Bergman minimal model of blood glucose/insulin dynamics.}
\label{fig4}
\end{figure}

To understand the importance of this model, it is interesting to consider its origin. Initially developed as a means to quantify the insulin sensitivity of individual patients\textcolor{blue}{\cite{bergman1979}}, the BMM was the result of early experiments to identify the structure of blood glucose-insulin interactions by obtaining frequent measurements of both plasma glucose and plasma insulin concentrations through the duration of a 1 hour intravenous glucose tolerance test (IVGTT) applied following a single bolus injection of glucose. Blood samples were taken first before injection, to obtain basal levels, and following injection at specific intervals (i.e., rapidly at first but reducing in frequency over the hour). The results of these blood tests were fitted by nonlinear least squares to various pre-established nonlinear ODE model candidates, which were evaluated according to the number and physiological meaning of model parameters, as well as the accuracy of the parameter estimates and each model's overall goodness-of-fit. Two of 7 possible models (one linear, the other nonlinear) passed this initial testing, both of which included insulin-dependent glucose utilization - 3 of the models were insulin-independent and were rejected for their poor prediction of the glucose kinetics. Ultimately, the nonlinear model was found to reproduce known physiological responses with higher fidelity. Thus, the BMM was selected as the 3-compartment nonlinear system of ODEs with 5 parameters shown schematically in \textcolor{blue}{{Fig. }\ref{fig4}}\textcolor{blue}{\cite{bergman1979}}. Notice, this model has 1 glucose compartment and 2 insulin compartments (one plasma compartment and one remote compartment).

Since the time of this original model, many adaptations have been made to account for additional dynamics, such as the connection of a second glucose compartment, the addition of glucagon feedback, and the replacement of IVGTT by oral glucose tolerance testing (OGTT) to obtain a more physiological measure of insulin sensitivity\textcolor{blue}{\cite{caumo2000}}. However, even with these augmentations, the BMM must be further adjusted to be useful for closed-loop control. For example, one prevalent control-oriented model intended for T1DM control that utilizes the BMM is detailed in ref.\textcolor{blue}{\cite{fabietti2006}}. This model augments the original BMM with insulin-dependent glucose dynamics, renal glucose clearance, meal absorption, external insulin infusion and subcutaneous transport, circadian variations, and an adjusted hepatic balance.

While many of the equations in this model are linear first-order ODEs (e.g., all of the insulin and the meal absorption dynamics), there are several nonlinearities including circadian variations in insulin sensitivity in the interstitial glucose compartment and hard nonlinearities in the endogenous blood glucose production dynamics. Further, any effects caused by glucagon or physical activity have been excluded. Note, the original model also includes intravenous routes for insulin and glucose, which have been omitted above.

Due to its questionable accuracy and remaining complexity, the BMM is not always the first choice for model-based closed-loop control. However, it is frequently adapted for this purpose, and has provided an important clinical tool for establishing an individual's level of $\beta$-cell function through the insulin sensitivity index\textcolor{blue}{\cite{bergman2005,schiavon2014,caumo2000}}. Like the BMM, other models used in controller design for closed-loop insulin therapy generally trade physiological complexity for model structure, identifiable parameters, and appropriate input-output behaviors -- typically by linearization. This trade-off with traditional physiological models is -- at least partially -- responsible for the increasing role of data-driven models for personalized control of diabetes\textcolor{blue}{\cite{cinar2018turksoy-modeling}}.

\subsubsection{UVa/Padova Models and Simulations}
Since the time of the BMM, several detailed models have been developed, or modified, to analyze the glucoregulatory behavior of diabetic patients. However, to date, there is only one family of models that has been approved by the United States Food and Drug Administration (FDA) as a replacement for pre-clinical trials in the validation of a closed-loop control design. These are the so-called UVa/Padova models\textcolor{blue}{\cite{kovatchev2009silico}} developed primarily between collaborators at the University of Virginia in the USA and the University of Padova in Italy.

In contrast to the simplicity of minimal models, the UVa/Padova model is a complex and hierarchically organized model that integrates high and low-level details into submodels of organ systems and metabolic pathways, to obtain a comprehensive, physiological simulation of human patients\textcolor{blue}{\cite{dallaman2007}}. This model, sometimes referred to as a maximal model due to its maximal level of detail\textcolor{blue}{\cite{cobelli2014}}, contains over 30 parameters\textcolor{blue}{\cite{wilinska2008}} that must be identified to match a specific patient and is generally too complex to produce desirable controllers without simplification. Conversely, this complex model can accurately reproduce many aspects of the glucoregulatory dynamics of healthy and diabetic individuals, which has led to the development of individualized patient simulation models known as in silico subjects. Using high quality clinical data, researchers have developed a large collection of patient models representing numerous individuals from each of various demographics, enabling studies on the entire population as well as particular cohorts of interest\textcolor{blue}{\cite{dallaman2007, kovatchev2009silico, man2014}}. Such detailed high-fidelity modeling is the foundation for an emerging symbiotic relationship, wherein simulations are not only useful for analyzing the effects of treatments on known (patho)physiology, but also for testing hypotheses regarding the (patho)physiology itself and models thereof \textcolor{blue}{\cite{cobelli2014}}.

Presently, a distributed version of the UVa/Padova type 1 diabetes metabolic simulator (T1DMS) for use in MATLAB/Simulink can be obtained by academic and commercial entities through The Epsilon Group\textcolor{blue}{\cite{theepsilongroup2017}}. In addition to its early capabilities, the current version of the UVa/Padova simulator includes dynamic models of the bi-hormonal (i.e., insulin and glucagon) glucoregulatory dynamics, an improved patient cohort, and updated depiction of regulatory feedback during hypoglycemia\textcolor{blue}{\cite{man2014,theepsilongroup2017}}. As of 2018, researchers from UVa/Padova have announced further updates to their simulation platform that introduce, among various other extensions and improvements, a set of time-varying patient models to account for diurnal variations in patient dynamics associated with overnight rise in blood glucose (i.e., the dawn phenomena) and patterns of varying insulin sensitivity at morning, afternoon, and evening meals\textcolor{blue}{\cite{visentin2018}}. While the previous updates\textcolor{blue}{\cite{man2014}} are FDA approved, approval for the more recent time-varying simulator\textcolor{blue}{\cite{visentin2018}} is still pending at the time of writing this article.

The development of the UVa/Padova model is well documented, and interested readers are referred to the relevant literature for extensive details on the underlying model, in silico population, recent updates, and application to the validation of control algorithms. Note, despite its complexity, several closed-loop control algorithms have been designed using reduced order approximations of these models\textcolor{blue}{\cite{colmegna2014, magni2009, incremona2018}}.

\subsubsection{Cambridge (Hovorka) Models}
Similar to the UVa/Padova model, the Cambridge model\textcolor{blue}{\cite{hovorka2004,wilinska2014, wilinska2010}} has frequently been used for in silico validation studies of closed-loop insulin systems and has its own MATLAB/Simulink based simulation environment and cohort of 18 clinically validated virtual subjects\textcolor{blue}{\cite{colmegna2014, wilinska2010}}. However, unlike the UVa/Padova simulator, results from this software are not FDA approved substitutes for pre-clinical trials. While this may lead to several additional steps prior to clinical studies, the Cambridge model has significantly fewer mathematical states and parameters than the UVa/Padova model\textcolor{blue}{\cite{wilinska2008}}, while maintaining physiological significance and modeling diurnal variations, which has allowed some researchers to modify and adapt it more readily for their specific needs\textcolor{blue}{\cite{boiroux2018}}.

Despite its complexity, the Cambridge model is frequently used in internal model or model predictive control schemes\textcolor{blue}{\cite{oviedo2017, hovorka2004}} and has been the basis for several clinically tested investigatory closed-loop devices\textcolor{blue}{\cite{wilinska2014}}. 

\subsubsection{Alternative and Data-Driven Models}
In addition to the more analytical and physiologically guided virtual patient models, numerous simplified control-oriented models have been proposed that attempt to leverage various modeling paradigms. Of particular interest, the use of linear parameter varying (LPV) type models\textcolor{blue}{\cite{colmegna2018linear}} has been shown to provide a simplified means to accurately reproduce the complex time-varying and nonlinear dynamics of maximal diabetes models. Using appropriate techniques, reduction of maximal models to LPV models may provide a superior model-based and control-oriented approach to blood glucose controller synthesis.

Noting that even the best physiological blood-glucose models are, at present, phenomenological and difficult to individualize for optimal real-time control, many researchers have instead adopted data-driven modeling paradigms such as autoregressive exogenous (ARX) and autoregressive moving average exogenous (ARMAX) models. Provided with a predefined model structure, these techniques utilize online system identification to obtain and, potentially, update models to achieve a best approximation of the input-output behaviors of the glucose-insulin system based on a given performance index\textcolor{blue}{\cite{cinar2018turksoy-modeling}}. Like LPV models, it is unlikely that data-driven models will maintain any detailed physiological information about a given patient, but instead provide a useful foundation for implementing established controller synthesis approaches.

\subsubsection{Models of Therapeutic Agents and CGM Sensors}
Following previous discussions on sensors and treatment approaches, it is natural that these components will contribute dynamics to the overall system. Models for CGM sensor accuracy\textcolor{blue}{\cite{huyett2018}} and insulin formulations and kinetics\textcolor{blue}{\cite{nucci200,schiavon2018}} have been developed by various investigators. These models enable simulations that provide a complete picture of blood glucose control under specific therapy strategies and allow the investigation of closed-loop systems while taking sensor and actuator performance limitations into account.

\subsection{Automated Therapy for Diabetes}
Attempts to develop fully closed-loop insulin delivery systems date back more than 50 years\textcolor{blue}{\cite{kadish1963}}, and many companies and academic researchers have contributed significantly in these developments\textcolor{blue}{\cite{cobelli2011}}. While early work focused on sugar and insulin infusion by intravenous methods, the success of these systems along with the development of both CSII pumps and CGM sensors has shifted the modern focus toward minimally invasive therapies applied via subcutaneous routes\textcolor{blue}{\cite{hovorka2011}}. Within the last 5 years, there has been -- and continues to be -- rapid progress toward the commercialization of fully automated insulin therapy systems\textcolor{blue}{\cite{cinar2018, turksoy2017}}, frequently referred to as artificial $\beta$-cell, artificial pancreas (AP), or automated insulin delivery (AID) systems. The primary goal of these automated therapy devices is to alleviate the burden and risks associated with multiple daily injections from T1DM patients and caregivers by reducing diabetes treatment to a combination of proper lifestyle modifications (perhaps even less stringent ones once on AP therapy) and device maintenance (to ensure proper AP function and patient safety). Although most current work is focused on T1DM, there are parallel efforts to investigate the use of these systems for specific cases such as insulin dependent T2DM\textcolor{blue}{\cite{kumareswaran2014, bally2018}}, T1DM during pregnancy\textcolor{blue}{\cite{murphy2011, stewart2018}} (but not yet GDM as far as we are aware), and blood glucose regulation for patients in intensive care units (ICUs)\textcolor{blue}{\cite{preiser2016}}. 

\subsubsection{Challenges to Automatic Control}
While the virtual patient models discussed previously can reproduce many aspects of normal and diabetic blood glucose regulation, there remain several challenges that prevent these models from fully predicting the dynamics of real patients in real-time and limit the accuracy of designs for real-world control performance. The most notable among these challenges are the blood glucose disturbances caused by meals and exercise and the intrapersonal variations -- particularly in insulin sensitivity and glucose production -- that may occur within a single patient over their course of treatment.

While many meal absorption and gastric emptying models have been proposed to describe how meal contents -- generally carbohydrates -- act as sources of exogenous glucose\textcolor{blue}{\cite{elfathi2018}}, the timing, duration, and nutritional content of meals and snacks are generally unknown for patients in an uncontrolled, or free living, environment. As a substitute for such detailed information, many current algorithms resort to manual meal announcement strategies, through which patients or caregivers can provide nutrition and schedule information to the controller regarding an upcoming meal. However, even with meal announcement and approximate nutritional content, there may still be unpredicted digestive variations. Thus, some schemes -- with or without meal announcement -- implement meal detection algorithms that identify when a meal has occurred and may also estimate carbohydrate content\textcolor{blue}{\cite{elfathi2018, cinar2018turksoy-multivariable, ramkissoon2018, cameron2012, colmegna2018automatic}}. At the current time, very few controllers can provide tight glycemic control when subjected to unannounced meals as it generally takes around 70 min after administration for peak insulin action to occur\textcolor{blue}{\cite{colmegna2018automatic}}. This delay in insulin action can provide a window for hyperglycemia to occur following even a well identified meal and may result in either hyperglycemia or hypoglycemia if meal properties are estimated incorrectly.

Like meals, periods of physical activity are known to disturb the nominal behavior of the blood glucose regulatory system. However, while meals primarily act by regulating exogenous glucose, physical activity can simultaneously affect numerous aspects of a closed-loop therapy system including physiological requirements\textcolor{blue}{\cite{riddell2015, bondia2018, moser2018}}, internal bio-mechanical pressure differentials \cite{moser2018}, and CGM accuracy\textcolor{blue}{\cite{moser2018}}. In fact, it is well known that different types, intensities, and durations of physical activity may provoke contrasting physiological responses (e.g., light aerobic exercise can cause hypoglycemia and brief stints of intense combined aerobic-anaerobic exercise can cause prolonged hyperglycemia\textcolor{blue}{\cite{riddell2015, turksoy2017}}). Thus, it is not only important to detect physical activity, but also to obtain enough data to classify -- and perhaps continuously reclassify, track, or predict -- the kinds of activity or exercises a patient performs. Further complicating this problem, certain kinds of physical activity can increase the rate of insulin absorption from the subcutaneous compartment to the blood stream\textcolor{blue}{\cite{riddell2015, moser2018}} and the corresponding increase in insulin action can cause hypoglycemia without any infusion of additional insulin. As noted previously, insulin does not act immediately and will not reach peak effectiveness for about an hour; thus, even if insulin has been regulated correctly for the current level of physical activity, variations in physical activity can change the internal utilization of the existing insulin dosage and affect hypoglycemic excursions that cannot be corrected by a single-hormone insulin therapy system\textcolor{blue}{\cite{riddell2015}}. To counteract these effects, many researchers are considering more advanced control systems that may utilize dual-hormone (i.e., glucagon and insulin) infusion sets\textcolor{blue}{\cite{moser2018, peters2018dual}} or multiple sensors and algorithms to detect and characterize physical activity\textcolor{blue}{\cite{cinar2018turksoy-multivariable, riddell2015, yavelberg2018}}. Despite these promising developments, there are currently no accurate models for simulations of exercise in diabetics\textcolor{blue}{\cite{cinar2018turksoy-multivariable, bondia2018}}, and physical activity remains a difficult challenge for fully automated free-living diabetes therapy.

Additionally, the effects of many other environmental factors and lifestyle behaviors on localized and body-wide dynamics including glucose production, insulin secretion, and insulin sensitivity are still poorly understood. For example, recent studies have suggested that insulin sensitivity decreases significantly in both healthy\textcolor{blue}{\cite{buxton2010}} and diabetic individuals\textcolor{blue}{\cite{donga2010, farabi2016}} with reduced duration and quality of sleep. Further, the effects of circadian phasing, misalignment, and behaviors (i.e., timing of meals as hours after waking up) were recently shown to produce 15-20\% changes in insulin sensitivity and $\beta$-cell function in healthy adults\textcolor{blue}{\cite{qian2018}}. To the authors' knowledge, similar results have yet to be reported in diabetics with T1DM or T2DM. Combined with other influential factors (e.g., sensor and physiological interference from outside substances, user error in attachment and maintenance, and electromechanical faults\textcolor{blue}{\cite{bertachi2018}}), any model or controller that has been individualized for a specific patient will almost certainly be subject to significant time-varying uncertainty, depending on patient behaviors and operating environment.

While these challenges apply to most diabetics, the extension of AP systems to T2DM may benefit from updated models that include compartments and inputs reflecting the action of noninsulin medications such as metformin, sulfonylureas, and GLP-1 agonists or additional hormones such as amylin (e.g., by the amylin analog pramlintide), which could also benefit T1DM patients and researchers\textcolor{blue}{\cite{weinzimer2012}}. Considering the apparent importance of glucagon for avoidance of hypoglycemia during exercise and the potential benefits of amylin for meal-time blood glucose regulation, it is not surprising that there has been recent interest in the development of closed-loop insulin-glucagon-amylin therapy\textcolor{blue}{\cite{nih2011, nih2012}}. However, to the authors' knowledge, such complex hormone and combination therapy models are not yet common in closed-loop diabetes therapy devices but may provide significant avenues for system augmentation and a more physiological replication of $\beta$-cell function as the relevant technologies mature.

\subsubsection{Proportional-Integral-Derivative Control Methods}
Proportional-Integral-Derivative (PID) control is one of the most versatile and commonly applied methods of control for both linear and nonlinear systems and has been likened to the physiological behavior of the pancreatic $\beta$-cells\textcolor{blue}{\cite{steil2013, steil2004}}. Numerous variations on PID control have been studied in simulations of T1DM\textcolor{blue}{\cite{huyett2015, mohammadridha2018}} and clinical trials. Most of the recent clinical studies using PID for closed-loop insulin delivery are related to the development of the PID-based external physiological insulin delivery (ePID) strategy\textcolor{blue}{\cite{steil2004, palerm2011}} including the introduction\textcolor{blue}{\cite{palerm2011, steil2011}} and refinement\textcolor{blue}{\cite{ruiz2012}} of insulin-feedback (IFB) to a form near the one currently implemented in the commercially available Medtronic 670G hybrid closed-loop (HCL) therapy system\textcolor{blue}{\cite{debock2017}}. In addition to the PID-based closed-loop control of basal insulin, these systems provide feedforward control action by meal-time boluses in response to manually entered announced carbohydrate loads\textcolor{blue}{\cite{debock2017}}. Systems that implement this kind of manual, feedforward control are considered hybrid rather than fully closed-loop.

Another recent algorithm, known as the safety auxiliary feedback element (SAFE) has been developed that implements a two-step approach to augment therapy controllers with safety constraints in order to decouple the problems of achieving nominal control performance and abiding system limitations\textcolor{blue}{\cite{revert2013}}. This auxiliary algorithm runs in a rapidly sampled (i.e., much faster than the 1 sample/min in the current core control algorithms) outer-loop that uses sliding mode reference conditioning to shape the desired insulin infusion profile to avoid violating safety constraints without affecting the nominal performance of the main control scheme. The SAFE algorithm can be used with a variety of controllers and constraints of known, constant relative degree, and has been used to enhance the meal-time performance of standard PID\textcolor{blue}{\cite{revert2013}} and modified ePID algorithms\textcolor{blue}{\cite{fushimi2018}}.

In addition to these single-hormone systems, PID has also been implemented in several dual-hormone (i.e., glucagon and insulin) systems where it has been used as the control algorithm for glucagon micro-bolusing (in conjunction with a separate insulin control algorithm) as well as the control of both insulin and glucagon\textcolor{blue}{\cite{cinar2018turksoy-dual}}.

\subsubsection{Model Predictive Control Methods}
For many, model predictive control (MPC) has become the standard formulation for insulin delivery algorithms because of its intuitive treatment of constraints and ease of modification. Like the many implementations of PID variants, numerous MPC algorithms have been reported that attempt to leverage new or existing MPC techniques to improve the compatibility of these algorithms with the requirements and constraints of diabetes treatment\textcolor{blue}{\cite{cobelli2011}}. Among the more recent developments in artificial pancreas systems are the extended\textcolor{blue}{\cite{cameron2011}}, multiple model\textcolor{blue}{\cite{cameron2012, cobelli2011, forlenza2018}}, run-to-run tuning\textcolor{blue}{\cite{cobelli2011, magni2009run}}, zone\textcolor{blue}{\cite{bequette2013,grosman2010}} and periodic-zone\textcolor{blue}{\cite{gondhalekar2016}}, event-triggered\textcolor{blue}{\cite{chakrabarty2018}}, and velocity-weighted \& velocity-penalized\textcolor{blue}{\cite{gondhalekar2018}} MPC algorithms. Each method of MPC attempts to overcome one or more of the challenges discussed previously including practical/safety restrictions\textcolor{blue}{\cite{toffanin2013artificial}}, interpatient variability, intrapatient variability, and unannounced disturbances such as meals or exercise. Some of these methods have already become quite successful and provide important guides for researchers and engineers seeking to refine diabetes control strategies.\\

Safety Considerations:
Among the first challenges overcome by AP algorithms were the restrictions imposed by limitations on insulin infusion. In principle there are limitations on both (i) the rate of insulin infusion, for instance due to saturations in pump displacement and the inability to remove insulin from the bloodstream once injected, and (ii) the amount of insulin in a patient's bloodstream at any given time. Notice, however, that while these problems are analogous to issues present in other industries, and general tools like anti-windup are readily available, the management of insulin is safety critical and MPC provides a means to design control systems that not only acknowledge these limitations but anticipate and avoid them while providing optimal control action. In fact, even predictive infusion shut-off systems\textcolor{blue}{\cite{buckingham2009}}, which temporarily halt basal insulin therapy when continued treatment is predicted to cause hypoglycemia, without any further feedback capabilities have been shown to drastically improve system safety by decreasing the risk of overnight hypoglycemia. However, the effectiveness of these simple predictive shut-off algorithms can be diminished if the patient already has significant insulin in their subcutaneous or plasma compartments. This point is of particular concern when insulin is infused according to feedback control laws, which may frequently exceed basal rates. Thus, many recent systems use insulin action curves to implement estimates of the patient's "insulin-on-board" (IOB)\textcolor{blue}{\cite{ellingsen2009}} to account for delayed insulin effects as part of their predictive algorithm and apply infusion rate limits to avoid dangerously aggressive profiles when glucose concentrations are low.

Similar to IOB estimates, another recent and promising MPC algorithm uses a more accurate estimate of the patient's plasma insulin concentration (PIC) to perform online adaptation of dosage constraints for individual patients. This method has performed well in tests on the UVa/Padova simulator without using any meal announcement but has not yet gone to clinical trials\textcolor{blue}{\cite{hajizadeh2018}}.

While IOB and PIC estimates can be useful for minimizing unnecessary control actions, the zone MPC approach attempts to resolve this issue by replacing the usual desired blood glucose set-point with a larger region, or zone, of acceptable concentrations\textcolor{blue}{\cite{grosman2010}}. Thus, the objective of the controller is better aligned with the normal physiological regulation of blood glucose to the euglycemic range, rather than a single specific concentration, and control action is only applied inside this range if the predictive algorithm anticipates an excursion. When no excursion is predicted, the basal infusion rate is maintained, greatly reducing unwanted blood glucose variations and the risk of insulin induced hypoglycemia. Variations of the zone MPC approach include periodic zone-MPC control\textcolor{blue}{\cite{gondhalekar2016}}, which accounts for diurnal variations in activity and reduced hyperglycemic risks overnight, and zone MPC with adaptive cost functions\textcolor{blue}{\cite{gondhalekar2018, shi2018}}.

Considering the non-negative nature of blood glucose/insulin concentrations and infusion rates, the performance limitations of the diabetes control problem, without glucagon, has been likened to recent results regarding fundamental trade-offs limiting the achievable response in linear positive systems\textcolor{blue}{\cite{goodwin2015}}. Despite nonlinearities in the UVa/Padova simulator, this approach may provide a means to obtain (in the ideal numerical case) the control signal that would give a near ideal combination of safety and performance, which has been suggested as a potential benchmarking tool for simulation studies on AP systems. One recent MPC variant\textcolor{blue}{\cite{gondhalekar2018}} seeks to achieve this ideal response by applying (i) a "velocity-weighted" optimization wherein the cost of blood glucose deviations is a function of the rate of change of the measured blood glucose and (ii) a "velocity-penalization" term in the cost function that penalizes actions that would allow blood glucose to increase (i.e., have a positive rate of change) when blood glucose is tending toward hyperglycemia. In combination, the velocity-weighting and velocity-penalty MPC scheme acts aggressively against disturbances by switching on the velocity-penalty at the first signs of impending hyperglycemia, as evidenced by elevated blood glucose levels (above some threshold) with a positive rate of change, while avoiding hypoglycemia due to over corrections by effectively tapering the control action as the blood glucose rate of change decreases. Notice, this combined action is reminiscent of a nonlinear PD control with predictive capabilities and appears to provide some benefits of both MPC and PD control strategies.\\

Model Individualization:
Recalling that every patient affected by type 1 diabetes may respond to treatment differently, there is no single model or set of model parameters that can capture every patient's physiology with acceptable accuracy. One method to eliminate the uncertainty (and conservatism) associated with the presence of such interpatient variability is to generate patient specific models prior to system deployment or during an initial training period. To this end, research into individualization of models\textcolor{blue}{\cite{messori2016, garcia2018, fischer1987, messori2018}} has been an active subject for many years, and remains an open problem in diabetes control. Despite this, individualized models alone may not be sufficient to account for the intrapatient variations discussed previously, and adaptive control schemes may provide a more robust answer to both kinds of variability.\\

Adaptive Control:
Unlike the model individualization techniques discussed previously, adaptive online identification techniques can be used to overcome both interpatient and intrapatient variability. Clinical uses of online model identification with MPC date back to the transformative Advanced Insulin Infusion using a Control Loop (ADICOL) trials of 2004\textcolor{blue}{\cite{hovorka2004}}, where model parameters were updated at every control interval using Bayesian methods over a predetermined learning window. Other implementations of adaptive control in MPC include the run-to-run tuning methods in\textcolor{blue}{\cite{magni2009run}}, the cost function updating schemes in\textcolor{blue}{\cite{gondhalekar2018, shi2018}}, and several of the adaptive schemes discussed in\textcolor{blue}{\cite{turksoy2014}}.

Furthermore, generalized predictive control (GPC) algorithms\textcolor{blue}{\cite{bequette2013, cinar2018turksoy-various}} have been used successfully in both clinical and simulation studies. Like MPC, GPC strategies utilize an internal model to predict future states and determine the next immediate optimal control signal. However, in GPC, a simplified model is used for prediction, and this model is updated (i.e., adapted) at every control interval to minimize the difference between the predicted and actual responses. Recent implementations of GPC in diabetes are reported in\textcolor{blue}{\cite{turksoy2014, turksoy2018}} using recursive time-series models and in\textcolor{blue}{\cite{hajizadeh2018}} using recursive predictor-based subspace identification.

\subsubsection{The Control to Range Method}
Similar to the two-step approach taken by the SAFE algorithm\textcolor{blue}{\cite{revert2013}}, the control-to-range (CTR) method was proposed as a modular two-element strategy consisting of (i) a range correction module (RCM) that operates at a longer (e.g., 1-15 minutes) sampling interval to maintain blood glucose within a specified target range and (ii) a safety supervisor module (SSM) that operates at a much higher sampling rate to predict and prevent dangerous hypoglycemic excursions\textcolor{blue}{\cite{kovatchev2009control, patek2012}}. Although this concept may be applied without regard for the control method used, the CTR strategy has developed around  the use of MPC-based RCM algorithms\textcolor{blue}{\cite{patek2012}}.

\subsubsection{Fuzzy-Logic Control Methods}
While PID and MPC have historically dominated the field of diabetes control, more recently, fuzzy logic controllers\textcolor{blue}{\cite{cinar2018turksoy-various, atlas2010, mauseth2013, grant2007}} have been quite successful in generating AP systems that closely mimic expert clinical decision making in the face of complex environmental factors and incomplete information. 

Among the most prominent fuzzy AP systems used in clinical trials is the MD-Logic system\textcolor{blue}{\cite{atlas2010, nimri2017}}, which includes information on the patient's clinical physiology and therapy guidelines obtained from a set of training data to provide fully automated individualized care (without meal announcement) using a combination of control to range and control to target (standard set-point control) methodologies. Treatment decisions are made using past, present, and projected future data (based on trends), forced to abide safety constraints similar to those discussed previously, and applied according to a variable basal infusion plus bolus strategy.

Alternatively, another clinically tested fully closed-loop fuzzy-logic controller (without meal announcement) known as the dose safety controller (DSC)\textcolor{blue}{\cite{mauseth2013}} eliminates the basal insulin infusion entirely and instead applies a micro-bolus strategy based on quantized sets of blood glucose level, rate (i.e., first time derivative of blood glucose), and acceleration (i.e., second time derivative of blood glucose). To individualize treatment, each patient is given an initial "personalization factor" (PF), which scales the dosage associated with each level-rate-accelerate combination, and the PF is adjusted to provide more or less aggressive treatment depending on the occurrence of hypoglycemic events. More information on recent updates to the DSC is available online at\textcolor{blue}{\cite{Dose2016}}. 

\subsubsection{Run-to-Run Control}
Apart from the more well-known PID, MPC, and fuzzy-logic controllers, a somewhat unique class of industrial process control techniques known as run-to-run algorithms\textcolor{blue}{\cite{doyle2001}} have been used for a variety of tasks in diabetes care. Run-to-run methods utilize a measurement-based algorithm to improve the effectiveness of a controller between distinct iterations of a repeated process. For diabetes, this process may be the 24-hour/3-meal cycle\textcolor{blue}{\cite{doyle2001, zisser2005, owens2006}}, in which case the control law is updated before the start of each day to account for some intrapatient variability and improve the rejection of blood glucose disturbances due to meals. Alternatively, similar run-to-run methods have been used to adjust basal insulin infusion rates in simulated trials\textcolor{blue}{\cite{palerm2008run}} and to improve the accuracy of CGM measurements based on week-to-week recalibration\textcolor{blue}{\cite{lee2015}}.

\subsubsection{Alternative Control Methods}
While clinical studies have been dominated by PID, MPC, and Fuzzy-Logic controllers (with several using CTR), a variety of additional paradigms have been studied in academic simulations. These methods include artificial intelligence methods\textcolor{blue}{\cite{contreras2018}}, H$\infty$ control\textcolor{blue}{\cite{colmegna2014}}, linear parameter varying (LPV) control\textcolor{blue}{\cite{colmegna2018linear}}, and sliding mode control\textcolor{blue}{\cite{ahmad2017}}. However, recent clinical studies implementing these techniques are sparse, if available at all.

\subsection{Outlook for Automated Diabetes Treatment}
The insulin-based treatment of type 1 diabetes is currently in the midst of an international transformation from manual to automated care. While more complete evidence of the benefits and burdens of closed-loop therapy will become available as these systems are adopted, the momentum behind this transition is bolstered by the significant well-known improvements in safety and accuracy of closed-loop control over sensor augmented pump (SAP) therapy and manual multiple daily insulin injection (MDI) approaches during overnight blood glucose regulation. Furthermore, both hybrid closed-loop (e.g., requiring meal announcement) and fully closed-loop treatments have demonstrated promising additional improvements in more general simulated and clinical scenarios including ambulatory and free-living conditions.

\begin{figure}
\centering
\includegraphics[width=80mm,scale=0.5]{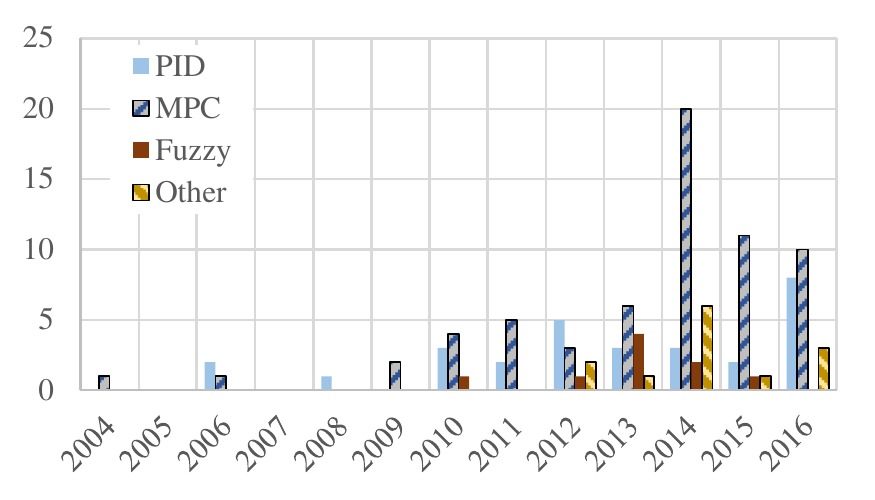}
\caption{Clinical AP trials from 2004-2016 grouped according to controller type.}
\label{fig5}
\end{figure}

Since the initiation of modern closed-loop insulin therapy clinical trials with the 2004 ADICOL studies\textcolor{blue}{\cite{hovorka2004}}, there have been over 100 publications concerning new clinical trials. A collection of available data regarding recent (2004-2016) clinical trials obtained from several studies\textcolor{blue}{\cite{bertachi2018, turksoy2017}} is shown in \textcolor{blue}{{Fig. }\ref{fig5}}. This data was cross-referenced against the AP clinical trial publication database\textcolor{blue}{\cite{thedoylegroup2015}} and shows a trend toward the use of MPC in clinical studies. Among the attractive features of MPC is its predictive adherence to complicated constraints, which cannot be offered by PID control or controllers that implement reference conditioning such as the SAFE algorithm. 

Despite interest in MPC, the first and only commercially available hybrid closed-loop system to-date (MiniMed 670G by Medtronic), released in 2016, uses a PID controller with insulin feedback (IFB)\textcolor{blue}{\cite{iturralde2017,foerster2017}}. While recent studies regarding the effectiveness and patient reception of this device are promising\textcolor{blue}{\cite{iturralde2017, boughton2018}}, many patients expected the device to have a higher degree of autonomy and found the system difficult to wear. However, as a larger variety of commercial systems and fully automated AP systems become available over the next few years\textcolor{blue}{\cite{trevitt2016}}, patients will be able to select between systems to find one that best meets their needs and expectations. With the widespread adoption of these technologies will come new safety concerns and hazards\textcolor{blue}{\cite{ramkissoon2017}} associated with the unfavorable conditions that may be encountered during continued, long-term, free-living use. Early action to preemptively address these obstacles may provide substantial benefits to patient safety, ergonomics, and perception of closed-loop artificial pancreas systems. Such actions might include adoption and development of standards relevant to commercial artificial pancreas systems, such as IEC 60601-1-10\textcolor{blue}{\cite{medical2007}}, and familiarization with relevant FDA guidance (e.g.,\textcolor{blue}{\cite{food2014infusion, food2012}}) and regulatory considerations arising from public discussions (e.g.\textcolor{blue}{\cite{parvinian2018}}).

In addition to commercial systems, opensource solutions to the treatment of type 1 diabetes, such as the open artificial pancreas system (OpenAPS)\textcolor{blue}{\cite{lewis2018what}}, have been gaining popularity in recent years. These systems are not regulated by the FDA and must be designed and constructed by diabetic patients themselves. While there is a growing repository of source code and community feedback for interested patients, and the results of the higher-end patient-driven studies\textcolor{blue}{\cite{lewis2016real,lewis2016poster}} are impressive, the safety and performance of opensource medical systems raise serious concerns. Implemented across a broad and inconsistent scale, these systems are vulnerable to algorithmic, communication, and electromechanical errors, which may arise when devices are combined and used in a manner inconsistent with their design. While some projects, such as OpenAPS provide a thorough reference guide and code that incorporates safety limitations, the essence of opensource implies that there is inevitably little preventing the end-user from tampering with these constraints and accidentally creating dangerous devices. This becomes an issue of further concern when considering the potential for proliferation of improper designs and the estimated 1/3 of OpenAPS users who are children operating systems designed by their parents\textcolor{blue}{\cite{lewis2018frequently}} (however this may not be representative of the larger opensource community). The availability of simple and reliable AP systems may help to significantly reduce the number of patients who choose to develop their own AP systems, and in fact the desire for such systems to come-to-market appears to underlie the goals of OpenAPS. Even with the availability of name-brand, FDA certified HCL and fully closed-loop systems, curiosity surrounding their operation -- as well as the desire to replicate and improve it -- will undoubtedly drive continued interest in opensource solutions.

Finally, it is important to keep in mind that poor glycemic control is a consequence of type 1 diabetes and not the underlying cause. As medical practice and technology progress, the artificial pancreas may be able to grow beyond insulin infusion and implement more comprehensive and physiologically appropriate therapies. Steps in this direction include the ongoing work to facilitate bi- and tri-hormonal systems augmented with glucagon and amylin. Furthermore, there may be benefits to the investigation of closed-loop therapies that are capable of affecting both glucose/insulin concentrations and insulin action (i.e., insulin sensitivity) -- perhaps through combination therapy with non-insulin treatments -- to provide personalized treatment options for the large and diverse population of diabetics with forms of diabetes other than T1DM. Of course, economic viability, education for clinicians and new users of AP devices, and research into a cure for diabetes may ultimately affect the impact and longevity of automated AP technology.

\section{Automated Anesthesia}\label{automated anesthesia}
The word anesthesia means the loss of sensation and can be defined as the lack of response to and recall of noxious stimuli. The first application of anesthesia was performed by Crawford Williamson Lang in 1842 using inhaled ether\textcolor{blue}{\cite{bibian2006}}. In clinical settings, anesthesia is induced by the administration of anesthetic drugs into a patient's body to achieve localized or general anesthetic effects. In the latter case, the term "general anesthesia" refers to a lack of movement across the whole body while also maintaining an unconscious state. In general anesthesia, patients may be given two types of anesthetic drugs: (i) inhaled anesthetics (gases or vapor drugs) and (ii) intravenous anesthetics (injection drugs). A concise comparison between these two methods is described well by Nascu:

"Intravenous medications are given directly into a vein and are commonly used to induce anesthesia, as induction is usually smoother and more rapid than that associated with most of the inhalational agents. Intravenous anesthetics administered as repeated bolus doses or by continuous infusion may also be used for maintenance of anesthesia, either alone or in combination with inhalational agents. An advantage of inhaled anesthetics is that measuring the difference between inhaled and exhaled concentrations allows an accurate estimation of plasma or brain drug uptake$\cdots$"\textcolor{blue}{\cite{nascu2017explicit}}.

\begin{figure}
\centering
\includegraphics[width=80mm,scale=0.5]{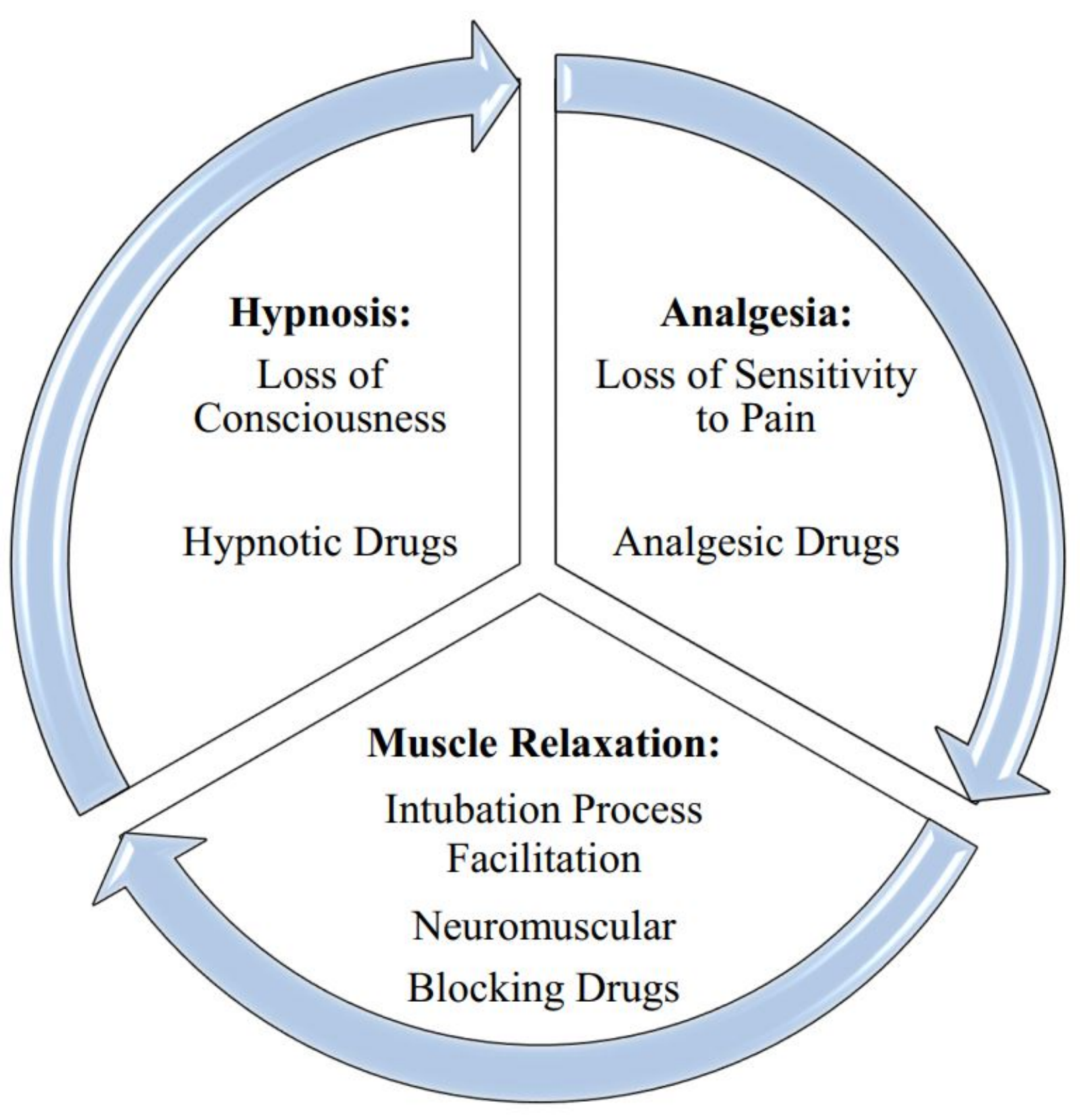}
\caption{The three functional components of clinical anesthesia: hypnosis, analgesia and immobility.}
\label{fig6}
\end{figure}

Generally, intravenous anesthesia drugs can be categorized into three different classes based on their physiological effects: (i) hypnotic drugs, (ii) analgesic drugs, and (iii) neuromuscular blocking (NMB) drugs. The first kind of anesthetic, called hypnotics, are used to numb the brain, keeping the patient unconscious during surgery. Propofol is currently the most common type of intravenous hypnotic drug due to its fast redistribution and metabolism inside the body, its lack of accumulation in tissues\textcolor{blue}{\cite{nascu2017explicit}}, and its lower chance of negative side effects compared to other hypnotic drugs\textcolor{blue}{\cite{borgeat1992}}. The second class of anesthetics, called analgesics, attenuate the sensation of pain. Currently, opioid analgesics (e.g. remifentanil) are the most common type of analgesic drug. Finally, the third class of anesthetics, known as NMB drugs, interrupt the transmission of nerve impulses in the neuromuscular junction and cause paralysis of the affected skeletal muscles\textcolor{blue}{\cite{bibian2006}}. NMB can be used to facilitate the processes of endotracheal intubation and mechanical ventilation\textcolor{blue}{\cite{mcmanus2001neuromuscular, zhusubaliyev2015}}. As shown in \textcolor{blue}{{Fig. }\ref{fig6}}, each of the 3 classes of drugs used in anesthesia contribute, respectively, to achieving the 3 main targets of general anesthesia: (i) hypnosis, or loss of consciousness (due to hypnotic drugs), (ii) analgesia, or loss of sensitivity to pain (due to analgesic drugs), and (iii) immobility (due to NMB drugs). 

General anesthesia is performed in 3 phases: (i) induction, (ii) maintenance, and (iii) emergence. Induction is most critical phase in anesthesia because all the three types of drugs -- hypnotic, analgesic, and NMB -- have to be administered at high enough doses to induce all 3 main aspects of anesthesia and allow the anesthesiologist to insert an endotracheal tube for mechanical ventilation to help the anesthetized patient breathe (i.e., endotracheal intubation).  

Maintenance follows the induction phase. In this phase, the anesthesiologist balances anesthesia with the use of anesthetics agents, manages the infusion of intravenous fluids, and regulates the administration of different drugs to maintain hemodynamic stability (i.e., stable blood flow), normal function of body organs, and adequate depth of hypnosis\textcolor{blue}{\cite{abdulla2012thes}}.

Emergence is the last stage of anesthesia and occurs at the end of surgery when the patient is removed from anesthesia and brought back to normal conditions as drug administration is tapered off. The goal of emergence is the full reversal of neuromuscular blockade and the smooth return of spontaneous ventilation and reflexes while maintaining the stability of hemodynamics and other physiological functions. Thus, the return of consciousness can occur in a safe and comfortable environment\textcolor{blue}{\cite{abdulla2012thes}}.

\subsection{Monitoring} \label{monitoring}
To achieve satisfactory anesthesia, the amount and timing of drug infusions must be estimated in each phase during surgery. Several clinical indicators and physiological parameters effect the anesthesiologist's estimation. Clinical indicators such as heart rate, blood pressure, lacrimation, sweating, and papillary dilatation are monitored during surgery and provide anesthesiologists valuable information about the anesthetic adequacy\textcolor{blue}{\cite{sinha2007}}. However, physiological monitors (i.e., electromyography (EMG), electrocardiography (ECG), blood pressure (BP), electroencephalography (EEG), and oxyhemoglobin saturation (SpO2)), and quantified indexes such as bispectral index (BIS), entropy, auditory evoked potential (AEP) and surgical stress index (SSI) can provide anesthesiologists more accurate information about a patient's status\textcolor{blue}{\cite{lan2012}}. The monitors and quantified indexes corresponding to several important clinical signs and physiological parameters are shown in \textcolor{blue}{{Table }\ref{table5}} for reference. 
\begin{table*}[]
\centering
\caption{Monitors and indices for clinical signs and physiological parameters (adapted from\textcolor{blue}{\cite{lan2012}}).}
\label{table5}
\begin{tabular} {ccccc}
\hline \hline
\multicolumn{2}{c} {\begin{tabular}[c]{@{}c@{}}Clinical signs of inadequate \\ general anesthesia\end{tabular}} &  {Monitors} & Signal processing & \begin{tabular}[c]{@{}c@{}}Evaluation \\ indices\end{tabular} \\ \hline \hline 
Somatic responses &  &  &  &  \\ \hline \hline

{Motor} & {\begin{tabular}[c]{@{}c@{}}Movement \\ Withdrawal\end{tabular}} & EMG Observation & \begin{tabular}[c]{@{}c@{}}Power spectral analysis, \\ BIS,\\  Entropy\end{tabular} & \begin{tabular}[c]{@{}c@{}}BIS\\  Entropy\\  AEP\end{tabular} \\ \cline{1-3}

{\begin{tabular}[c]{@{}c@{}}Consciousness \\ Nociception\end{tabular} } & \begin{tabular}[c]{@{}c@{}}Awareness \\ Pain\end{tabular} & \begin{tabular}[c]{@{}c@{}}EEG\\ Subjective experience\end{tabular} &  &  \\ \hline \hline
Autonomic responses &  &  &  &  \\ \hline \hline
\multicolumn{1}{c}{Breathing} & {\begin{tabular}[c]{@{}c@{}}Breathing pattern \\ change\end{tabular}} & {\begin{tabular}[c]{@{}c@{}}Observation Respiration \\ rate Volume\end{tabular}} & {\begin{tabular}[c]{@{}c@{}}Coherence Cross-power \\ Spectrum\end{tabular} } & {\begin{tabular}[c]{@{}c@{}}CPC(cardiopulmonary coupling) \\ combined with ECG\end{tabular}} \\ \hline

\multicolumn{1}{c}{Hemodynamic} & Tachycardia & ECG & \begin{tabular}[c]{@{}c@{}}HRV (heart rate variability)\\ RRI (R peak interval)\\ FFT (fast Fourier transform)\\ LFP (low frequency power)\\ HFP (high frequency power)\end{tabular} & \begin{tabular}[c]{@{}c@{}}LHR (low/high frequency ratio)\\ {\begin{tabular}[c]{@{}c@{}}DFA (detrended fluctuation  \\ algorithm, $\alpha$ index)\end{tabular}}\\ MSE (multi scale entropy)\\ CI (complex index)\\ SSI (surgical stress index)\\ PVI (pleth variability index)\end{tabular} \\ \cline{2-4}
\multicolumn{1}{c}{} & Volume & SPO2 & PPGA (plethysmograph amplitude) &  \\ \cline{2-5} 
\multicolumn{1}{c}{} & Hypertension & BP & BPV (blood pressure variability) &  \\ \hline

\multicolumn{1}{c}{Sudomotor} & Sweating & \begin{tabular}[c]{@{}c@{}}Skin\\ conductivity\end{tabular} &  &  \\ \cline{2-3}
\multicolumn{1}{c}{} & Papillary dilatation & Pupillometry &  &  \\ \cline{2-3}
\multicolumn{1}{c}{} & Lacrimation & Observation &  &  \\ \hline
\multicolumn{1}{c}{Hormonal} & \begin{tabular}[c]{@{}c@{}}Catecholamines\\ Corticosteroids\end{tabular} & {\begin{tabular}[c]{@{}c@{}}Blood drawing \\ and lab analysis\end{tabular}} &  &  \\ \hline
\end{tabular}
\end{table*}

Analyzing the electroencephalogram (EEG) is one of the most common approaches for precise evaluation of anesthetic delivery. EEG signals depict the electrical activity in the cerebral cortex and the characteristics of the EEG waveforms (i.e., frequency and amplitude) vary by the type of drug that has been used and the amount of its dosage\textcolor{blue}{\cite{chang2015}}. There are a number of signal monitoring devices such as the Bispectral Index (BIS) monitor, Narcotrend monitor, Cerebral state monitor (CSM), and AEP monitor\textcolor{blue}{\cite{mantzaridis1997}} that measure and quantify the EEG in order to derive a surrogate measurement of anesthetic adequacy\textcolor{blue}{\cite{musizza2010, bibian2003}}. 

\begin{figure}
\centering
\includegraphics[width=80mm,scale=0.5]{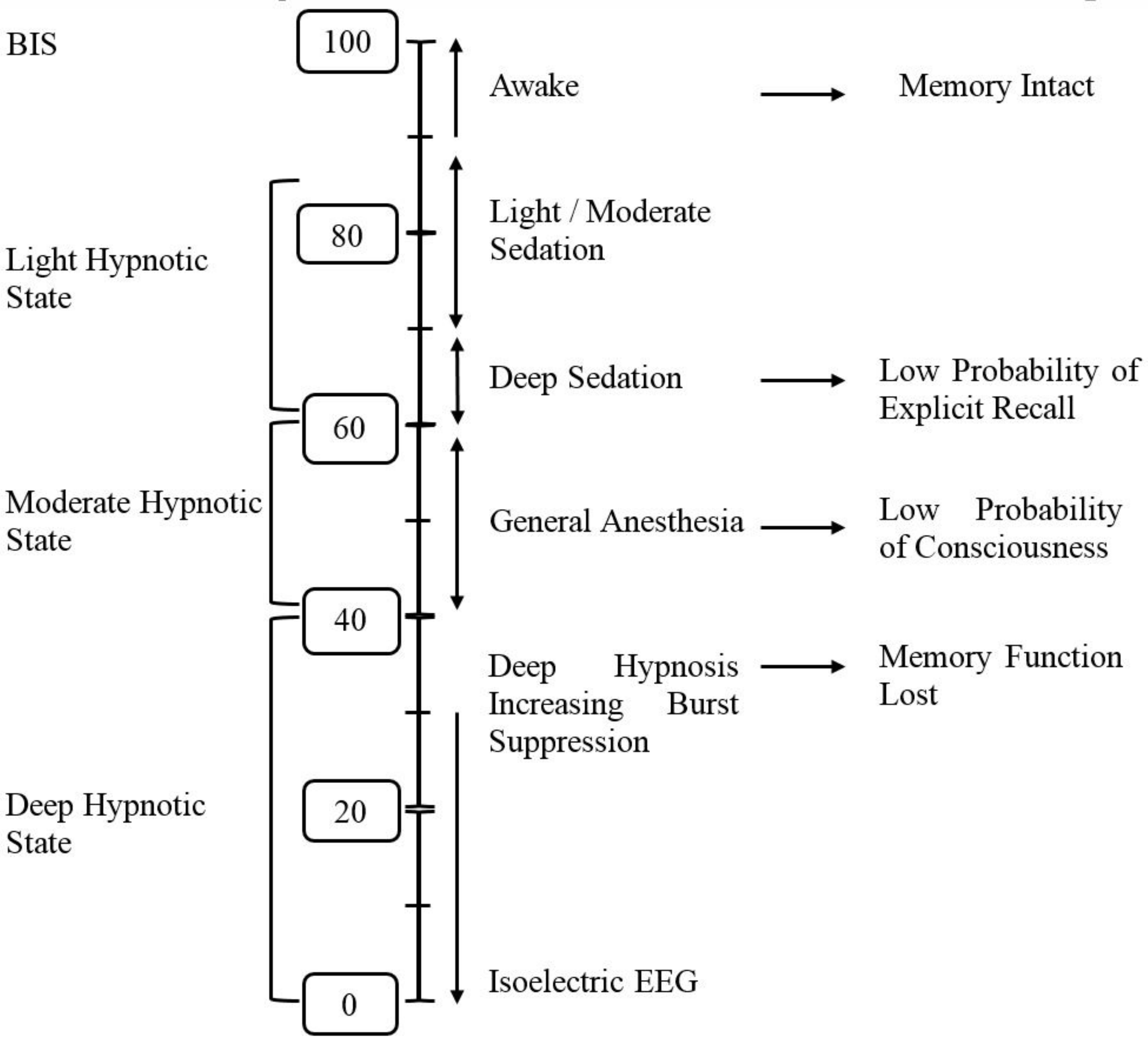}
\caption{Patient states corresponding to BIS indices (adapted from\textcolor{blue}{\cite{bibian2005, zikov2002}}.}
\label{fig7}
\end{figure}

Bispectral index (BIS) is the most notable index and is closely related to the level of consciousness\textcolor{blue}{\cite{bailey2005}}. Measurements of the BIS provide anesthesiologists important information during anesthesia, and clinical trials have shown its potential to increase the patient safety.\textcolor{blue}{\cite{struys2004, struys2003}}. The BIS index is scaled between 0 and 100, where a value of 100 represents the fully conscious state and value of zero depicts an isoelectric EEG. With increasing concentrations of anesthetic drugs, the BIS index decreases\textcolor{blue}{\cite{zikov2002}}. \textcolor{blue}{{Fig. }\ref{fig7}} shows the patient state for ranges of BIS indices along with the corresponding effects on memory of surgical events. Acceptable BIS scores for general anesthesia range from 40-60\textcolor{blue}{\cite{bailey2005}}.

Despite its wide-spread use, recent experimental results have shown that the BIS index is not reliable when patients have been infused with NMB drugs. In one pivotal study\textcolor{blue}{\cite{schuller2015response}}, fully awake subjects were shown to achieve BIS scores below 50 (i.e., corresponding with BIS during general anesthesia) when dosed with NMB drugs alone. Furthermore, additional research has suggested that BIS values should be interpreted with caution when patients are children\textcolor{blue}{\cite{malviya2007effect}}, elderly\textcolor{blue}{\cite{yamashita2008age}}, have known neurological disorders, or are taking psychoactive medications\textcolor{blue}{\cite{cascella2016mechanisms}}.  Beyond BIS, it has been noted that, "the use of DoA monitors remains a major controversy in anesthesiology. Indeed, many factors, such as age, race, gender, low core body temperature, acid-base imbalances, low blood glucose, drugs administered to the patient (e.g. neuromuscular blocking agents), and brain ischaemia, have a significant effect on raw EEG data. Additionally, DoA monitors are limited by their calibration range and inter patient variability in dose response curves..."\textcolor{blue}{\cite{cascella2016mechanisms}}. These limitations and interactions can degrade the performance of manual infusion systems and will no-doubt impose challenges to high-quality closed-loop control of anesthesia. Recently, patient state index (PSI) algorithm-based devices have been considered as alternatives to BIS monitors for their high sensitivity to variations in sedation/hypnosis and uniform response to various anesthetic agents\textcolor{blue}{\cite{prichep2004patient, cascella2016mechanisms}}. Although the PSI devices have some advantages to classical EEG monitors, including reduced sensitivity to noise, the PSI index is more sensitive to interpatient variability than BIS\textcolor{blue}{\cite{chen2002comparison}} and may be affected by EMG signals\textcolor{blue}{\cite{lee2015patient, cascella2016mechanisms}}. In general, there remains a need for studies investigating the effects of drug combinations and EMG signals on sedation indices. For more detailed information on current challenges and limitations of monitoring during anesthesia, readers are referred to the extensive coverage in the literature (e.g.,\textcolor{blue}{\cite{cascella2016mechanisms}}).

\subsection{Modeling}\label{modeling}
An accurate infusion of anesthesia drugs and continuous monitoring of brain states during each phase can help decrease the frequency and severity of dosage errors, improve the efficiency of anesthesia, and reduce surgical costs. As mentioned in \textcolor{blue}{Section \ref{monitoring}}, the BIS index is the most common monitoring metric available for displaying a patient's anesthesia level. However, to fully utilize the BIS and other measures for optimal drug dosing, a mathematical model is needed that represents all aspects of anesthesia including the effective physiological parameters and various dynamics (i.e., corresponding to the hypnotic, analgesic, and paralytic states). Various such models exist and can be categorized as single-input single-output (SISO), multi-input single-output (MISO), or multi-input multi-output (MIMO) models. In what follows, anesthesia models from each of these three classes will be briefly discussed.

\subsubsection{SISO Model}\label{siso}
The pharmacokinetic/pharmacodynamic (PK/PD) model is a compartmental model that illustrates both the drug distribution in the human body and its effect on certain physiological systems. For anesthesia, the pharmacokinetics are represented by a linear compartmental model that defines the concentration of the administered drug in different parts of human body, while the pharmacodynamics are represented by a static nonlinear function (i.e., Hill curve) relating the resulting drug concentration to the measured effect (e.g., BIS index)\textcolor{blue}{\cite{derendorf1999,chang2015}}. 

\begin{figure}
\centering
\includegraphics[width=80mm,scale=0.5]{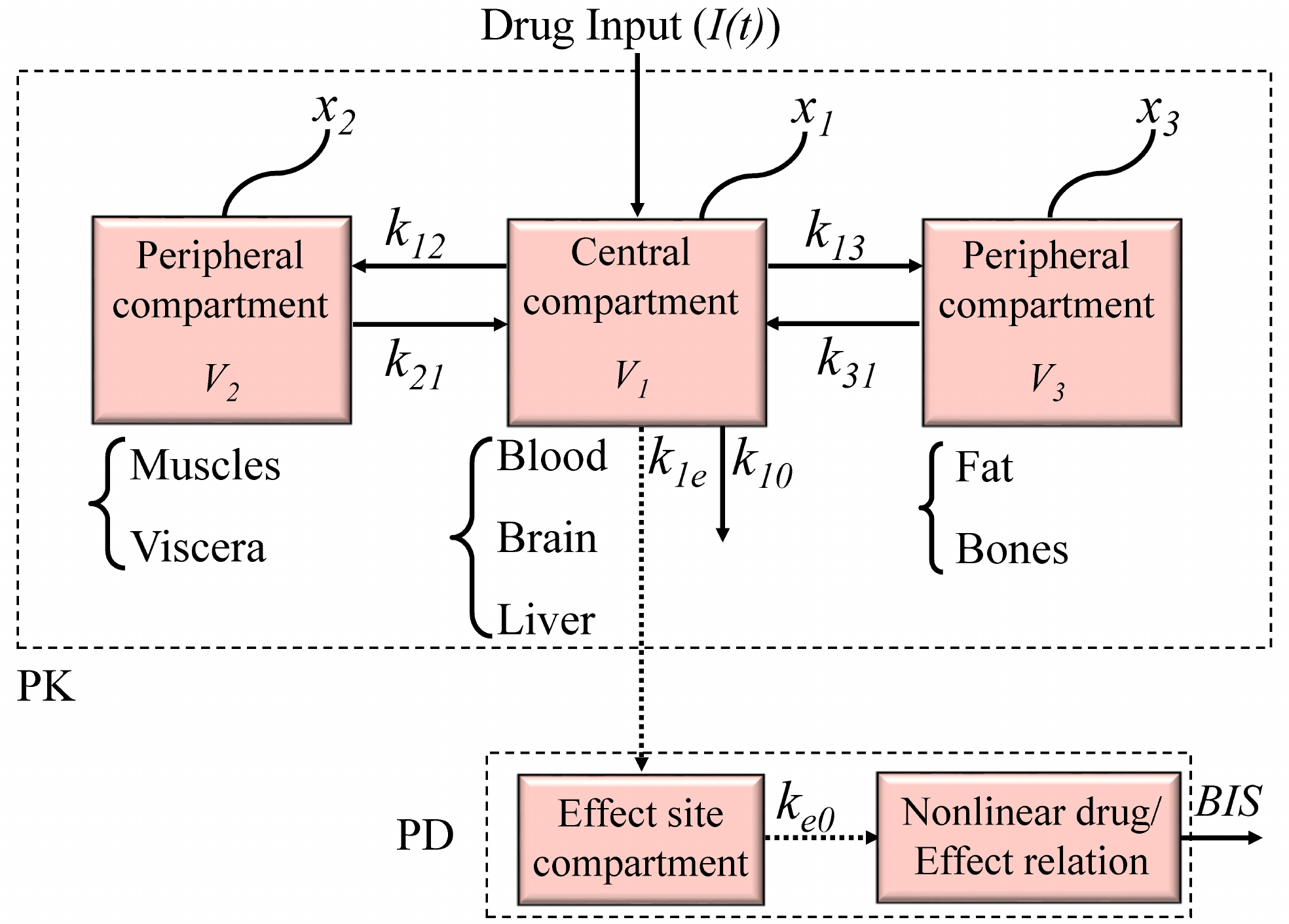}
\caption{Patient states corresponding to BIS indices (adapted from\textcolor{blue}{\cite{bibian2005, zikov2002}}.}
\label{fig8}
\end{figure}

The current PK/PD model is a SISO model with either hypnotic drugs (e.g., propofol) or analgesic drugs (e.g., remifentanil) as the input and either patient consciousness or degree of pain as the output. Such SISO models are used mainly in anesthesia modeling by control engineers and clinicians. As shown in \textcolor{blue}{{Fig. }\ref{fig8}}, the PK/PD model of anesthesia is a compartmental model with three main compartments and one effect-site compartment where $V_1$ describes the central compartment including arterial blood, brain, and liver volumes; $V_2$ denotes the fast peripheral compartment, which has fast dynamics and contains the well perfused body tissues such as muscles and viscera; $V_3$ represents the slow peripheral dynamics, which include the dynamics of poorly perfused body tissues like fat and bone; and the effect-site compartment is used to account for the time lag between drug concentration in blood plasma and its effect\textcolor{blue}{\cite{schnider1998}}. \\

Pharmacokinetic Model:
A pharmacokinetic model of a drug is a mathematical expression relating the concentration $C_p (t)$ (mg/ml) of the drug in the blood plasma to the infusion rate of the drug into the central compartment I(t) (mg/min)\textcolor{blue}{\cite{bailey2005}}. Pharmacokinetic compartmental models are derived by establishing balance equations for the amount of drug ($x_i$ in mg) distributed in each compartment. Accordingly, the governing equations for the PK model in \textcolor{blue}{{Fig. }\ref{fig8}} are given by
\begin{equation}
    {{\dot{x}}_{1}}=\frac{I(t)}{{{V}_{1}}}-{{k}_{12}}{{x}_{1}}-{{k}_{13}}{{x}_{1}}-{{k}_{10}}{{x}_{1}}+{{k}_{21}}{{x}_{2}}+{{k}_{31}}{{x}_{3}}\label{eq5}
\end{equation}
\begin{equation}
    {{\dot{x}}_{1}}={{k}_{12}}{{x}_{1}}-{{k}_{21}}{{x}_{2}}\label{eq6}
\end{equation}
\begin{equation}
    {{\dot{x}}_{1}}={{k}_{13}}{{x}_{1}}-{{k}_{31}}{{x}_{3}}\label{eq7}
\end{equation}
where the constants $k_{ij}$ ($min^{-1}$) indicate the transfer rate of the drug from the $i$-th compartment to the $j$-th compartment; the constant $k_{10}$ is the rate at which the drug is metabolized; and $V_1$ is the volume of the first compartment. These equations can be written in state space as
\begin{equation}
    \begin{matrix}
   \left[ \begin{matrix}
   {{{\dot{x}}}_{1}}  \\
   {{{\dot{x}}}_{2}}  \\
   {{{\dot{x}}}_{3}}  \\
\end{matrix} \right]=\left( \begin{matrix}
   -\left( {{k}_{12}}+{{k}_{13}}+{{k}_{10}} \right) & {{k}_{21}} & {{k}_{31}}  \\
   {{k}_{12}} & -{{k}_{21}} & 0  \\
   {{k}_{13}} & 0 & -{{k}_{31}}  \\
\end{matrix} \right)  \\
   \times \left[ \begin{matrix}
   {{x}_{1}}  \\
   {{x}_{2}}  \\
   {{x}_{3}}  \\
\end{matrix} \right]+\left[ \begin{matrix}
   {}^{1}/{}_{{{V}_{1}}}  \\
   0  \\
   0  \\
\end{matrix} \right]I(t)  \\
\end{matrix}
\label{eq8}
\end{equation}

\begin{equation}
  {{C}_{p}}=\left[ \begin{matrix}
   1 & 0 & 0  \\
\end{matrix} \right]\left[ \begin{matrix}
   {{x}_{1}}  \\
   {{x}_{2}}  \\
   {{x}_{3}}  \\
\end{matrix} \right]
\label{eq9}
\end{equation}

Then, the transfer function for the PK model from infusion rate $I(t)$ to concentration $C_p(t)=x_1(t)$ is
\begin{equation}
PK(s)=\frac{{{C}_{p}}(s)}{I(s)}=\frac{1}{{{V}_{1}}}\frac{\left( s+{{k}_{21}} \right)\left( s+{{k}_{31}} \right)}{\left( s+{{p}_{1}} \right)\left( s+{{p}_{2}} \right)\left( s+{{p}_{3}} \right)}
\label{eq10}
\end{equation}
where every $p_i$ is a function of the constants $k_{ij}$ as given by the conditions
\begin{equation}
    \left\{ \begin{matrix}
       {{p}_{1}}+{{p}_{2}}+{{p}_{3}}={{k}_{10}}+{{k}_{12}}+{{k}_{13}}+{{k}_{21}}+{{k}_{31}}  \\
       {{p}_{1}}{{p}_{2}}+{{p}_{1}}{{p}_{3}}+{{p}_{2}}{{p}_{3}}=\left( {{k}_{10}}+{{k}_{12}} \right)\left( {{k}_{21}}{{k}_{31}} \right)+{{k}_{13}}{{k}_{21}}  \\
       {{p}_{1}}{{p}_{2}}{{p}_{3}}={{k}_{10}}{{k}_{21}}{{k}_{31}}  \\
    \end{matrix} \right.
    \label{eq11}
\end{equation}

The parameters $k_{ij}$ are different for each person and can be calculated as\textcolor{blue}{\cite{schnider1998}}
\begin{equation}
    \left[ \begin{matrix}
       {{k}_{10}}  \\
       \begin{matrix}
       {{k}_{12}}  \\
       {{k}_{21}}  \\
    \end{matrix}  \\
       {{k}_{31}}  \\
       {{k}_{13}}  \\
    \end{matrix} \right]=\left( \begin{matrix}
       \begin{matrix}
       \frac{1}{{{V}_{1}}} & 0 & 0  \\
    \end{matrix}  \\
       \begin{matrix}
       \begin{matrix}
       0 & \frac{1}{{{V}_{1}}} & 0  \\
    \end{matrix}  \\
       \begin{matrix}
       0 & 0 & \frac{1}{{{V}_{1}}}  \\
    \end{matrix}  \\
    \end{matrix}  \\
       \begin{matrix}
       0 & \frac{1}{{{V}_{2}}} & 0  \\
    \end{matrix}  \\
       \begin{matrix}
       0 & 0 & \frac{1}{{{V}_{3}}}  \\
    \end{matrix}  \\
    \end{matrix} \right)\left[ \begin{matrix}
       C{{l}_{1}}  \\
       C{{l}_{2}}  \\
       C{{l}_{3}}  \\
    \end{matrix} \right]
    \label{eq12}
\end{equation}
where $V_i$ and $Cl_i$ denote, respectively, the volume and clearance rate (i.e., the rate at which the drug is cleared from the body or removed from the central compartment) of the $i$-th compartment. For propofol, the clearance rate and volume parameters can be obtained as
\begin{equation}
    \begin{matrix}
       \left[ \begin{matrix}
       C{{l}_{1}}  \\
       C{{l}_{2}}  \\
       C{{l}_{3}}  \\
    \end{matrix} \right]=\left[ \begin{matrix}
       \begin{matrix}
       0.0456  \\
       0  \\
       0  \\
    \end{matrix} & \begin{matrix}
       0.0264  \\
       0  \\
       0  \\
    \end{matrix} & \begin{matrix}
       -0.0681  \\
       0  \\
       0  \\
    \end{matrix} & \begin{matrix}
       0  \\
       -0.024  \\
       0  \\
    \end{matrix}  \\
    \end{matrix} \right]  \\
       \times \left[ \begin{matrix}
       weight  \\
       height  \\
       lbm  \\
       age  \\
    \end{matrix} \right]+\left[ \begin{matrix}
       -2.271  \\
       0.018  \\
       0.836  \\
    \end{matrix} \right]  \\
    \end{matrix}
\label{eq13}
\end{equation}

\begin{equation}
\left[ \begin{matrix}
   {{V}_{1}}  \\
   {{V}_{2}}  \\
   {{V}_{3}}  \\
\end{matrix} \right]=\left[ \begin{matrix}
   0  \\
   -0.391  \\
   0  \\
\end{matrix} \right]\left[ age \right]+\left[ \begin{matrix}
   4.27  \\
   39.623  \\
   238  \\
\end{matrix} \right]
\label{eq14}
\end{equation}

It should be noted that although the above expressions are the most commonly used relations for calculating the values of $V_i$ and $Cl_i$, there are alternative, generally more complex, relations used to calculate these parameters for the infusion of propofol, such as those developed by\textcolor{blue}{\cite{schuttler2000population}}.

For the infusion of remifentanil, the PK parameters can be calculated as\textcolor{blue}{\cite{minto1997influence, minto1997anesthesiology}}
\begin{equation}
\left[ \begin{matrix}
   C{{l}_{1}}  \\
   C{{l}_{2}}  \\
   C{{l}_{3}}  \\
\end{matrix} \right]=\left[ \begin{matrix}
   \begin{matrix}
   -0.0162  \\
   0.0301  \\
   -0.00113  \\
\end{matrix} & \begin{matrix}
   0.0191  \\
   0  \\
   0  \\
\end{matrix}  \\
\end{matrix} \right]\left[ \begin{matrix}
   age  \\
   lbm  \\
\end{matrix} \right]+\left[ \begin{matrix}
   2.197  \\
   3.254  \\
   0.1212  \\
\end{matrix} \right]
    \label{eq15}
\end{equation}
\begin{equation}
\left[ \begin{matrix}
   {{V}_{1}}  \\
   {{V}_{2}}  \\
   {{V}_{3}}  \\
\end{matrix} \right]=\left[ \begin{matrix}
   \begin{matrix}
   -0.0201  \\
   -0.0811  \\
   0  \\
\end{matrix} & \begin{matrix}
   0.072  \\
   0.108  \\
   0  \\
\end{matrix}  \\
\end{matrix} \right]\left[ \begin{matrix}
   age  \\
   lbm  \\
\end{matrix} \right]+\left[ \begin{matrix}
   2.043  \\
   7.124  \\
   5.42  \\
\end{matrix} \right]
    \label{eq16}
\end{equation}

Note, in \eqref{eq13}-\eqref{eq16}, the patient's height, weight, and age are given in units of cm, kg, and years, respectively. Furthermore, the lean body mass (lbm) is calculated differently for male and female patients and can be obtained as\textcolor{blue}{\cite{hallynack1981}}
\begin{equation}
    \left[ \begin{matrix}
       lb{{m}_{m}}  \\
       lb{{m}_{f}}  \\
    \end{matrix} \right]=\left( \begin{matrix}
       m & 0  \\
       0 & f  \\
    \end{matrix} \right)\left( \begin{matrix}
       1.1 & -128  \\
       1.07 & -148  \\
    \end{matrix} \right)\left[ \begin{matrix}
       weight  \\
       {{\left( \frac{weight}{height} \right)}^{2}}  \\
    \end{matrix} \right]
    \label{eq17}
\end{equation}
where $m$ and $f$ are set equal to one and zero, respectively, for male patients and vice versa for female patients. \\

Pharmacodynamic Model:
Pharmacodynamic models are used to describe the relationship between the concentration of a drug and its effect on the body. As shown in \textcolor{blue}{{Fig. }\ref{fig8}}, the general PD model has two parts. The first part is a compartment called the effect-site, introduced by\textcolor{blue}{\cite{sheiner1979}}, that represents the lag between drug concentration and its effect. In the case of anesthesia, since the effect site compartment is small in comparison with the central compartment, its effect is assumed to be negligible\textcolor{blue}{\cite{shafer1998}}. In steady-state, the concentration of the effect-site can be related to the plasma concentration by
\begin{equation}
    {{\dot{C}}_{e}}={{\dot{x}}_{e}}={{k}_{1e}}{{x}_{1}}-{{k}_{e0}}{{x}_{e}}
    \label{eq18}
\end{equation}
where $k_{e0}$  and $k_{1e}$ are constants and $x_e$ is the amount of the drug in the effect compartment. The rate at which drugs enter and exit this compartment is assumed to be constant and equal to $k_{e0}$ (e.g., $k_{1e}=k_{e0}=0.456$ for propofol)\textcolor{blue}{\cite{schnider1998}}. For the infusion of remifentanil, $k_{e0}$ can be calculated as
\begin{equation}
    {{k}_{e0}}=0.595-0.007\left( age-40 \right)
    \label{eq19}
\end{equation}

Consequently, \eqref{eq19} is often used as
\begin{equation}
    {{\dot{C}}_{e}}={{k}_{e0}}\left( {{C}_{p}}-{{C}_{e}} \right)
    \label{eq20}
\end{equation}
where $C_e$ is the concentration within the effect-site compartment. Thus, for propofol, the effect site works like a low-pass filter that relates the plasma propofol concentration (i.e., in the blood) to the effect-site propofol concentration. The transfer function describing the relation from the plasma concentration to the effect-site concentration is given by 
\begin{equation}
    \frac{{{C}_{e}}(s)}{{{C}_{p}}(s)}=\frac{{{k}_{e0}}}{s+{{k}_{e0}}}
    \label{eq21}
\end{equation}
The second part of the PD model, shown as the "nonlinear drug/effect relation" in \textcolor{blue}{{Fig. }\ref{fig8}}, is often comprised of a sigmoid Hill equation
\begin{equation}
    E={{E}_{0}}-{{E}_{\max }}\frac{C_{e}^{\gamma }}{C_{e}^{\gamma }+EC_{50}^{\gamma }}
    \label{eq22}
\end{equation}
where $E_0$ indicates the baseline value (obtained from the awake state without propofol), which is typically set to 100; $E_{max}$ is the maximum effect achievable by the drug infusion; $EC_{50}$ denotes the drug concentration at half maximal effect, which represents the patient's sensitivity to the drug and should be measured experimentally; and $\gamma$ determines slope of the sigmoid curve (i.e., the receptiveness of the patient to the drug)\textcolor{blue}{\cite{nino2009}}. In \eqref{eq22}, the value of the effect $E$ is obtained in terms of the BIS scale. In \textcolor{blue}{{Table }\ref{table6}}, the values of these parameters are shown for a cohort of 12 representative patients. These values are among the most frequently used for designing and simulating anesthesia control systems. Additional data sets are available in\textcolor{blue}{\cite{bibian2006}}.
\begin{table*}[]
\caption{Hill equation characteristic variables for propofol (adapted from\textcolor{blue}{\cite{nino2009}}).}
\label{table6}
\centering
\begin{tabular}{ccccccccc}
\hline \hline
Patient $\#$ & Age & Height (cm) & Weight (kg) & Gender & $EC_{50}$ & $E_{0}$ & $E_{max}$ & $\gamma$ \\ \hline \hline
1 & 40 & 163 & 54 & F & 6.33 & 98.80 & 94.10 & 2.24 \\
2 & 36 & 163 & 50 & F & 6.76 & 98.60 & 86.00 & 4.29 \\
3 & 28 & 164 & 52 & F & 8.44 & 91.20 & 80.70 & 4.1 \\
4 & 50 & 163 & 83 & F & 6.44 & 95.90 & 102.00 & 2.18 \\
5 & 28 & 164 & 60 & M & 4.93 & 94.70 & 85.30 & 2.46 \\
6 & 43 & 163 & 59 & F & 12.10 & 90.20 & 147.00 & 2.42 \\
7 & 37 & 187 & 75 & M & 8.02 & 92.00 & 104.00 & 2.10 \\
8 & 38 & 174 & 80 & F & 6.56 & 95.50 & 76.40 & 4.12 \\
9 & 41 & 170 & 70 & F & 6.15 & 89.20 & 63.80 & 6.89 \\
10 & 37 & 167 & 58 & F & 13.70 & 83.10 & 151.00 & 1.65 \\
11 & 42 & 179 & 78 & M & 4.82 & 91.80 & 77.90 & 1.85 \\
12 & 34 & 172 & 58 & F & 4.95 & 96.20 & 90.80 & 1.84 \\
Mean & 38 & 169 & 65 & F & 7.42 & 93.1 & 96.6 & 3 \\ \hline
\end{tabular}
\end{table*}

Although the aforementioned models are used quite frequently, they are in need of some refinements. Among other issues, the most critical weaknesses of the models presented thus far include a lack of standard parameters for other kinds of hypnotic drugs, important neglected factors such as cardiac output and mode of drug administration (e.g., bolus or infusion), and unmodeled drug interactions\textcolor{blue}{\cite{chang2015}}. It is also worth noting that the commonly used values in \textcolor{blue}{{Table }\ref{table2}} are not representative for a variety of important demographics. In fact, in the case of infants and adolescents, different PK/PD models are required, which have been described completely in\textcolor{blue}{\cite{constant2010, dahmani2010, mani2010, rigouzzo2010}}.

It should be noted that the SISO models discussed above incorporate the infusion of hypnotic and analgesic drugs but do not consider neuromuscular blocking drugs because neuromuscular blockade has no effective interactions with anesthetic drugs and opioids. In some studies, separate (different) SISO models are used to represent the dynamics of NMB\textcolor{blue}{\cite{zhusubaliyev2013bifurcation, dasilva2012nonlinear}}. Furthermore, a different PK/PD model structure, comprised of one central compartment (representing both blood and organs heavily supplied with blood), one peripheral compartment (representing the parts of the body with a lower blood flow), and the effect compartment (representing the PD model) is presented in the literature\textcolor{blue}{\cite{ward1983,weatherley1983}}. Alternatively, a second order linear dynamic model is presented for NMB in\textcolor{blue}{\cite{hodrea2013}}.

\subsubsection{MISO Model}
While the SISO model in \textcolor{blue}{Section \ref{siso}} can be used separately for both hypnotic and analgesic drugs, it is not capable of fully representing the combined administration and effects of these drugs. As a work around, some studies have used parallel SISO models of each drug to obtain a simple MISO model. In this arrangement, the WAVCNS index used by the NeuroSense NS-701 (NeuroWave Systems, Cleveland Heights, USA) monitor appears to be a more suitable measurement index than BIS for measuring the anesthesia level because it is obtained through a linear time-invariant relation that does not cause an additional delay in the closed-loop system\textcolor{blue}{\cite{soltesz2012simulated, zikov2006}}.
The interaction of propofol and remifentanil is described in\textcolor{blue}{\cite{kern2004}} as
\begin{equation}
    E\left( {{v}_{p}},{{v}_{r}} \right)=\frac{{{\left( {{v}_{p}}+{{v}_{r}}+\alpha {{v}_{p}}{{v}_{r}} \right)}^{\gamma }}}{{{\left( {{v}_{p}}+{{v}_{r}}+\alpha {{v}_{p}}{{v}_{r}} \right)}^{\gamma }}+1}
    \label{eq23}
\end{equation}
where $\nu_p=(c_e^p)/(EC_{50}^p )$, $\nu_r=(c_e^r)/(EC_{50}^r)$, and the parameter $\alpha$ (identified from clinical data by\textcolor{blue}{\cite{kern2004}}) determines the degree of interaction between propofol and remifentanil. This interaction is clasified as synergistic if $\alpha>0$; additive if $\alpha=0$; and antagonistic if $\alpha<0$. The parameters $EC_{50}^p$ and $EC_{50}^r$ are the propofol and remifentanil concentrations at half maximal effect.

As an alternative to using the PK/PD models of propofol and remifentanil in parallel, a new MISO Weiner model called the parameter parsimonious model (PPM) has recently been introduced by\textcolor{blue}{\cite{Dasilva2010}}. In this model, the effect-site concentrations of propofol and remifentanil (i.e., $c_e^p$ and $c_e^r$, respectively) due to their infusion rates (i.e., $u^p$ and $u^r$, respectively) are given in the Lapalce domain as
\begin{equation}
    C_{e}^{p}\left( s \right)=\frac{{{k}_{1}}{{k}_{2}}{{k}_{3}}{{\alpha }^{3}}}{\left( {{k}_{1}}\alpha +s \right)\left( {{k}_{2}}\alpha +s \right)\left( {{k}_{3}}\alpha +s \right)}{{u}^{p}}\left( s \right)
    \label{eq24}
\end{equation}
\begin{equation}
    C_{e}^{r}\left( s \right)=\frac{{{l}_{1}}{{l}_{2}}{{l}_{3}}{{\eta }^{3}}}{\left( {{l}_{1}}\eta +s \right)\left( {{l}_{2}}\eta +s \right)\left( {{l}_{3}}\eta +s \right)}{{u}^{r}}\left( s \right)
    \label{eq25}
\end{equation}
where $\alpha$ and $\eta$ are dependent on patient parameters and do not have any physiological meaning. The parameters $k_i$ and $l_j$ are dimensionless constants that were obtained from clinical data by\textcolor{blue}{\cite{silva2011prediction}} as $k_1=10$ , $k_2=9$, $k_3=1$, $l_1=3$, $l_2=2$, and $l_3=1$. The corresponding BIS value $z(t)$ is approximated by\textcolor{blue}{\cite{Dasilva2010}} as
\begin{equation}
    z\left( t \right)=\frac{97.7}{1+U{{\left( t \right)}^{\gamma }}}
    \label{eq26}
\end{equation}
where
\begin{equation}
    U\left( t \right)=\mu \frac{C_{e}^{p}}{EC_{50}^{p}}+\frac{C_{e}^{r}}{EC_{50}^{r}}
    \label{eq27}
\end{equation}

In these equations, the parameters $\mu$ and $\gamma$ do not have any physiological meaning and are dependent on each patient. This model can be expressed in state space as\textcolor{blue}{\cite{nogueira2016}}
\begin{equation}
    \dot{x}\left( t \right)=Ax\left( t \right)+Bu\left( t \right)
    \label{eq28}
\end{equation}
\begin{equation}
        \left[ \begin{matrix}
       C_{e}^{p}\left( t \right)  \\
       C_{e}^{r}\left( t \right)  \\
    \end{matrix} \right]=\left[ \begin{matrix}
       \begin{matrix}
       \begin{matrix}
       0  \\
       0  \\
    \end{matrix} & \begin{matrix}
       0  \\
       0  \\
    \end{matrix}  \\
    \end{matrix} & \begin{matrix}
       \begin{matrix}
       1  \\
       0  \\
    \end{matrix} & \begin{matrix}
       0  \\
       0  \\
    \end{matrix}  \\
    \end{matrix} & \begin{matrix}
       \begin{matrix}
       0  \\
       0  \\
    \end{matrix} & \begin{matrix}
       0  \\
       1  \\
    \end{matrix}  \\
    \end{matrix}  \\
    \end{matrix} \right]x\left( t \right)
    \label{eq29}
\end{equation}
\begin{equation}
    U\left( t \right)=Cx\left( t \right)
    \label{eq30}
\end{equation}
where
\begin{equation}
    \begin{matrix}
   A=\left[ \begin{matrix}
   {{A}^{p}} & 0  \\
       0 & {{A}^{r}}  \\
    \end{matrix} \right], & B=\left[ \begin{matrix}
       {{B}^{p}} & 0  \\
       0 & {{B}^{r}}  \\
    \end{matrix} \right]  \\
    \end{matrix}
    \label{eq31}
\end{equation}

\begin{equation}
      \begin{matrix}
       {{A}^{p}}=\left[ \begin{matrix}
       -{{k}_{1}}\alpha  & 0 & 0  \\
       {{k}_{2}}\alpha  & -{{k}_{2}}\alpha  & 0  \\
       0 & {{k}_{3}}\alpha  & -{{k}_{3}}\alpha   \\
    \end{matrix} \right],  \\
       {{A}^{r}}=\left[ \begin{matrix}
       -{{l}_{1}}\eta  & 0 & 0  \\
       {{l}_{2}}\eta  & -{{l}_{2}}\eta  & 0  \\
       0 & {{l}_{3}}\eta  & -{{l}_{3}}\eta   \\
    \end{matrix} \right]  \\
    \end{matrix}
    \label{eq32}
\end{equation}
\begin{equation}
    \begin{matrix}
   {{B}^{p}}=\left[ \begin{matrix}
   {{k}_{1}}\alpha   \\
   0  \\
   0  \\
    \end{matrix} \right], & {{B}^{r}}=\left[ \begin{matrix}
       {{l}_{1}}\eta   \\
       0  \\
       0  \\
    \end{matrix} \right]  \\
    \end{matrix}
    \label{eq33}
\end{equation}
\begin{equation}
    C=\left[ \begin{matrix}
   \begin{matrix}
   0 & 0  \\
    \end{matrix} & \begin{matrix}
       \frac{\mu }{EC_{50}^{p}} & 0  \\
    \end{matrix} & \begin{matrix}
       0 & \frac{1}{EC_{50}^{r}}  \\
    \end{matrix}  \\
    \end{matrix} \right]
    \label{eq34}
\end{equation}

The common values for $\alpha$, $\gamma$, $\mu$ and $\eta$ are identified by\textcolor{blue}{\cite{mendoncca2012comparing}} and are given in \textcolor{blue}{{Table } \ref{table7}}.
\begin{table*}[]
\caption{Common PPM parameter values (adapted from\textcolor{blue}{\cite{nogueira2016}}).}
\label{table7}
\centering
\begin{tabular}{ccccccccc}
\hline \hline
Patient $\#$ & Gender & Age & Height (cm) & Weight (kg) & $\alpha$ & $\eta$ & $\gamma$ & $\mu$ \\ \hline \hline
1 & F & 56 & 160 & 88 & 0.0667 & 0.3989 & 1.7695 & 2.1502 \\
2 & F & 48 & 158 & 52 & 0.0874 & 0.0670 & 0.9365 & 4.7014 \\
3 & F & 51 & 165 & 55 & 0.0693 & 0.0482 & 2.8186 & 1.1700 \\
4 & F & 56 & 160 & 65 & 0.0590 & 0.0425 & 2.7594 & 1.4077 \\
5 & F & 64 & 146 & 60 & 0.0489 & 0.1269 & 1.5627 & 1.4171 \\
6 & F & 59 & 159 & 110 & 0.0677 & 0.3373 & 4.1247 & 1.1444 \\
7 & F & 29 & 163 & 59 & 0.0737 & 0.2793 & 0.7812 & 0.8986 \\
8 & F & 45 & 155 & 58 & 0.0860 & 0.0212 & 0.9780 & 1.4203 \\
9 & F & 51 & 163 & 55 & 0.0701 & 0.2837 & 1.0956 & 1.2164 \\
10 & F & 32 & 172 & 56 & 0.1041 & 0.1038 & 1.2165 & 1.9085 \\
11 & F & 68 & 160 & 64 & 0.0343 & 3.5768 & 1.7097 & 2.5451 \\
12 & F & 50 & 161 & 68 & 0.0467 & 0.1254 & 2.4877 & 1.4884 \\
13 & F & 68 & 158 & 113 & 0.0687 & 4.5413 & 1.0859 & 2.3951 \\
14 & F & 70 & 161 & 78 & 0.0774 & 0.0397 & 1.4038 & 1.5460 \\
15 & F & 73 & 160 & 75 & 0.0995 & 0.0377 & 1.3706 & 2.0485 \\
16 & F & 34 & 162 & 57 & 0.0929 & 0.1205 & 4.5194 & 1.5565 \\
17 & F & 43 & 155 & 62 & 0.0811 & 0.1033 & 2.1978 & 2.0338 \\
18 & F & 66 & 155 & 74 & 0.1336 & 0.2307 & 1.0849 & 1.2061 \\ \hline
\end{tabular}
\end{table*}

Although the PPM model is not entirely physiological, it is better suited to model-based control than the general PK/PD model because it has fewer parameters that need to be identified\textcolor{blue}{\cite{nogueira2016}}. However, both the PK/PD and PPM models have been used in many controller design studies. Some important factors for choosing between these models include the availability of measurement devices, the type of desired control structure, the types of drugs being administered, and the costs associated with the overall design and implementation. In the following sections, some recent studies that utilize these models will be discussed. 

\subsubsection{MIMO Model}
While the previous models have a single measurable output (e.g., BIS or WAVCNS), there are several other variables that can be considered as additional system outputs. Two MIMO models have been reported prevalently in the literature, both of which take the dosage of hypnotic and analgesic drugs as inputs but differ in the quantities used as outputs. When considering MIMO systems with m inputs and n outputs, we can define $m\times{n}$ transfer functions that describe the effects of each input on each output. 

The first of these MIMO models\textcolor{blue}{\cite{tan2010}}, which is used less frequently in the literature, includes some clinical metrics like heart rate and blood pressure as system outputs. By assuming the analgesic drugs have negligible effects on the BIS value\textcolor{blue}{\cite{nakaya2002}} and considering depth of hypnosis (DoH) and blood pressure as the system outputs, this model is given by
\begin{equation}
    {{x}_{B}}={{e}^{-{{\tau }_{B}}s}}\frac{{{K}_{B}}}{{{T}_{B}}s+1}{{U}_{B}}\left( s \right)
    \label{eq35}
\end{equation}
\begin{equation}
    {{x}_{P}}={{e}^{-{{\tau }_{P}}s}}\frac{{{K}_{P}}}{{{T}_{P}}s+1}{{U}_{P}}\left( s \right)
    \label{eq36}
\end{equation}
\begin{equation}
    {{y}_{B}}=100-{{f}_{B}}\left( {{x}_{B}}\left( t \right) \right)+{{d}_{B}}
    \label{eq37}
\end{equation}
\begin{equation}
    {{y}_{P}}=100-{{f}_{P}}\left( {{x}_{P}}\left( t \right) \right)+{{d}_{P}}
    \label{eq38}
\end{equation}
where indexes $B$ and $P$ denote hypnotic and analgesic drugs, respectively; $\tau_B$ and $\tau_P$ are initial time delays; $K_B$ and $K_P$ are defined as drug sensitivities; $T_B$ and $T_P$ are time-constants representing the response speed of the patient; $f_B$ and $f_P$ are defined in\textcolor{blue}{\cite{tan2011}} as nonlinear sensitivity functions; $d_B$ and $d_P$ are external disturbances; $U_B(s)$ and $U_P(s)$ are the Laplace transforms of drug infusion rates; $x_B$ and $x_P$ are the effect-site concentrations of the anesthetic drugs; and $y_B$ and $y_P$ are the DoH and mean blood pressure, respectively. According to \eqref{eq27}, this representation consists of four (i.e., $2\times{2}$) different transfer functions, which are reported in\textcolor{blue}{\cite{tan2010}}.

\begin{figure}
\centering
\includegraphics[width=80mm,scale=0.5]{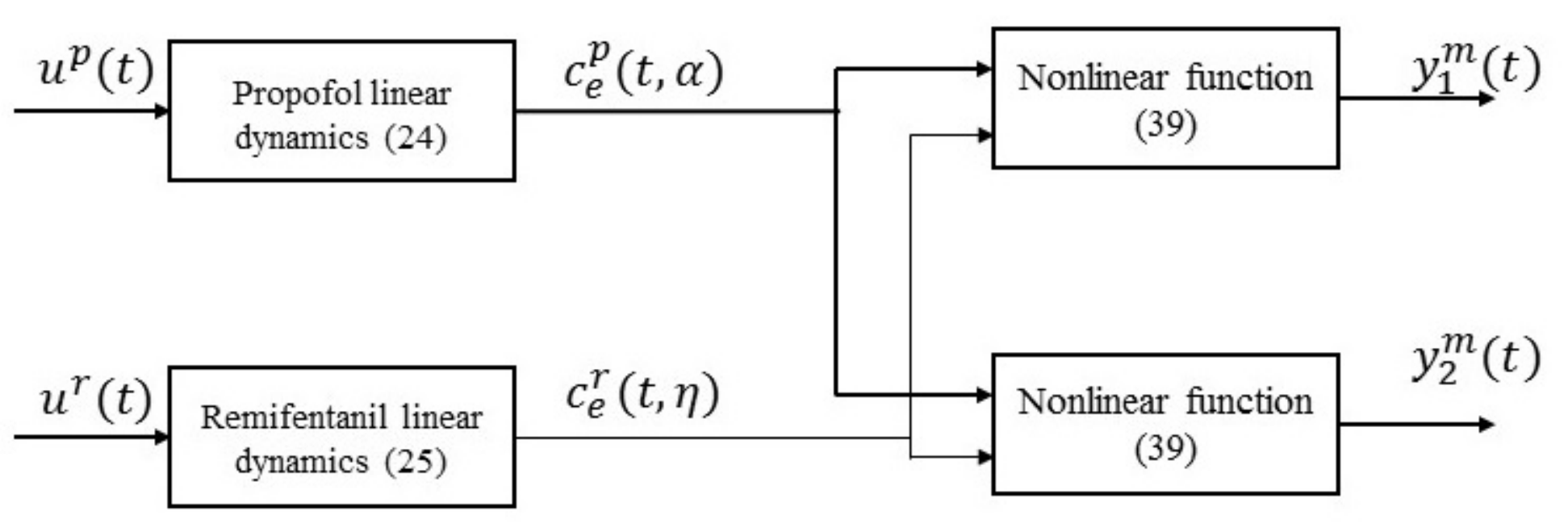}
\caption{MIMO Wiener model for the depth of anesthesia (adapted from\textcolor{blue}{\cite{silva2014a}}).}
\label{fig9}
\end{figure}

The second common MIMO model, depicted in \textcolor{blue}{{Fig. }\ref{fig9}}, defines the infusion rates of propofol (i.e., the hypnotic drug) and remifentanil (i.e., the analgesic drug) as the inputs, and the patient's depth of hypnosis and degree of pain are used as the system outputs. Thus, the depth of hypnosis and degree of pain are interpreted from BIS and state entropy (SE) measurements, which consequently may be considered alternative model outputs.
The state-space of this model is similar to the MISO model of \eqref{eq24}-\eqref{eq33}; the main difference between these models is the output matrix, which is described as\textcolor{blue}{\cite{silva2014a}}
\begin{equation}
    y=\left[ \begin{matrix}
   {{y}_{1}}  \\
   {{y}_{2}}  \\
    \end{matrix} \right]=\left[ \begin{matrix}
   \frac{{{y}_{01}}}{1+{{\left( \frac{C_{e}^{r}}{EC_{50}^{r}}+{{\rho }_{1}}\frac{C_{e}^{p}}{EC_{50}^{p}} \right)}^{{{\gamma }_{1}}}}}  \\
   \frac{{{y}_{02}}}{1+{{\left( \frac{C_{e}^{r}}{EC_{50}^{r}}+{{\rho }_{2}}\frac{C_{e}^{p}}{EC_{50}^{p}} \right)}^{{{\gamma }_{2}}}}}  \\
\end{matrix} \right]
    \label{eq39}
\end{equation}

where $y_1$ denotes the BIS value; $y_2$ denotes the SE value; $y_{01}$ and $y_{02}$ are equal to 97.7 and 91, respectively; and $\rho_1$, $\rho_2$, $\gamma_1$ and $\gamma_2$ are parameters that are estimated by clinicians. 

\subsection{Automation in Anesthesia}
In 2010, 48.3 million surgical operations were performed in the U.S.\textcolor{blue}{\cite{hall2017}}. Many of these surgeries require general anesthesia, yet ensuring optimal and on-time infusion of anesthetic drugs remains an important challenge that is complicated by the unique physiology of each patient and numerous known and unknown environmental factors. Further complicating this problem, patients' anesthetic requirements are known to change during surgery in both predictable and unpredictable ways, and once administered, anesthetics cannot be removed. Poor regulation of anesthetics, such as too little hypnotic or too much analgesic, can result in complications including anesthesia awareness (i.e., unintended consciousness during surgery), pain during surgery, prolonged emergence period, and increased risks of other side effects\textcolor{blue}{\cite{abdulla2012Depth}}. As an example of the regularity of these complications, \textcolor{blue}{{Table }\ref{table8}} shows data collected on the incidence of anesthesia awareness over the last half-century.
\begin{table}[]
\caption{Incidence of awareness during surgery (adapted from\textcolor{blue}{\cite{bruhn2006}}).}
\label{table8}
\centering
\begin{tabular}{ccc}
\hline \hline
Year & Sample size & Awareness (percent of patients) \\ \hline \hline
1960 & 656 & 1.2 \\
1971 & 120 & 1.6 \\
1973 & 200 & 1.5 \\
1975 & 490 & 0.8 \\
1990 & 1000 & 0.2 \\
1997-1998 & 11785 & 0.15 \\
1993-2000 & 10811 & 0.11 \\
2001-2002 & 19575 & 0.13 \\
2003 & 7826 & 0.18 \\
2004 & 1238 & 0.9 \\
2005 & 150000 & 0.07 \\ \hline
\end{tabular}
\end{table}

Although the statistics above show a near steady improvement, there are still many faults occurring during anesthesia administration which may be avoidable by the development of an automated anesthesia system that improves the administration of anesthetics and mitigates the hazards of system faults. In fact, besides a reduced risk of surgical complications, the use of an automatic control system may provide significant additional benefits such as a reduction in the cognitive workload of the anesthesiologist (who must be present to supervise any automated system), a decreased volume of drugs administered (which implies a faster and better recovery time of the patient in the post-operative phase), and a more robust performance with fewer over- and under-dosing episodes and better rejection of surgical disturbances. Of course, the main goal is increased safety and comfort for the patient\textcolor{blue}{\cite{padula2017}}. The advantages of automatic anesthesia systems over manual anesthesia have recently been validated by experimental results\textcolor{blue}{\cite{neckebroek2018}}.

Any automated anesthesia system will likely include two main phases: first, the desired depth of anesthesia (DoA) (i.e. desired BIS index) must be reached, and second the DoA needs to be maintained by the application of various anesthetic drugs. During the first phase, the system should calculate the sufficient drug dose and regulate its infusion to move from the initial BIS value $E_0$ of the patient during the induction phase to a final desired BIS value. Note, the determination of a desired DoA means setting final values for all the three types of anesthesia (hypnotic, analgesic, and neuromuscular blocking drugs). However, to date, most closed-loop studies have been focused only on achieving the depth of hypnosis (DoH), which is the most important type of anesthesia, while the analgesic and neuromuscular blocking drugs must still be infused manually by clinicians. 

During the second phase, the system should infuse the drugs continuously while monitoring the aforementioned physiological parameters and clinical signs and also rejecting disturbances that occur during the maintenance phase. The main goal during this phase is to maintain the patients at the final BIS value throughout the surgery until anesthesia is stopped during the emergence phase.

\begin{figure}
\centering
\includegraphics[width=80mm,scale=0.5]{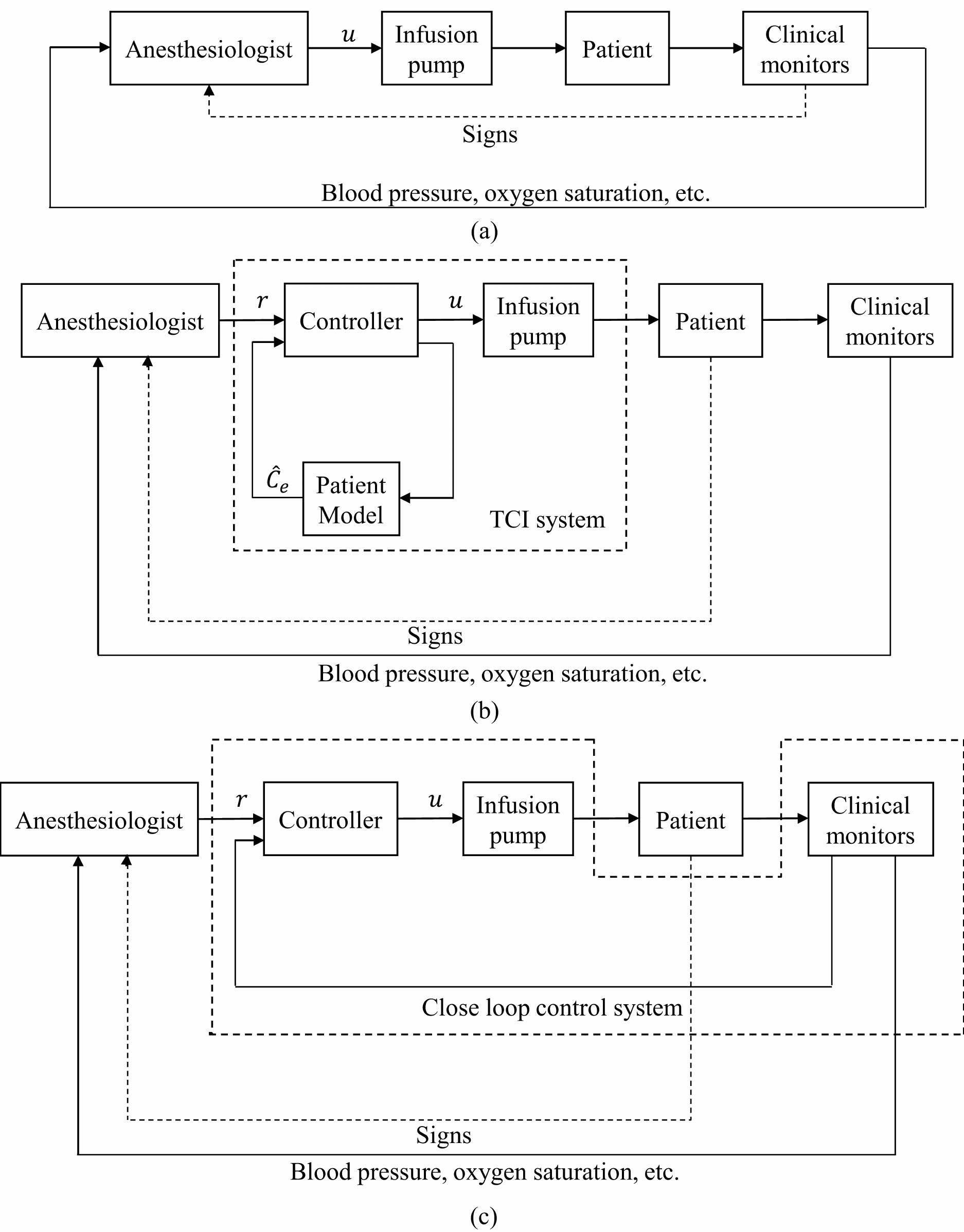}
\caption{Schematic overview of drug dosage calculation and infusion used in clinical anesthesia with (a) manual control, (b) open-loop feedforward control, also known as TCI, and (c) closed-loop control system (adapted from\textcolor{blue}{\cite{soltesz2013automation}})}
\label{fig10}
\end{figure}

Control applications in general anesthesia began receiving attention several decades ago\textcolor{blue}{\cite{biebuyck1992, Asbury1997}} from both control engineers and clinicians. Throughout this period, a variety of methods have been proposed with varying degrees of automation. As illustrated in \textcolor{blue}{{Fig. }\ref{fig10}}, most systems can be categorized as manual, open-loop feed-forward, or closed-loop controllers based on the approach used to calculate the drug infusion rates in general anesthesia.

Manual systems (i.e., without automation), shown in \textcolor{blue}{{Fig. }\ref{fig10}a}, are the current standard practice. In these systems, an anesthesiologist first sets the final value of the DoA or DoH and then monitors the anesthetic state of the patient and regulates the drug dosages accordingly\textcolor{blue}{\cite{soltesz2013automation}}. The success of this method is highly dependent on the expertise of the individual anesthesiologist and is not reliable in some critical situations (e.g., irregular events during surgery). 

In 2003, a large step toward achieving fully automated anesthesia was taken with the introduction of target controlled infusion (TCI) systems\textcolor{blue}{\cite{absalom2003}}. The Diprifusor system was the earliest commercial TCI pump for propofol\textcolor{blue}{\cite{glen1998}} and since its introduction several other TCI systems have become available from various manufacturers. In fact, developments and modifications to TCI systems has made them a standard among infusion systems for the administration of opioids, propofol, and other anesthetics in clinical practice\textcolor{blue}{\cite{glen1998}}. Considering our categorization, TCI systems operate as open-loop feed-forward controllers, shown in \textcolor{blue}{{Fig. }\ref{fig10}b}, which rely on both an anesthesiologist to assess the patient's state and the output of a PK/PD model to estimate an exact drug dosage and infusion time. In practice, the anesthesiologist sets a target drug concentration, and the TCI system determines an infusion rate and sends this value to the infusion pump. Although these systems are used in many countries and have some advantages over manual anesthesia systems, the success of TCI still depends heavily on the expertise of the anesthesiologist and does not have its own means of identifying the patient's current needs when tuning the infusion rate. Since TCI lacks real-time feedback, its performance depends on the accuracy of the patient model and is susceptible to disturbances caused by surgical stimulation and hypnotic-opioid synergy (i.e., the combined effects of these drugs)\textcolor{blue}{\cite{soltesz2013automation}}. 

Closed-loop controllers are the most advanced of the 3 categories and consider the individual patient's physiological parameters in addition to the clinical metrics (used by anesthesiologists in manual and TCI approaches) in the process of calculating infusion rates. As mentioned in \textcolor{blue}{Section \ref{monitoring}}, such physiological parameters may already be available as the measured output of advanced monitors that record and display values including the patient's EMG, EEG, ECG, and BIS. In this case, feedback control is used instead of an anesthesiologist to frequently adjust the infusion profile or target concentration in closed-loop controllers; however, anesthesiologists are still needed to set the desired DoA or DoH and supervise the patient's state via clinical metrics. The result is a reduction in the anesthesiologist's workload, potentially helping to avoid problems associated with distractions and fatigue, as well as increased safety for the patient due to continuous monitoring, the possibility of lower administered drug dosages with a faster postoperative recovery, and a reduction in the occurrence of drug-induced side effects\textcolor{blue}{\cite{merigo2017}}. For these reasons, fully automated systems may have the potential to one-day outperform manual infusion dosing\textcolor{blue}{\cite{liu2006, liu2011}}.

\begin{figure}
\centering
\includegraphics[width=80mm,scale=0.5]{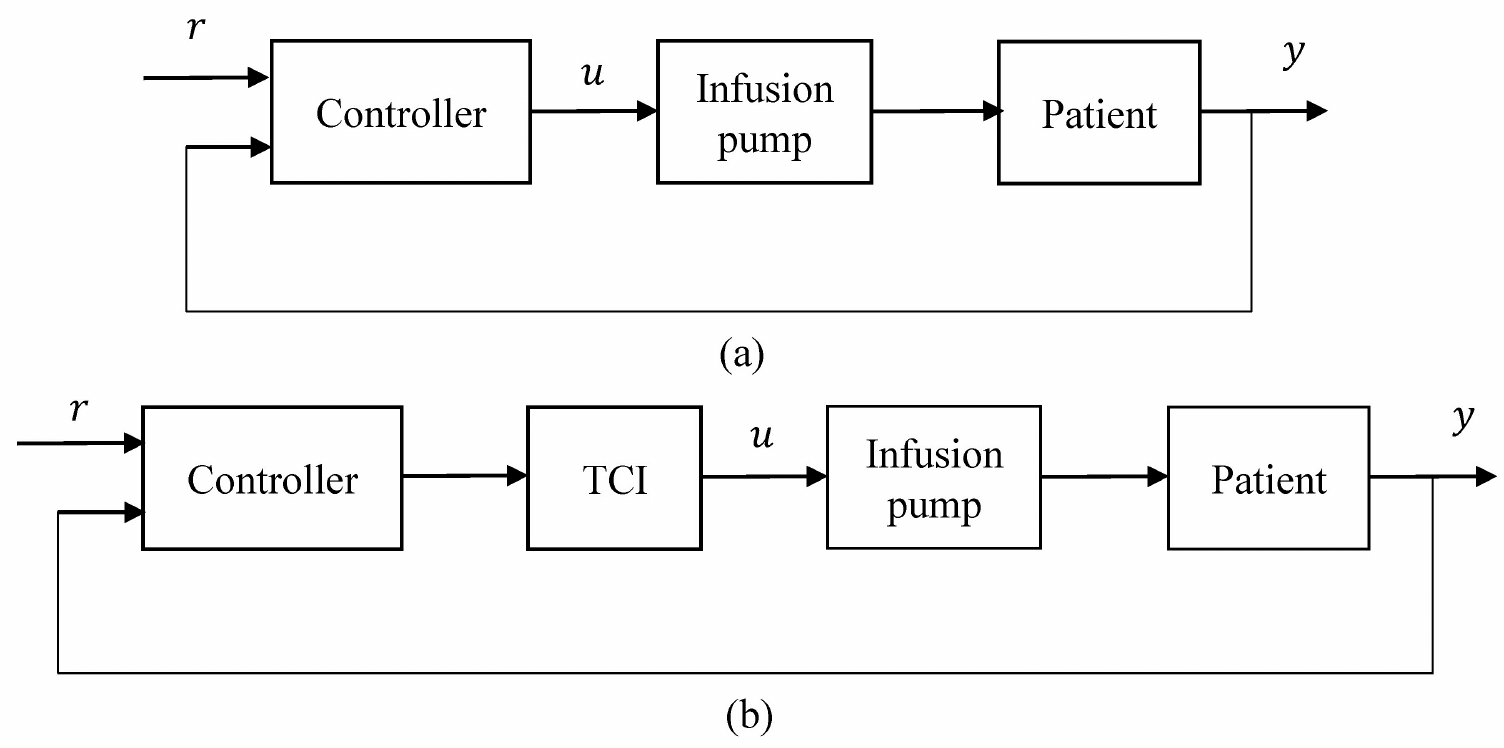}
\caption{Two commonly used closed-loop control architectures in anesthesia drug delivery with (a) direct control of the infusion rate, and (b) cascaded control of TCI system (adapted from\textcolor{blue}{\cite{soltesz2013automation}})}
\label{fig11}
\end{figure}

Two different approaches for implementing closed-loop controllers in anesthesia are shown in \textcolor{blue}{{Fig. }\ref{fig11}}. The first, in \textcolor{blue}{{Fig. }\ref{fig11}a}, is based on a scheme that directly identifies the anesthetic infusion rates and sends control signals directly to the infusion pump (e.g.\textcolor{blue}{\cite{ionescu2008}}). The second, in \textcolor{blue}{{Fig. }\ref{fig11}b}, continuously adjusts the target value for a downstream TCI system, which in turn continuously sets the rate of the infusion pump (e.g.\textcolor{blue}{\cite{desmet2008, struys2001}}). The second method can be considered a special case of the first, and so the first (direct) approach is less restrictive and may be preferable for control design as the TCI dynamics can be replaced with any effective algorithm. 

Generally, automation in anesthesia consists of three main components: a patient model, a measurement system (i.e., sensors or monitors), and a controller. Each of these components can have significant effects on the performance of the other components and the overall system. For example, as mathematical models become more accurate and include more detail on variables and effective parameters, controllers can be designed to provide higher quality drug infusion.  Furthermore, as measurement devices become more accurate, individualized physiological model parameters can be identified more accurately and controllers will have access to a higher quality and quantity of relevant information via feedback and observation - ultimately resulting in improved drug infusion. Note, the quality of monitoring directly effects the controller performance in implementation.

\begin{figure}
\centering
\includegraphics[width=80mm,scale=0.5]{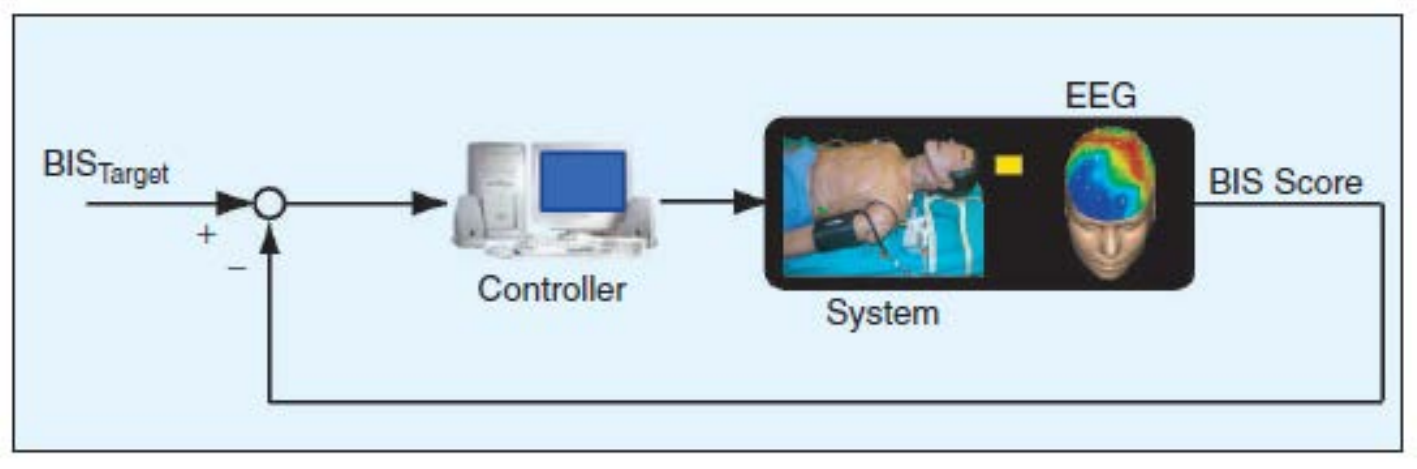}
\caption{Closed-loop control for drug administration in anesthesia (taken from\textcolor{blue}{\cite{bailey2005}}) with permission pending.}
\label{fig12}
\end{figure}

While \textcolor{blue}{Sections \ref{monitoring}} and \textcolor{blue}{\ref{modeling}} have focused on modeling and monitoring, controller design is among the largest remaining challenges to automated anesthesia, and we will now review several prominent strategies. Many scientists and control engineers have explored various control approaches such as PID, MPC, fuzzy-logic, adaptive, and neural networks as well as hybridizations of these controllers to design and produce automatic anesthesia infusion systems for clinical use following the general structure in \textcolor{blue}{{Fig. }\ref{fig12}}. Some comprehensive reviews on controller designs can be found in\textcolor{blue}{\cite{dumont2012, dumont2013, chang2015, ilyas2017, chilcoat1980}}.

While the majority of recent publications in the physiological closed-loop control of anesthesia are related to the use of intravenous anesthesia, several closed-loop studies have specifically considered the application of inhalational anesthesia\textcolor{blue}{\cite{sieber2000model, struys2004time, lortat2009assessing}}. One benefit of the inhalational approach to anesthesia is that it may be considered safer than intravenous anesthesia because end tidal gas measurements are available to monitor the drug concentration within the brain and the vessel rich group (VRG) at steady state. However, these sensors provide a poor approximation of the VRG concentration during common dynamic situations such as initial uptake and emergence\textcolor{blue}{\cite{connor2019optimizing}}. Thus, some researchers have developed mathematical models to represent these dynamics (e.g., for purposes of concentration estimation) with the aim of investigating, by simulation and clinical studies, the closed-loop control of inhalational anesthesia using model predictive control\textcolor{blue}{\cite{mahfouf2003unconstrained, krieger2013modeling, krieger2014}}. For detailed information on modeling and closed-loop control of inhalational anesthesia, interested readers are referred to the extensive coverage in the literature (e.g.,\textcolor{blue}{\cite{krieger2013modellingb}}).

\subsubsection{Challenges to Automated Anesthesia}
During the design and simulation phases of the controller development, there are several important constraints that must be taken into account. First, each drug has a maximum effective dosage, which means higher doses will not have any greater effect on the patient. Second, because the control signal defines the infusion rate, it cannot have a negative value (i.e., the drug cannot be extracted once infused) and must further be within the operating range of available infusion pumps. Third, excessive doses infused over a short time can shock the human body and may harm organ systems\textcolor{blue}{\cite{das2014}}; this can be avoided by maintaining lower infusion rates. The selection of proper drug infusion rates is determined as a trade-off between the side-effects of rapid infusion and the desire to reach the required BIS value within a certain time-frame. In current surgical operations, the optimal time-frame has been reported as around 15 minutes\textcolor{blue}{\cite{nascu2017explicit}}, while shorter periods are frequently obtained in simulation studies found in the literature (e.g., 4-8 minutes in\textcolor{blue}{\cite{padula2017}}). Lastly, the designed controller must work in the presence of both interpatient and intrapatient variability.

\begin{figure}
\centering
\includegraphics[width=80mm,scale=0.5]{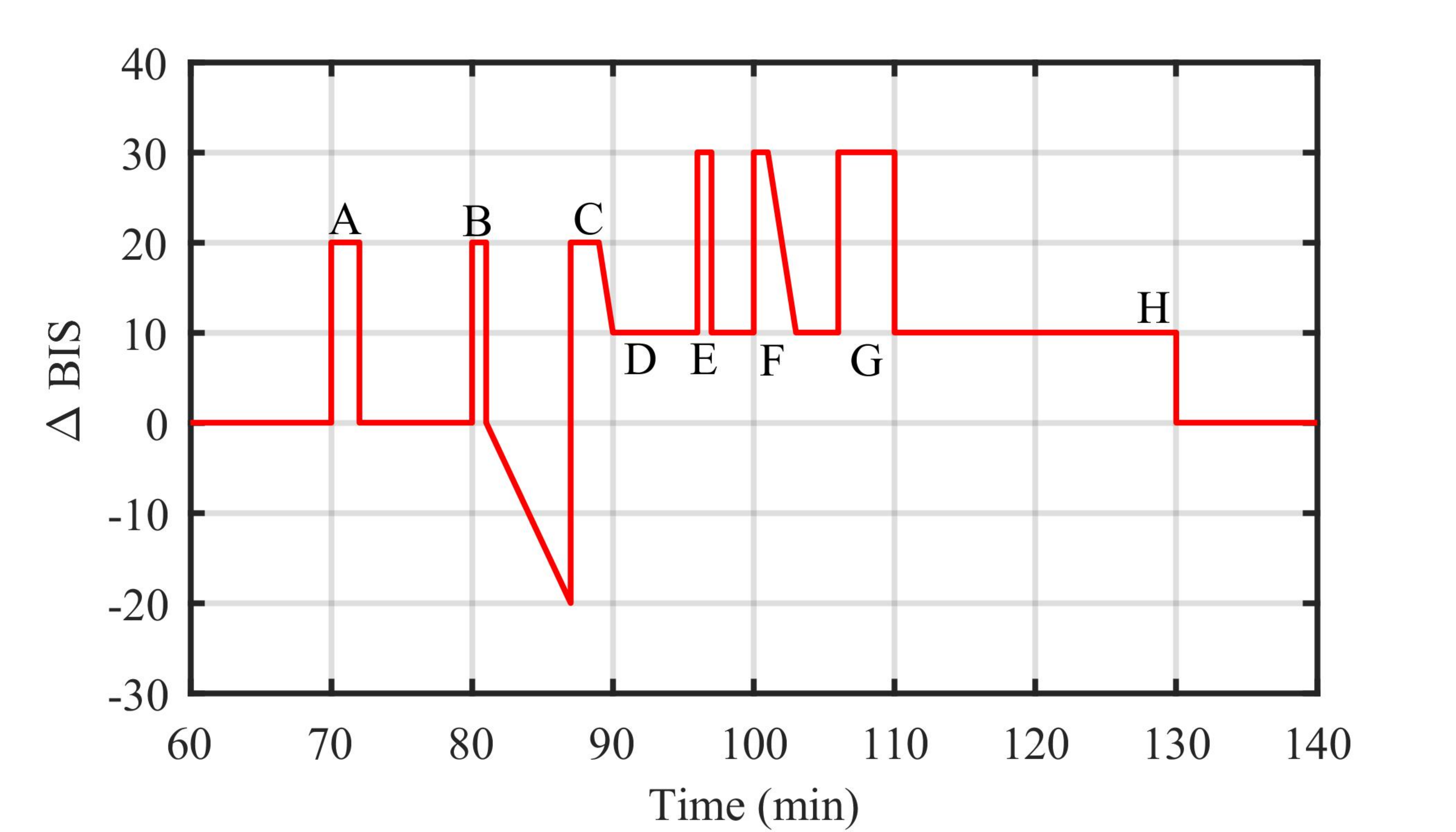}
\caption{A standard surgical stimulation profile (adapted from\textcolor{blue}{\cite{struys2004}}).}
\label{fig13}
\end{figure}

In addition to these systematic constraints, different sources of disturbances and noise -- such as poor signal quality and surgical stimulations -- should be considered in the system modeling and controller design activities. Of these disturbances, surgical stimulations are among the most challenging to manage and are a subject of current investigations by both clinicians and control engineers. As an example of how these stimulations may arise, \textcolor{blue}{{Fig. }\ref{fig13}} shows a typical timeline of the variations in BIS index, caused by surgical stimulations, during the time-course of an archetypal surgical procedure. In this figure, stimulus A shows the arousal due to laryngoscopy/intubation; B represents surgical incision followed by a period of no surgical stimulation (e.g., waiting for laboratory results); C represents an abrupt stimulus after a period of low level stimulation; D represents the onset of continuous normal surgical stimulation; E, F, and G simulate high amplitude short-term stimulations within the surgical period; and H simulates the withdrawal of stimulation during the closing period\textcolor{blue}{\cite{struys2004}}.

As discussed in \textcolor{blue}{Section \ref{monitoring}}, there is currently a need for more accurate, reliable, and robust sensing techniques that provide real-time monitoring of a patient's depth of anesthesia (DoA) without reliance on demographic characteristics or undesirable interference from EMG signals, NMB drugs, or interactions with other substances (e.g., opioids affect AEP measurements\textcolor{blue}{\cite{higgen2018mid}}). The present limitations to these sensing technologies may represent a fundamental barrier to accurate closed-loop control. Consider, for instance, the recent results regarding NMB drugs and BIS\textcolor{blue}{\cite{schuller2015response}}, which have a direct consequence on closed-loop control. According to this study, an anesthetic PCLC device can -- under the right circumstances -- achieve the desired BIS value without actually reaching the desired DoA (e.g., potentially resulting in anesthesia awareness). Even in the case where DoA is generally well represented by the sensing paradigm, automated anesthesia remains challenged by sensor noise, measurement error, and sensing delays.

In the next sections, the most common controllers used in closed-loop control of anesthesia will be discussed, and the recent developments and remaining challenges to each approach are presented. 

\subsubsection{Proportional-Integral-Derivative Control}
Proportional-integral-derivative (PID) control is one of the most used control approaches in industrial applications due to its simple mathematics, well-known gain tuning methods, and broad capabilities. Several successful clinical and simulation studies of automated anesthesia have been performed using PID based controllers over the past two decades such as\textcolor{blue}{\cite{kenny1999, moreley2000, absalom2003, hemmerling2010, liu2012, west2013, vanheusden2014}}. The simulated and experimental results of these studies have generally indicated that PID control may have some utility in the closed-loop control of anesthesia. One of the most common PID controllers used anesthesia is given by the Laplace domain transfer function\textcolor{blue}{\cite{padula2017}}
\begin{equation}
    \frac{U\left( s \right)}{E\left( s \right)}={{k}_{p}}\left( 1+\frac{1}{s{{T}_{i}}}+{{T}_{d}}s \right)\frac{1}{{{\left( {{T}_{f}}s+1 \right)}^{2}}}
    \label{eq40}
\end{equation}

where $U(s)$ and $E(s)$ are, respectively, the Laplace domain control and error signals; $k_p$ is the proportional gain; $T_i$ is the integral time-constant; $T_d$ is the derivative time-constant; and $T_f$ is the time-constant of a second-order filter used to attenuate measurement noise.

In the relevant works, the proposed PID controllers have been shown to track desired anesthesia levels (e.g., desired BIS values); however, anesthetists and control engineers agree that this simple PID algorithm suffers from major deficiencies because it is not a robust and adaptive controller and does not perform well in the presence of noise and physiological variability. As a result, simple PID gives poor disturbance rejection, unwanted oscillations in the BIS response, and can suffer from windup of the integral control term during the induction phase. Therefore, recent studies are trying to address these problems to achieve comparable performance to other more advanced controllers. For instance, in\textcolor{blue}{\cite{simalatsar2016cascaded}} it is shown that the low-pass filter, included in \eqref{eq40}, reduces the sensitivity of the derivative term to random changes. Furthermore, introducing zero reference weight for the proportional and derivative control terms (i.e. no reference input for proportional and derivative parts) results in better disturbance rejection and avoids unwanted oscillation in the  BIS response\textcolor{blue}{\cite{soltesz2013individualized}}. Other investigations have shown benefits from applying a tracking time-constant as the geometric mean of the integral and derivative time-constants of the PID controller\textcolor{blue}{\cite{astrom2006}} and introducing reference shaping (i.e. the desired BIS index is changed from a step input to a specialized profile)\textcolor{blue}{\cite{zhusubaliyev2015}} and integrator anti-windup\textcolor{blue}{\cite{ejaz2004}} to prevent integrator windup during the induction phase. Some alternative approaches that have also been shown to improve the PID controller performance include using an event-based control scheme, which decreases the variations of controller signal and optimally cancels the noise and disturbances, and using an inversion-based methodology, which increases patient safety by reducing BIS overshoot and producing a smoother drug infusion rate\textcolor{blue}{\cite{padula2016}}. From the clinical point of view, event-based control is similar to manual infusion and is more transparent to anesthesiologists because they can easily monitor the drug infusion profile\textcolor{blue}{\cite{merigo2017}}. 

Additional challenges to controller design stem from the presence of nonlinearities (i.e., a Hill equation) in the anesthesia dynamics. One common approach to handle this nonlinearity is to linearize \eqref{eq22} around the nominal operating point (e.g., a BIS value of 50) in the maintenance phase\textcolor{blue}{\cite{anna2010, hahn2012}}. This linearization converts \eqref{eq22} into a constant coefficient $k_m$ given by
\begin{equation}
    {{k}_{m}}=-\frac{BI{{S}_{0\gamma }}}{4E{{C}_{50}}}
    \label{eq41}
\end{equation}
where $BIS_0$ is the nominal operating point in the maintenance phase. However, this approach is not feasible in the induction phase because the lack of a well-defined nominal operating point causes a large overshoot in the response due to the integral term of the controller. Recently, a new approach has been proposed, which is applicable to both induction and maintenance phases. This approach uses a linearizing filter in series with the inverse of the nonlinear part of the Hill function\textcolor{blue}{\cite{soltesz2013individualized, merigo2018a}}. 

Since the structure of PID controllers is generally well-defined, the selection or tuning of controller gains is very important. This task is challenging in closed-loop control of anesthesia, since the gains should be tuned according to the physiological parameters of each patient\textcolor{blue}{\cite{dumont2009, bhole2015}}. While trial and error is common in tuning PID controllers\textcolor{blue}{\cite{hahn2012}}, optimization methods, such as genetic algorithms\textcolor{blue}{\cite{astrom2006, merigo2018optimized, panagopoulos2002}} aimed at minimizing the integrated error, can increase the performance significantly\textcolor{blue}{\cite{ejaz2004, castillo2005}}. Nevertheless, the physiological parameters of patients vary based on age, weight, disease, and type of surgery being performed, and presently the available patient data is limited and does not adequately depict the physiological parameters of all patients. Therefore, online identification of patient parameters can be useful for improving the controller performance\textcolor{blue}{\cite{soltesz2011individualized}}. Additional studies have shown that the use of gain scheduling techniques may also be beneficial. In this case, the gains should be tuned different for different phases of surgery. During the induction phase, the gains are tuned to follow the set-point with a reasonable setting-time and low overshoot. Then, during the maintenance phase, the gains should be tuned for disturbance rejection. Therefore, when DoH or BIS levels reach the target value and have stabilized around the set-point for an adequate time interval (i.e., signaling the end of induction phase) the controller parameters should be switched to those tuned for the maintenance phase\textcolor{blue}{\cite{padula2017}}. 

Another important point is that PID controllers are frequently used in different configurations (e.g., PD or PI only) for different dynamics. However, recent studies have proved that all three parts are necessary in anesthesia control. For example, in\textcolor{blue}{\cite{padula2015}} the derivative part is eliminated to prevent noise amplification, but the performance in the induction phase and disturbance rejection in maintenance phase were less satisfactory for this PI controller than for the PID case, which had a shorter settling-time and less undershoot in the BIS response.

\subsubsection{Model Predictive Control}
Model predictive control (MPC) is a robust and optimal control technique used to handle systems with constraints while optimizing the controller action and predicting system outputs for robustness against noise and disturbances. The applicability of MPC and related strategies (e.g., generalized predictive control (GPC)) to medical systems -- especially in closed-loop control of anesthesia -- has been verified by both simulation and clinical results\textcolor{blue}{\cite{sawaguchi2003model, furutani2005, bamdadian2008generalized, goodwin2019}}. The outstanding features of this controller have recently encouraged control engineers to further develop MPC controllers to address the complex and nonlinear behavior of closed-loop anesthesia control.

One caveat with standard MPC is the complexity of the computations used to solve the online optimization problem, which prevents its use in real-time anesthesia control. This limitation can be overcome by using a multi-parametric MPC (mp-MPC) technique, which reduces the complexity of the computations by solving an offline optimization problem on an analytical function instead of numerical solutions\textcolor{blue}{\cite{ingole2015, nacscu2016a}}. Further, to improve rejection of BIS noise and surgical disturbances, MPC has been implemented in combination with advanced state estimation techniques such as Kalman filtering and moving horizon estimation (MHE). Simulation results show that the MHE method has better accuracy and reduced overshoot compared to Kalman filtering\textcolor{blue}{\cite{darby2007, nacscu2016b}}. Likewise, studies have also shown that the addition of event-based input and state output correction features to MPC can increase its robustness against noise and decrease the amount of anesthetic drugs administered\textcolor{blue}{\cite{nascu2017modeling, pawlowski2017event}}.

Another strong point of MPC is its ability to handle intrapatient and interpatient variability through online estimation of pharmacodynamic parameters\textcolor{blue}{\cite{sawaguchi2008, krieger2014}}, estimation of the slope of the linearized Hill equation at each time-step\textcolor{blue}{\cite{robayo2010estimating, sartori2006line}}, and the use of offset-free and state output correction strategies\textcolor{blue}{\cite{ingole2017, nascu2017modeling}}. 

Piecewise linear PK/PD models can be used to address the problem of nonlinearities in the PD model, which represent one of the main challenges in closed-loop control of anesthesia. This technique defines several linear parts in the hill function to achieve a more accurate linear approximation of this function\textcolor{blue}{\cite{ingole2017, nino2009}}. Likewise, multiplying the inverse of the Hill function
\begin{equation}
    {{C}_{e}}\left( t \right)=W{{C}_{50}}{{\left( \frac{{{E}_{0}}-BIS\left( t \right)}{{{E}_{\max }}-{{E}_{0}}+BIS\left( t \right)} \right)}^{\frac{1}{\gamma }}}
    \label{eq42}
\end{equation}
into the input command and the feedback signal is a common approach to cancel the nonlinear part of the anesthesia system\textcolor{blue}{\cite{nascu2016advanced}}. Other techniques such as linear model predictive control (LMPC)\textcolor{blue}{\cite{sopasakis2014}}. switching the control strategy of mp-MPC to apply bolus treatments in critical BIS indexes\textcolor{blue}{\cite{nacscu2016b}}, extended prediction self-adapting control (EPSAC)\textcolor{blue}{\cite{nino2009}}, and online identification of time delays\textcolor{blue}{\cite{sawaguchi2008}} can improve the performance of closed-loop anesthesia as evidenced by reduced settling-times (i.e., obtaining a faster response), decreased overshoot, and the elimination of unwanted oscillations from the BIS response.

\subsubsection{Adaptive Control}
Adaptive controllers are frequently applied to systems with variable or unknown parameters, and implement algorithms that retune or restructure the effective controller during operation (i.e., online)\textcolor{blue}{\cite{haddad2003adaptive, haddad2003nonlinear}}. As mentioned, the PK/PD model of anesthesia is subjected to large (possibly time varying) uncertainties and its coefficients are different from patient to patient. Thus, adaptive controllers show promise for estimating parameters of the PK/PD model and improving the regulation of anesthesia.

Adaptive controllers that adjust their control action by directly changing controller gains (rather than an internal model) are known as direct adaptive controllers. One such method of direct adaptive control is to implement an adaptive proportional-integral-derivative control algorithm, which allows adaptive control of anesthesia without directly using the PK/PD model\textcolor{blue}{\cite{absalom2002}}. However, in practice, this controller produced undesirable oscillations and its performance was considered unacceptable. Some studies have also shown improvements to the performance of these controllers by considering the nonlinear uncertainties in the PK model and designing nonlinear adaptive controllers\textcolor{blue}{\cite{haddad2006}}. Additionally, neural networks have been used to improve the performance of nonlinear adaptive controllers for dynamic systems with complicated uncertainties\textcolor{blue}{\cite{hayakawa2004}}. In these controllers, states of the physical system remain in nonnegative orthant of the state space\textcolor{blue}{\cite{haddad2011}}. The effectiveness of this type of controller has been investigated by using several experiments and clinical evaluation\textcolor{blue}{\cite{bailey2006}}. 

As an alternative to direct adaptive control, indirect adaptive control uses parameter updating laws to identify unknown parameters in an internal reference model and thereby retune the effective controller gains. Several methods are used to obtain such control laws\textcolor{blue}{\cite{schwilden1987, schwilden1989, struys2001}}, and studies have shown that variations between simulated and real-world control performance can be attenuated by using a minimally parameterized model with an online identification strategy (e.g., an Extended Kalman Filter) to design adaptive nonlinear controllers\textcolor{blue}{\cite{silva2011prediction}}.

Other types of adaptive feedback controllers have also been designed for anesthesia, including model reference adaptive controller (MRAC). However, MRAC controllers based on standard models do not show significant improvements in performance over non-adoptive techniques. Alternatively, MRACs may also be implemented using fractional order models\textcolor{blue}{\cite{navarro2015}}, and the results of studies using fractional order models have confirmed the effectiveness and robustness of fractional order model reference adaptive control (FOMRAC), which can also compensate for the time delays in the anesthesia system\textcolor{blue}{\cite{navarro2017}}. Similarly, L1-adaptive methods can achieve faster adaptation than MRAC\textcolor{blue}{\cite{ralph2011}} and have demonstrated good performance and interpatient robustness\textcolor{blue}{\cite{kharisov2012}}. In addition, system identification methods can reduce the order of patient models while also improving performance over standard L1-adaptive control\textcolor{blue}{\cite{kharisov2013}}. Furthermore, L1-adaptive controllers can be used to ensure the safe switching between human-in-the-loop and automated close-loop modes of operation\textcolor{blue}{\cite{kharisov2015}}. 

\subsubsection{Fuzzy-Logic Control}
Since the compartmental models of anesthesia are not exact representations of the human body and are subject to variability in system parameters, closed-loop control of anesthesia can benefit from fuzzy modeling and control. In fact, fuzzy-logic has been applied successfully to all three compartments of anesthesia (i.e., hypnosis\textcolor{blue}{\cite{simanski2009, chou2010, abdulla2011, mendez2018}}, analgesia\textcolor{blue}{\cite{shieh2007an, shieh2007a, schubert2008fuzzy}} and immobility\textcolor{blue}{\cite{linkens1991, fazendeiro2002}}). Some comprehensive reviews can be found in\textcolor{blue}{\cite{abbod1998, lan2012}}.

The main benefit of using fuzzy-logic for anesthesia is its ability to express patient models by clustering real patient data without any knowledge of the underlying physiology. Most fuzzy anesthesia systems use regular (i.e., type-1) fuzzy sets, which have members that take crisp values in [0 1]; however, some systems use type-2 fuzzy sets where the membership values of each element are in another fuzzy set in [0 1]\textcolor{blue}{\cite{liu2015}}. Finding the best membership function (MF) for such fuzzy sets in [0 1] is one of the main problems with using fuzzy models. To this end, genetic and neural network algorithms are commonly used to improve performance and optimize the type-2 fuzzy sets' parameters (e.g., footprint of uncertainty (FOU), centroid, and scaling factors)\textcolor{blue}{\cite{shieh2006, muthuswamy1999, shieh2002, taheriyan2018applying}}.

To develop fuzzy control laws, sets of rules are collected according to the expert opinions of anesthesiologists regarding the best corrective actions to take under sets of circumstances. Then, these sets of rules are implemented with an aim to create a closed-loop system that mimics their expertise at manual infusion\textcolor{blue}{\cite{zhang2002}}. It should be mentioned that even with these rules, fuzzy-logic controllers have not demonstrated suitable performance without also being combined with genetic and neural network algorithms. Using these algorithms improves the fuzzy controller by allowing adaptation in the presence of interpatient and intrapatient variability\textcolor{blue}{\cite{marrero2017}}.

Frequently, fuzzy-logic controllers are changed to so-called self-organized fuzzy-logic controllers that are capable of adapting to changes in the system\textcolor{blue}{\cite{shieh2009}}. However, these controllers are known to cause steady-state error when used with bolus type therapy. To overcome this, simple fuzzy-logic can be used during bolus treatment and the controller can be switched to the self-organized fuzzy-logic type once the system is operating near the desired set-point\textcolor{blue}{\cite{chou2010}}. As mentioned before, genetic algorithm can then be used for optimizing the fuzzy-logic model\textcolor{blue}{\cite{abbod1998}}.

The majority of the fuzzy-logic controllers are based on the type-1 set which cannot handle model uncertainties, and controllers that are based on type-1 sets suffer from steady-state error in the fuzzy estimation of the desired set-points. However, these problems can be handled by changing to type-2 fuzzy sets\textcolor{blue}{\cite{elbardini2011}}. Moreover, fuzzy neural network controllers can be used to handle uncertainties in anesthesia\textcolor{blue}{\cite{allen2001, nunes2005, mahfouf2005, tosun2010anesthetic}}, but this approach has limited effects on the type-1 fuzzy sets. So, type-2 fuzzy neural network controllers are used to solve this problem\textcolor{blue}{\cite{elnegar2014}}. Furthermore, controllers based on type-2 fuzzy-logic sets can be improved by switching to a self-organized strategy, resulting in self-organized type-2 fuzzy-logic controllers, which can compensate for control uncertainties\textcolor{blue}{\cite{doctor2016}}. By adding genetic algorithms to find the best membership function for the self-organized type-2 fuzzy-logic controller, better results can be obtained in noisy (i.e., real) environments at the cost of increased error in the noise-free environment\textcolor{blue}{\cite{liu2015}}.

As mentioned, type-2 self-organizing fuzzy-logic controllers are effective. However, signal noise and dynamic uncertainties -- such as changes in the PD and PK systems -- can degrade controller performance\textcolor{blue}{\cite{yu2018an}}. In most cases, merging fuzzy-logic controllers with other control schemes can ameliorate the final results. For instance, hybridization of fuzzy-logic with MPC can help to predict the effects of uncertainties in system\textcolor{blue}{\cite{mendez2016, wang2017}}. 

\subsection{Outlook for Automated Anesthesia}
As mentioned previously, there are four different types of controllers that are commonly used in the closed-loop control of anesthesia: proportional-integral-derivative, model predictive, adaptive, and fuzzy-logic controllers. Besides these controllers, strategies such as observer-controllers\textcolor{blue}{\cite{nogueira2016}}, robust and robust deadbeat controllers\textcolor{blue}{\cite{abdulla2012Depth, dumont2009, sadati2018multi}}, nonlinear H-infinity controllers\textcolor{blue}{\cite{rigatos2016nonlinear}}, non-overshooting tracking controller\textcolor{blue}{\cite{padmanabhan2019nonovershooting}}, sliding mode controllers\textcolor{blue}{\cite{iskrenovic2017, abood2019design}}, and other types of nonlinear controllers\textcolor{blue}{\cite{khaqan2016}} have been studied for use in anesthesia. Some of these studies have employed performance metrics such as settling-time, overshoot, undershoot, lowest observed BIS value (BIS-NADIR), performance error (PE), median performance error (MDPE), median absolute performance error (MDAPE), total variation (TV), integrated absolute error (IAE), and WOBBLE (an index of response variations over time) to compare their results and controller effectiveness to other control strategies in attempts to demonstrate advantages over other schemes and find the most suitable class of controllers for anesthesia\textcolor{blue}{\cite{khaqan2017, hodrea2014, yelneedi2009, bamdadian2008generalized}}.

From a control systems perspective, most of the controllers discussed have shown adequate performance. Additionally, some important studies recently have been done regarding safety systems for closed-loop of anesthesia\textcolor{blue}{\cite{yousefi2017formally, yousefi2019formalized}}. However, due to the safety critical nature of anesthesia and consequently restrictive laws, from the clinical point of view, no controller has been completely acceptable for use in a fully closed-loop anesthesia system yet. In fact, none of the four controllers discussed are by themselves able to overcome the complex problem of anesthesia due to the presence of intra- and inter-patient variability, surgical disturbances, and nonlinear dynamics. As recent papers indicate\textcolor{blue}{\cite{mendez2016, wang2017, khodaei2018}}, combinations of these controllers -- optimized to leverage their individual strengths -- show promise for improving the performance of closed-loop anesthesia and achieving acceptable simulation results. 

The development of automated anesthesia systems capable of achieving broad international regulatory approval may benefit from improved physiological models designed to include the interactions of more organ systems, physiological parameters, and clinical signs for an appropriately diverse cohort of subjects and surgical scenarios. If such a model further included all three modes of anesthesia, simulation results might provide a higher degree of confidence in the safety and performance of control algorithms at an early stage. Concurrently, it is also apparent that there is a need for control relevant anesthesia models with reduced model complexity while maintaining the ability to capture intra- and inter-patient variability. The development of such models will require collaboration between clinicians, anesthesiologists, mathematicians, and control engineers.

Furthermore, automated anesthesia would also benefit from developments in clinical monitoring aimed at achieving significant reductions in measurement noise and time-delays as well as the realization of methods to quantify specific aspects of anesthesia. To this end, quantification of the depth of anesthesia may provide interesting opportunities to investigate the use of sensor fusion to develop reliable metrics based on multimodal (and perhaps redundant) monitoring. Since the nonlinear pharmacodynamics of anesthesia are difficult to model, the development of methods to measure the drug concentration in the effect compartment may also help to remove the need for the corresponding nonlinear model equations and introduce a new set-point to complement or replace BIS values. Again considering the shortcomings of current monitoring technologies, it appears that fully autonomous (unsupervised) anesthesia should not be employed, until a reliable real-time measure of DoA is discovered. Under further consideration of the rapidly developing pharmacological and medical fields, the challenge presented by possible as-of-yet unknown drug interactions and patient variability (with respect to DoA metrics) may, in general, provide sufficient cause to opt for moderated levels of automation (LOA) in PCLC anesthesia devices (i.e., always incorporating the anesthesiologist in a supervisory role).

Moving forward, research into the optimal combination of control strategies has the potential to introduce controllers that are feasible for real-world use. Of course, any commercial controller for anesthesia should be able to perform properly in the presence of infusion limitations (e.g., the amount, rate, and frequency of drug infusion), intra- and inter-patient variability, noise, disturbances, and nonlinearities, while also meeting standards for safety and regulatory approval.

\section{Discussion and Future of Physiological Closed-Loop Control} \label{discussion}
Considering the historical and current investigations in PCLCs, including the artificial pancreas and automated anesthesia, there are several common issues that must be handled when implementing control on physiological systems. Chief among these are the ever-present intra- and inter-patient variations that come with age, health, genetics, hormones, medications, physical activity, and a variety of other inherent and environmental factors. The combination of this variability, its many pathways, and the complex nature of physiology leads to challenges in both controller design and system validation. The presence of interpatient variability, or differences in the responses observed between different patients, introduces the need for controller robustness and a large collection of high quality clinical data for validation. However, even with a large collection of data, there remains a question of patient dynamics that are not represented by the data or have not yet been encountered. Furthermore, the presence of intrapatient variability, or changes to the responses observed within a single patient, introduce time-varying dynamics which may further introduce either parametric or structural uncertainties. The overwhelming potential for plant-model mismatch has led to a wide acceptance of adaptive techniques in PCLC formulations, used in combination with robust or predictive control strategies.

Another key issue in PCLC is the presence of strict constraints on therapy, which often allow only positive valued control actions (e.g., substances like propofol and insulin-analogs cannot be removed once injected) and limit both the rates and amplitudes of applied treatments and estimated physiological states. Of course, one of the most important methods in designing systems that combine these requirements with adaptation and robustness are those based on predictive control approaches. Then, the application of prediction to physiological control places a significant burden on the observers and estimators to provide reliable approximations of constrained physiological quantities. While there are arguments for other control methodologies, predictive control methods have been gaining momentum in PCLC and are likely to continue as mathematical models of physiological and pharmaceutical systems develop.

Changing our focus to a different portion of the feedback loop, the ability to obtain accurate and on-time measurements is of utmost importance for both safety and performance. Thus, the development and availability of improved sensing technologies will, in part, dictate the success of PCLCs. Currently, direct access to many measurable physiological states requires highly invasive approaches, which often cannot be justified for long-term or ambulatory use. Thus, less accurate, and possibly delayed, measurements are being used as surrogates, for instance in diabetes, which may impose significant performance limitations. While weighing the risks of trade-offs in measurement quality vs. invasiveness provides a short-term solution in some fields, there is an immediate need for improved and miniaturized measurement techniques, perhaps by translation from other fields, that provide rapid and accurate results without jeopardizing patient comfort or health. If we consider the case of long-term PCLC solutions, worn by patients for months or years or in free-living conditions, other issues arise as well. For instance, many electrochemical sensors limited lifespans. This case is evident in the CGM systems used for artificial pancreas systems, which last at most around 2 weeks prior to replacement and may provide degraded performance at the early and late stages of use. Additional concerns with long-term use include irritation or infections at attachment sites. Further, the unrestricted environments presented by free-living conditions may lead to cases of unforeseen sensor failures or errors, similar to the effects of physical activity on subcutaneous glucose measurements, wherein physiological redistribution of glucose during exercise may bias subcutaneous measurements. On the contrary, the FDA recently approved the first fully implantable glucose sensor, which has a 90-day lifespan and operates on fluorescent sensing technology\textcolor{blue}{\cite{senseonics2018}}; new technologies such as this may provide solutions to some of the classical issues with GCM.

Finally, the potential for interactions between multiple drugs or between drugs and sensing elements, as in the case of Tylenol and some CGM systems, should be considered, to the extent possible, in future investigations. As PCLCs mature and data becomes available, it may be possible to implement such interactions within the in-silico models used for validation. These models have already vastly improved the rate of controller development in the field of diabetes control and been successful in many other areas as well\textcolor{blue}{\cite{colquitt2011}}. The addition of PK-PD models for multiple drugs and known drug interactions within these models may help identify risks or possible failure modes in ambulatory use.

Relaxing, slightly, the definition of physiological control to include systems with a primary goal of achieving a target drug concentration (rather than physiological effect), there is additional growing interest in applying control theory to regulate the concentration of pharmaceuticals within certain organs/tissues in the human body\textcolor{blue}{\cite{tucker2017, sopasakis2014}}. While dosage control alone may reduce the effects of variability in pharmacokinetics and indirectly improve regulation of physiological variables, it is also possible that such dosage control could be incorporated as a target in future MIMO physiological control systems. Of particular note in this adjacent field are the recent success of feedback control using electrochemical-aptamer-based (E-AB) sensors\textcolor{blue}{\cite{arroyocurras2018}}) and the development of long-acting bioresponsive pharmaceuticals\textcolor{blue}{\cite{yu2018advance}}, which implement a form of analog closed-loop control whereby tailored drugs remain inactive in the body until stimulated by biological feedback conditions.

Modern interest in physiological and pharmaceutical control have helped bring about the development of a new field distinguished as mathematical pharmacology\textcolor{blue}{\cite{vandergraaf2016}}, which focuses specifically on modeling the pharmacokinetic and pharmacodynamic properties of substances. Future work in this field is likely to provide direct support to advanced automation in medical practice in the form of new and refined pharmacological models as well as insights into the classes of models\textcolor{blue}{\cite{sopasakis2018}} (and therefore control paradigms) that are well matched to a given drug-dosage or physiological control problem. Furthermore, the integration of mathematical pharmacology with systems physiology has the potential to entrust modeling and control design activities with the respective experts in each area and alleviate the present bottleneck wherein control design and validation is faced with significant uncertainty in part due to the complexity of physiological and pharmaceutical interactions.

Clearly, the development of mathematical models for control design\textcolor{blue}{\cite{parker2001control}} and evaluation remain important areas of research in physiological closed-loop control. Further, while we have mentioned that some methods of control do not require an explicit mathematical model, these methods may be difficult to evaluate in a computational setting, and considering the risks involved with physiological systems the immediate future of such methods in PCLC -- especially with regard to regulatory approval -- is unclear\textcolor{blue}{\cite{parvinian2018}}. Of course, the effectiveness of model-based control is itself subject to the appropriateness of the model(s) used for controller design\textcolor{blue}{\cite{parker2001control}}, and a recent review\textcolor{blue}{\cite{parvinian2019credibility}} has shown that many published methods fall short of completely demonstrating their suitability to the proposed use (e.g., controller design, closed-loop performance evaluation, hardware-in-the-loop testing) in PCLC devices. This problem is exacerbated by the limited data available to many researchers as well as the extraordinary range of potential disturbances, interactions, and variability that affect physiological systems. Ongoing work in the area of credibility assessment for computational patient models\textcolor{blue}{\cite{parvinian2019credibility, mulugeta2018credibility}} and the recently published V\&V 40 standard for "assessing credibility of computational modeling" for medical devices\textcolor{blue}{\cite{asme2018}} from the American Society of Mechanical Engineers (ASME) seek to provide guidance for establishing model credibility based on the specific context of use (COU), model influence in decision making (e.g., selection of infusion rate), and consequences of suboptimal decisions\textcolor{blue}{\cite{parvinian2019credibility}}. Within this framework, researchers are able to qualify their modeling and control results with appropriate rigor and can establish the credibility of models and techniques as new areas of PCLC emerge. This framework may also have significant value for justifying comparisons between control and safety strategies.

With all of these ongoing efforts and the promise of future developments, it seems that the broad introduction of commercially available PCLCs is now imminent, and the coming decades will likely see many breakthroughs in this regard. Thus, pertinent discussions on appropriate regulatory policy are underway\textcolor{blue}{\cite{parvinian2018}}, and similar conversations are being held with regard to ethical implementation of automation in medicine\textcolor{blue}{\cite{virk2018, yang2017}}. The continuation of these discussions and refinement of regulatory and ethical guidelines is essential to the future of PCLCs, for establishing both methods of safe practice and guidelines for design, and also for well-informed selection and use of novel medical instruments by hospitals and physicians.

\section{Conclusion} \label{conclusion}
In the years to come, closed-loop medical devices, artificial intelligence, and networked medical instruments\textcolor{blue}{\cite{dey2018}} will begin to emerge along existing medical robotics\textcolor{blue}{\cite{crawford2016, beasley2012}}; and physiological closed-loop control systems will become visible within the clinical, home-care, and ambulatory environments. Ultimately, it is possible that many of these devices will be linked to an electronic health network, which could maintain critical patient data or even allow physicians to monitor and adjust treatment goals remotely.

The integration of feedback control and artificial intelligence in medical systems has the potential to improve adherence to prescribed treatment regimens and protocols, rapidly adapt to new or changing therapy strategies, provide access to treatment in remote areas, optimize the utilization of available resources in scenarios with supply limitations, and increase the capacity of the individual clinician in disaster response and humanitarian crises. The consistency of treatment and availability of data that follows automation may be a useful tool for evaluating the performance of certain clinical practices, validating new treatment options, creating innovative therapies, and enabling the wide-spread availability of personalized medicine. These achievements, and others yet to be conceived, will not be obtained without navigating substantial challenges but appear within reach during the 21st century.

\bibliographystyle{IEEEtran}
\bibliography{PCLCReferences.bib}

\begin{thebibliography}{100}
\providecommand{\url}[1]{#1}
\csname url@samestyle\endcsname
\providecommand{\newblock}{\relax}
\providecommand{\bibinfo}[2]{#2}
\providecommand{\BIBentrySTDinterwordspacing}{\spaceskip=0pt\relax}
\providecommand{\BIBentryALTinterwordstretchfactor}{4}
\providecommand{\BIBentryALTinterwordspacing}{\spaceskip=\fontdimen2\font plus
\BIBentryALTinterwordstretchfactor\fontdimen3\font minus
  \fontdimen4\font\relax}
\providecommand{\BIBforeignlanguage}[2]{{%
\expandafter\ifx\csname l@#1\endcsname\relax
\typeout{** WARNING: IEEEtran.bst: No hyphenation pattern has been}%
\typeout{** loaded for the language `#1'. Using the pattern for}%
\typeout{** the default language instead.}%
\else
\language=\csname l@#1\endcsname
\fi
#2}}
\providecommand{\BIBdecl}{\relax}
\BIBdecl

\bibitem{palerm2011}
C.~C. Palerm, ``Physiologic insulin delivery with insulin feedback: a control
  systems perspective,'' \emph{Computer Methods and Programs in Biomedicine},
  vol. 102, no.~2, pp. 130--137, 2011.

\bibitem{doctor2016}
F.~Doctor, C.-H. Syue, Y.-X. Liu, J.-S. Shieh, and R.~Iqbal, ``Type-2 fuzzy
  sets applied to multivariable self-organizing fuzzy logic controllers for
  regulating anesthesia,'' \emph{Applied Soft Computing}, vol.~38, pp.
  872--889, 2016.

\bibitem{luspay2015}
T.~Luspay and K.~Grigoriadis, ``Robust linear parameter-varying control of
  blood pressure using vasoactive drugs,'' \emph{International Journal of
  Control}, vol.~88, no.~10, pp. 2013--2029, 2015.

\bibitem{gaweda2008}
A.~E. Gaweda, A.~A. Jacobs, G.~R. Aronoff, and M.~E. Brier, ``Model predictive
  control of erythropoietin administration in the anemia of esrd,''
  \emph{American Journal of Kidney Diseases}, vol.~51, no.~1, pp. 71--79, 2008.

\bibitem{cunningham2018}
J.~J. Cunningham, J.~S. Brown, R.~A. Gatenby, and K.~Sta{\v{n}}kov{\'a},
  ``Optimal control to develop therapeutic strategies for metastatic castrate
  resistant prostate cancer,'' \emph{Journal of Theoretical Biology}, vol. 459,
  pp. 67--78, 2018.

\bibitem{alomari2012}
A.-H.~H. AlOmari, A.~V. Savkin, M.~Stevens, D.~G. Mason, D.~L. Timms, R.~F.
  Salamonsen, and N.~H. Lovell, ``Developments in control systems for rotary
  left ventricular assist devices for heart failure patients: a review,''
  \emph{Physiological Measurement}, vol.~34, no.~1, pp. R1--R27, 2012.

\bibitem{petrou2018}
A.~Petrou, J.~Lee, S.~Dual, G.~Ochsner, M.~Meboldt, and M.~Schmid~Daners,
  ``Standardized comparison of selected physiological controllers for rotary
  blood pumps: In vitro study,'' \emph{Artificial Organs}, vol.~42, no.~3, pp.
  E29--E42, 2018.

\bibitem{rinehart2013}
J.~Rinehart, C.~Lee, M.~Cannesson, and G.~Dumont, ``Closed-loop fluid
  resuscitation: robustness against weight and cardiac contractility
  variations,'' \emph{Anesthesia \& Analgesia}, vol. 117, no.~5, pp.
  1110--1118, 2013.

\bibitem{parvinian2018}
B.~Parvinian, C.~Scully, H.~Wiyor, A.~Kumar, and S.~Weininger, ``Regulatory
  considerations for physiological closed-loop controlled medical devices used
  for automated critical care: Food and drug administration workshop discussion
  topics,'' \emph{Anesthesia \& Analgesia}, vol. 126, no.~6, pp. 1916--1925,
  2018.

\bibitem{food2014food}
{Food and Drug Administration}, ``Food and drug administration modernization
  act of 1997: modifications to the list of recognized standards, recognition
  list number: 036,'' pp. 38\,910--38\,910, 2014.

\bibitem{health2019}
\BIBentryALTinterwordspacing
{Health Canada}, ``List of recognized standards for medical devices,'' 2019.
  [Online]. Available:
  \url{https://www.canada.ca/en/health-canada/services/drugs-health-products/medical-devices/standards/list-recognized-standards-medical-devices-guidance.html}
\BIBentrySTDinterwordspacing

\bibitem{European2017}
\BIBentryALTinterwordspacing
{European Commission}, ``Commission communication in the framework of the
  implementation of the council directive 93/42/eec concerning medical devices
  (publication of titles and references of harmonised standards under union
  harmonisation legislation) text with eea relevance,'' pp. 29--61, 2017.
  [Online]. Available:
  \url{https://eur-lex.europa.eu/legal-content/EN/TXT/PDF/?uri=OJ:C:2017:389:FULL\&from=EN}
\BIBentrySTDinterwordspacing

\bibitem{medical2007}
``Medical electrical equipment - part 1-10: General requirements for basic
  safety and essential performance - collateral standard: Requirements for the
  development of physiologic closed-loop controllers,'' 2007.

\bibitem{distefano2015}
J.~DiStefano~III, \emph{Dynamic systems biology modeling and simulation}.\hskip
  1em plus 0.5em minus 0.4em\relax Academic Press, 2015.

\bibitem{soliman2011}
S.~A.-H. Soliman and A.-A.~H. Mantawy, \emph{Modern optimization techniques
  with applications in electric power systems}.\hskip 1em plus 0.5em minus
  0.4em\relax Springer Science \& Business Media, 2011.

\bibitem{mamdani1975}
E.~H. Mamdani and S.~Assilian, ``An experiment in linguistic synthesis with a
  fuzzy logic controller,'' \emph{International Journal of Man-Machine
  Studies}, vol.~7, no.~1, pp. 1--13, 1975.

\bibitem{takagi1985}
T.~{Takagi} and M.~{Sugeno}, ``Fuzzy identification of systems and its
  applications to modeling and control,'' \emph{IEEE Transactions on Systems,
  Man, and Cybernetics}, vol. SMC-15, no.~1, pp. 116--132, 1985.

\bibitem{sugeno1988}
M.~Sugeno and G.~Kang, ``Structure identification of fuzzy model,'' \emph{Fuzzy
  Sets and Systems}, vol.~28, no.~1, pp. 15--33, 1988.

\bibitem{roder2016pancreatic}
P.~V. R{\"o}der, B.~Wu, Y.~Liu, and W.~Han, ``Pancreatic regulation of glucose
  homeostasis,'' \emph{Experimental \& Molecular Medicine}, vol.~48, no.~3, p.
  e219, 2016.

\bibitem{aronoff2004}
S.~L. Aronoff, K.~Berkowitz, B.~Shreiner, and L.~Want, ``Glucose metabolism and
  regulation: beyond insulin and glucagon,'' \emph{Diabetes Spectrum}, vol.~17,
  no.~3, pp. 183--190, 2004.

\bibitem{2association2018}
{American Diabetes Association}, ``2. classification and diagnosis of diabetes:
  Standards of medical care in diabetes{\textemdash}2018,'' \emph{Diabetes
  Care}, vol.~41, no. Supplement 1, pp. S13--S27, 2018.

\bibitem{punthakee2018}
Z.~Punthakee, R.~Goldenberg, and P.~Katz, ``Definition, classification and
  diagnosis of diabetes, prediabetes and metabolic syndrome,'' \emph{Canadian
  Journal of Diabetes}, vol.~42, pp. S10--S15, 2018.

\bibitem{ozougwu2013}
J.~Ozougwu, K.~Obimba, C.~Belonwu, and C.~Unakalamba, ``The pathogenesis and
  pathophysiology of type 1 and type 2 diabetes mellitus,'' \emph{Journal of
  Physiology and Pathophysiology}, vol.~4, no.~4, pp. 46--57, 2013.

\bibitem{dimeglio2018}
L.~A. DiMeglio, C.~Evans-Molina, and R.~A. Oram, ``Type 1 diabetes,'' \emph{The
  Lancet}, vol. 391, no. 10138, pp. 2449--2462, 2018.

\bibitem{stidsen2018}
J.~V. Stidsen, J.~E. Henriksen, M.~H. Olsen, R.~W. Thomsen, J.~S. Nielsen,
  J.~Rungby, S.~P. Ulrichsen, K.~Berencsi, J.~A. Kahlert, S.~G. Friborg
  \emph{et~al.}, ``Pathophysiology-based phenotyping in type 2 diabetes: A
  clinical classification tool,'' \emph{Diabetes/Metabolism Research and
  Reviews}, vol.~34, no.~5, p. e3005, 2018.

\bibitem{ahlqvist2018}
E.~Ahlqvist, P.~Storm, A.~K{\"a}r{\"a}j{\"a}m{\"a}ki, M.~Martinell, M.~Dorkhan,
  A.~Carlsson, P.~Vikman, R.~B. Prasad, D.~M. Aly, P.~Almgren \emph{et~al.},
  ``Novel subgroups of adult-onset diabetes and their association with
  outcomes: a data-driven cluster analysis of six variables,'' \emph{The Lancet
  Diabetes \& Endocrinology}, vol.~6, no.~5, pp. 361--369, 2018.

\bibitem{gedebjerg2018}
A.~Gedebjerg, T.~P. Almdal, K.~Berencsi, J.~Rungby, J.~S. Nielsen, D.~R. Witte,
  S.~Friborg, I.~Brandslund, A.~Vaag, H.~Beck-Nielsen \emph{et~al.},
  ``Prevalence of micro-and macrovascular diabetes complications at time of
  type 2 diabetes diagnosis and associated clinical characteristics: a
  cross-sectional baseline study of 6958 patients in the danish dd2 cohort,''
  \emph{Journal of Diabetes and its Complications}, vol.~32, no.~1, pp. 34--40,
  2018.

\bibitem{wexler2018}
D.~J. Wexler, C.~E. Powe, L.~A. Barbour, T.~Buchanan, D.~R. Coustan, R.~Corcoy,
  P.~Damm, F.~Dunne, D.~S. Feig, and A.~Ferrara, ``Research gaps in gestational
  diabetes mellitus: executive summary of a national institute of diabetes and
  digestive and kidney diseases workshop,'' \emph{Obstetrics \& Gynecology},
  vol. 132, no.~2, pp. 496--505, 2018.

\bibitem{6association2018}
{American Diabetes Association}, ``6. glycemic targets: Standards of medical
  care in diabetes{\textemdash}2018,'' \emph{Diabetes Care}, vol.~41, no.
  Supplement 1, pp. S55--S64, 2018.

\bibitem{8association2018}
------, ``8. pharmacologic approaches to glycemic treatment: standards of
  medical care in diabetes-2018,'' \emph{Diabetes Care}, vol.~41, no.
  Supplement 1, pp. S73--S85, 2018.

\bibitem{adofsson2018}
P.~Adolfsson, C.~G. Parkin, A.~Thomas, and L.~G. Krinelke, ``Selecting the
  appropriate continuous glucose monitoring system-a practical approach,''
  \emph{European Endocrinology}, vol.~14, no.~1, pp. 24--29, 2018.

\bibitem{sacks2011}
D.~B. Sacks, M.~Arnold, G.~L. Bakris, D.~E. Bruns, A.~R. Horvath, M.~S.
  Kirkman, A.~Lernmark, B.~E. Metzger, and D.~M. Nathan, ``Guidelines and
  recommendations for laboratory analysis in the diagnosis and management of
  diabetes mellitus,'' \emph{Clinical Chemistry}, vol.~57, no.~6, pp. e1--e47,
  2011.

\bibitem{leelarathna2018}
L.~Leelarathna and E.~Wilmot, ``Flash forward: a review of flash glucose
  monitoring,'' \emph{Diabetic Medicine}, vol.~35, no.~4, pp. 472--482, 2018.

\bibitem{wood2018}
A.~Wood, D.~O'neal, J.~Furler, and E.~I. Ekinci, ``Continuous glucose
  monitoring: a review of the evidence, opportunities for future use and
  ongoing challenges,'' \emph{Internal medicine journal}, vol.~48, no.~5, pp.
  499--508, 2018.

\bibitem{medtronic2014}
\BIBentryALTinterwordspacing
Medtronic, ``Enlite: Glucose sensor user guide,'' 2014. [Online]. Available:
  \url{https://www.medtronicdiabetes.com/sites/default/files/library/download-library/user-guides/Mp6025676-2AF1\_a\_pdf.pdf}
\BIBentrySTDinterwordspacing

\bibitem{medtronic2017}
\BIBentryALTinterwordspacing
------, ``Guardian sensor (3) performance,'' 2017. [Online]. Available:
  \url{https://www.medtronicdiabetes.com/sites/default/files/library/download-library/user-guides/MP6026113-2AF1DOC\_A\_FINAL.pdf}
\BIBentrySTDinterwordspacing

\bibitem{medtronic2018}
\BIBentryALTinterwordspacing
------, ``Guardian connect: System user guide,'' 2018. [Online]. Available:
  \url{https://www.medtronicdiabetes.com/sites/default/files/library/download-library/user-guides/guardian-connect-v3\_2/us-manual.pdf.}
\BIBentrySTDinterwordspacing

\bibitem{medtronic20183}
\BIBentryALTinterwordspacing
------, ``Guardian sensor(3): User guide,'' 2018. [Online]. Available:
  \url{https://www.medtronicdiabetes.com/sites/default/files/library/download-library/user-guides/Guardian\%20Sensor\%203\%20User\%20Guide\%20-\%20June-\%202018.pdf}
\BIBentrySTDinterwordspacing

\bibitem{dexcom2017}
\BIBentryALTinterwordspacing
Dexcom, ``User's guide: Dexcom {G}4\textregistered \ platinum continuous
  glucose monitoring system receiver with share,'' 2017. [Online]. Available:
  \url{https://s3-us-west-2.amazonaws.com/dexcompdf/LBL012528+Rev+004+User's+Guide\%2-C+G4+PLATINUM+with+Share+US+Web+with+cover.pdf}
\BIBentrySTDinterwordspacing

\bibitem{5dexcom2018}
\BIBentryALTinterwordspacing
------, ``User's guide: Dexcom {G}5\textregistered \ mobile continuous glucose
  monitoring system,'' 2018. [Online]. Available:
  \url{https://s3-us-west-2.amazonaws.com/dexcompdf/G5-Mobile-Users-Guide.pdf}
\BIBentrySTDinterwordspacing

\bibitem{6dexcom2018}
\BIBentryALTinterwordspacing
------, ``User's guide: Dexcom {G}6\textregistered \ continuous glucose
  monitoring system,'' 2018. [Online]. Available:
  \url{https://s3-us-west-2.amazonaws.com/dexcompdf/G6-CGM-Users-Guide.pdf}
\BIBentrySTDinterwordspacing

\bibitem{houlden2018}
R.~L. Houlden, L.~Berard, J.~M. Lakoff, V.~Woo, J.-F. Yale, and D.~C. C. P.
  G.~E. Committee, ``Diabetes and driving,'' \emph{Canadian Journal of
  Diabetes}, vol.~42, pp. S150--S153, 2018.

\bibitem{castle2010}
J.~R. Castle and W.~K. Ward, ``Amperometric glucose sensors: sources of error
  and potential benefit of redundancy,'' \emph{Journal of Diabetes Science and
  Technology}, vol.~4, no.~1, pp. 221--225, 2010.

\bibitem{basu2017}
A.~Basu, M.~Q. Slama, W.~T. Nicholson, L.~Langman, T.~Peyser, R.~Carter, and
  R.~Basu, ``Continuous glucose monitor interference with commonly prescribed
  medications: a pilot study,'' \emph{Journal of diabetes science and
  technology}, vol.~11, no.~5, pp. 936--941, 2017.

\bibitem{bruen2017}
D.~Bruen, C.~Delaney, L.~Florea, and D.~Diamond, ``Glucose sensing for diabetes
  monitoring: recent developments,'' \emph{Sensors}, vol.~17, no.~8, 2017.

\bibitem{kim2018}
J.~Kim, A.~S. Campbell, and J.~Wang, ``Wearable non-invasive epidermal glucose
  sensors: A review,'' \emph{Talanta}, vol. 177, pp. 163--170, 2018.

\bibitem{chatterjee2017}
S.~Chatterjee, K.~Khunti, and M.~J. Davies, ``Type 2 diabetes,'' \emph{The
  Lancet}, vol. 389, no. 10085, pp. 2239--2251, 2017.

\bibitem{tomaro2017}
E.~S. Tomaro, R.~Pippi, E.~Reginato, C.~Aiello, L.~Buratta, C.~Mazzeschi,
  C.~Perrone, C.~Ranucci, A.~Tirimagni, and A.~Russo, ``Intensive lifestyle
  intervention is particularly advantageous in poorly controlled type 2
  diabetes,'' \emph{Nutrition, Metabolism and Cardiovascular Diseases},
  vol.~27, no.~8, pp. 688--694, 2017.

\bibitem{4associaion2018}
``4. lifestyle management: Standards of medical care in
  diabetes{\textemdash}2018,'' \emph{Diabetes Care}, vol.~41, no. Supplement 1,
  pp. S38--S50, 2018.

\bibitem{moore2018}
L.~E. Moore, ``Insulin,'' in \emph{Diabetes in Pregnancy}.\hskip 1em plus 0.5em
  minus 0.4em\relax Springer, 2018, pp. 87--101.

\bibitem{pickup1978}
J.~Pickup, H.~Keen, J.~Parsons, and K.~Alberti, ``Continuous subcutaneous
  insulin infusion: an approach to achieving normoglycaemia,'' \emph{BMJ},
  vol.~1, no. 6107, pp. 204--207, 1978.

\bibitem{pozzilli2016}
P.~Pozzilli, T.~Battelino, T.~Danne, R.~Hovorka, P.~Jarosz-Chobot, and
  E.~Renard, ``Continuous subcutaneous insulin infusion in diabetes: patient
  populations, safety, efficacy, and pharmacoeconomics,''
  \emph{Diabetes/Metabolism Research and Reviews}, vol.~32, no.~1, pp. 21--39,
  2016.

\bibitem{oviedo2017}
S.~Oviedo, J.~Veh{\'\i}, R.~Calm, and J.~Armengol, ``A review of personalized
  blood glucose prediction strategies for t1dm patients,'' \emph{International
  Journal for Numerical Methods in Biomedical Engineering}, vol.~33, no.~6, p.
  e2833, 2017.

\bibitem{fritzen2018}
K.~Fritzen, L.~Heinemann, and O.~Schnell, ``Modeling of diabetes and its
  clinical impact,'' \emph{Journal of Diabetes Science and Technology},
  vol.~12, no.~5, pp. 976--984, 2018.

\bibitem{colmegna2014}
P.~Colmegna and R.~S. Pe{\~n}a, ``Analysis of three {T1DM} simulation models
  for evaluating robust closed-loop controllers,'' \emph{Computer Methods and
  Programs in Biomedicine}, vol. 113, no.~1, pp. 371--382, 2014.

\bibitem{cobelli2014}
C.~Cobelli, C.~Dalla~Man, M.~G. Pedersen, A.~Bertoldo, and G.~Toffolo,
  ``Advancing our understanding of the glucose system via modeling: a
  perspective,'' \emph{IEEE Transactions on Biomedical Engineering}, vol.~61,
  no.~5, pp. 1577--1592, 2014.

\bibitem{ajmera2013}
I.~Ajmera, M.~Swat, C.~Laibe, N.~Le~Novere, and V.~Chelliah, ``The impact of
  mathematical modeling on the understanding of diabetes and related
  complications,'' \emph{CPT: Pharmacometrics \& Systems Pharmacology}, vol.~2,
  no.~7, pp. 1--14, 2013.

\bibitem{wilinska2008}
M.~E. Wilinska and R.~Hovorka, ``Simulation models for in silico testing of
  closed-loop glucose controllers in type 1 diabetes,'' \emph{Drug Discovery
  Today: Disease Models}, vol.~5, no.~4, pp. 289--298, 2008.

\bibitem{bergman1979}
R.~N. Bergman, Y.~Z. Ider, C.~R. Bowden, and C.~Cobelli, ``Quantitative
  estimation of insulin sensitivity,'' \emph{American Journal of
  Physiology-Endocrinology And Metabolism}, vol. 236, no.~6, p. E667, 1979.

\bibitem{bergman2005}
R.~N. Bergman, ``Minimal model: perspective from 2005,'' \emph{Hormone Research
  in Paediatrics}, vol.~64, no. Suppl. 3, pp. 8--15, 2005.

\bibitem{man2014}
C.~D. Man, F.~Micheletto, D.~Lv, M.~Breton, B.~Kovatchev, and C.~Cobelli, ``The
  uva/padova type 1 diabetes simulator: new features,'' \emph{Journal of
  Diabetes Science and Technology}, vol.~8, no.~1, pp. 26--34, 2014.

\bibitem{wilinska2010}
M.~E. Wilinska, L.~J. Chassin, C.~L. Acerini, J.~M. Allen, D.~B. Dunger, and
  R.~Hovorka, ``Simulation environment to evaluate closed-loop insulin delivery
  systems in type 1 diabetes,'' \emph{Journal of Diabetes Science and
  Technology}, vol.~4, no.~1, pp. 132--144, 2010.

\bibitem{cobelli2009}
C.~Cobelli, C.~Dalla~Man, G.~Sparacino, L.~Magni, G.~De~Nicolao, and B.~P.
  Kovatchev, ``Diabetes: models, signals, and control,'' \emph{IEEE Reviews in
  Biomedical Engineering}, vol.~2, pp. 54--96, 2009.

\bibitem{caumo2000}
A.~Caumo, R.~N. Bergman, and C.~Cobelli, ``Insulin sensitivity from meal
  tolerance tests in normal subjects: a minimal model index,'' \emph{The
  Journal of Clinical Endocrinology \& Metabolism}, vol.~85, no.~11, pp.
  4396--4402, 2000.

\bibitem{fabietti2006}
P.~G. Fabietti, V.~Canonico, M.~O. Federici, M.~M. Benedetti, and E.~Sarti,
  ``Control oriented model of insulin and glucose dynamics in type 1
  diabetics,'' \emph{Medical and Biological Engineering and Computing},
  vol.~44, no. 1-2, pp. 69--78, 2006.

\bibitem{schiavon2014}
M.~Schiavon, C.~Dalla~Man, Y.~C. Kudva, A.~Basu, and C.~Cobelli, ``Quantitative
  estimation of insulin sensitivity in type 1 diabetic subjects wearing a
  sensor-augmented insulin pump,'' \emph{Diabetes Care}, vol.~37, no.~5, pp.
  1216--1223, 2014.

\bibitem{cinar2018turksoy-modeling}
A.~Cinar and K.~Turksoy, ``Modeling glucose and insulin concentration
  dynamics,'' in \emph{Advances in Artificial Pancreas Systems}.\hskip 1em plus
  0.5em minus 0.4em\relax Springer, 2018, pp. 33--50.

\bibitem{kovatchev2009silico}
B.~P. Kovatchev, M.~Breton, C.~D. Man, and C.~Cobelli, ``In silico preclinical
  trials: A proof of concept in closed-loop control of type 1 diabetes,''
  \emph{Journal of Diabetes Science and Technology}, vol.~3, no.~1, pp. 44--55,
  2009.

\bibitem{dallaman2007}
C.~Dalla~Man, R.~A. Rizza, and C.~Cobelli, ``Meal simulation model of the
  glucose-insulin system,'' \emph{IEEE Transactions on Biomedical Engineering},
  vol.~54, no.~10, pp. 1740--1749, 2007.

\bibitem{theepsilongroup2017}
\BIBentryALTinterwordspacing
{The Epsilon Group}, ``{T1DMS} - the epsilon group,'' 2017. [Online].
  Available: \url{https://tegvirginia.com/software/t1dms-2014/}
\BIBentrySTDinterwordspacing

\bibitem{visentin2018}
R.~Visentin, E.~Campos-N{\'a}{\~n}ez, M.~Schiavon, D.~Lv, M.~Vettoretti,
  M.~Breton, B.~P. Kovatchev, C.~Dalla~Man, and C.~Cobelli, ``The uva/padova
  type 1 diabetes simulator goes from single meal to single day,''
  \emph{Journal of Diabetes Science and Technology}, vol.~12, no.~2, pp.
  273--281, 2018.

\bibitem{magni2009}
L.~Magni, D.~M. Raimondo, C.~Dalla~Man, G.~De~Nicolao, B.~Kovatchev, and
  C.~Cobelli, ``Model predictive control of glucose concentration in type i
  diabetic patients: An in silico trial,'' \emph{Biomedical Signal Processing
  and Control}, vol.~4, no.~4, pp. 338--346, 2009.

\bibitem{incremona2018}
G.~P. Incremona, M.~Messori, C.~Toffanin, C.~Cobelli, and L.~Magni, ``Model
  predictive control with integral action for artificial pancreas,''
  \emph{Control Engineering Practice}, vol.~77, pp. 86--94, 2018.

\bibitem{hovorka2004}
R.~Hovorka, V.~Canonico, L.~J. Chassin, U.~Haueter, M.~Massi-Benedetti, M.~O.
  Federici, T.~R. Pieber, H.~C. Schaller, L.~Schaupp, and T.~Vering,
  ``Nonlinear model predictive control of glucose concentration in subjects
  with type 1 diabetes,'' \emph{Physiological Measurement}, vol.~25, no.~4, pp.
  905--920, 2004.

\bibitem{wilinska2014}
M.~E. Wilinska and R.~Hovorka, ``Simulation models for in-silico evaluation of
  closed-loop insulin delivery systems in type 1 diabetes,'' in
  \emph{Data-driven Modeling for Diabetes}.\hskip 1em plus 0.5em minus
  0.4em\relax Springer, 2014, pp. 131--149.

\bibitem{boiroux2018}
D.~Boiroux, A.~K. Duun-Henriksen, S.~Schmidt, K.~N{\o}rgaard, S.~Madsbad, N.~K.
  Poulsen, H.~Madsen, and J.~B. J{\o}rgensen, ``Overnight glucose control in
  people with type 1 diabetes,'' \emph{Biomedical Signal Processing and
  Control}, vol.~39, pp. 503--512, 2018.

\bibitem{colmegna2018linear}
P.~Colmegna, R.~S{\'a}nchez-Pe{\~n}a, and R.~Gondhalekar, ``Linear
  parameter-varying model to design control laws for an artificial pancreas,''
  \emph{Biomedical Signal Processing and Control}, vol.~40, pp. 204--213, 2018.

\bibitem{huyett2018}
L.~M. Huyett, E.~Dassau, H.~C. Zisser, and F.~J. Doyle~III, ``Glucose sensor
  dynamics and the artificial pancreas: The impact of lag on sensor measurement
  and controller performance,'' \emph{IEEE Control Systems}, vol.~38, no.~1,
  pp. 30--46, 2018.

\bibitem{nucci200}
G.~Nucci and C.~Cobelli, ``Models of subcutaneous insulin kinetics. a critical
  review,'' \emph{Computer methods and programs in biomedicine}, vol.~62,
  no.~3, pp. 249--257, 2000.

\bibitem{schiavon2018}
M.~Schiavon, C.~Dalla~Man, and C.~Cobelli, ``Modeling subcutaneous absorption
  of fast-acting insulin in type 1 diabetes,'' \emph{IEEE Transactions on
  Biomedical Engineering}, vol.~65, no.~9, pp. 2079--2086, 2018.

\bibitem{kadish1963}
A.~H. Kadish, ``Automation control of blood sugar a servomechanism for glucose
  monitoring and control,'' \emph{ASAIO Journal}, vol.~9, no.~1, pp. 363--367,
  1963.

\bibitem{cobelli2011}
C.~Cobelli, E.~Renard, and B.~Kovatchev, ``Artificial pancreas: past, present,
  future,'' \emph{Diabetes}, vol.~60, no.~11, pp. 2672--2682, 2011.

\bibitem{hovorka2011}
R.~Hovorka, ``Closed-loop insulin delivery: from bench to clinical practice,''
  \emph{Nature Reviews Endocrinology}, vol.~7, no.~7, pp. 385--395, 2011.

\bibitem{cinar2018}
A.~Cinar, ``Artificial pancreas systems: An introduction to the special
  issue,'' \emph{IEEE Control Systems}, vol.~38, no.~1, pp. 26--29, 2018.

\bibitem{turksoy2017}
K.~Turksoy, N.~Frantz, L.~Quinn, M.~Dumin, J.~Kilkus, B.~Hibner, A.~Cinar, and
  E.~Littlejohn, ``Automated insulin delivery-the light at the end of the
  tunnel,'' \emph{The Journal of pediatrics}, vol. 186, pp. 17--28, 2017.

\bibitem{kumareswaran2014}
K.~Kumareswaran, H.~Thabit, L.~Leelarathna, K.~Caldwell, D.~Elleri, J.~M.
  Allen, M.~Nodale, M.~E. Wilinska, M.~L. Evans, and R.~Hovorka, ``Feasibility
  of closed-loop insulin delivery in type 2 diabetes: a randomized controlled
  study,'' \emph{Diabetes Care}, vol.~37, no.~5, pp. 1198--1203, 2014.

\bibitem{bally2018}
L.~Bally, H.~Thabit, S.~Hartnell, E.~Andereggen, Y.~Ruan, M.~E. Wilinska, M.~L.
  Evans, M.~M. Wertli, A.~P. Coll, C.~Stettler \emph{et~al.}, ``Closed-loop
  insulin delivery for glycemic control in noncritical care,'' \emph{New
  England Journal of Medicine}, vol. 379, no.~6, pp. 547--556, 2018.

\bibitem{murphy2011}
H.~R. Murphy, K.~Kumareswaran, D.~Elleri, J.~M. Allen, K.~Caldwell,
  M.~Biagioni, D.~Simmons, D.~B. Dunger, M.~Nodale, and M.~E. Wilinska,
  ``Safety and efficacy of 24-h closed-loop insulin delivery in well-controlled
  pregnant women with type 1 diabetes: a randomized crossover case series,''
  \emph{Diabetes Care}, vol.~34, no.~12, pp. 2527--2529, 2011.

\bibitem{stewart2018}
Z.~A. Stewart, M.~E. Wilinska, S.~Hartnell, L.~K. O'Neil, G.~Rayman, E.~M.
  Scott, K.~Barnard, C.~Farrington, R.~Hovorka, and H.~R. Murphy,
  ``Day-and-night closed-loop insulin delivery in a broad population of
  pregnant women with type 1 diabetes: a randomized controlled crossover
  trial,'' \emph{Diabetes care}, vol.~41, no.~7, pp. 1391--1399, 2018.

\bibitem{preiser2016}
J.-C. Preiser, J.~G. Chase, R.~Hovorka, J.~I. Joseph, J.~S. Krinsley,
  C.~De~Block, T.~Desaive, L.~Foubert, P.~Kalfon, and U.~Pielmeier, ``Glucose
  control in the icu: a continuing story,'' \emph{Journal of Diabetes Science
  and Technology}, vol.~10, no.~6, pp. 1372--1381, 2016.

\bibitem{elfathi2018}
A.~El~Fathi, M.~R. Smaoui, V.~Gingras, B.~Boulet, and A.~Haidar, ``The
  artificial pancreas and meal control: An overview of postprandial glucose
  regulation in type 1 diabetes,'' \emph{IEEE Control Systems}, vol.~38, no.~1,
  pp. 67--85, 2018.

\bibitem{cinar2018turksoy-multivariable}
A.~Cinar and K.~Turksoy, ``Multivariable control of glucose concentration,'' in
  \emph{Advances in Artificial Pancreas Systems}.\hskip 1em plus 0.5em minus
  0.4em\relax Springer, 2018, pp. 65--82.

\bibitem{ramkissoon2018}
C.~M. Ramkissoon, P.~Herrero, J.~Bondia, and J.~Vehi, ``Unannounced meals in
  the artificial pancreas: Detection using continuous glucose monitoring,''
  \emph{Sensors}, vol.~18, no.~3, p. 884, 2018.

\bibitem{cameron2012}
F.~Cameron, G.~Niemeyer, and B.~W. Bequette, ``Extended multiple model
  prediction with application to blood glucose regulation,'' \emph{Journal of
  Process Control}, vol.~22, no.~8, pp. 1422--1432, 2012.

\bibitem{colmegna2018automatic}
P.~Colmegna, F.~Garelli, H.~De~Battista, and R.~S{\'a}nchez-Pe{\~n}a,
  ``Automatic regulatory control in type 1 diabetes without carbohydrate
  counting,'' \emph{Control Engineering Practice}, vol.~74, pp. 22--32, 2018.

\bibitem{riddell2015}
M.~C. Riddell, D.~P. Zaharieva, L.~Yavelberg, A.~Cinar, and V.~K. Jamnik,
  ``Exercise and the development of the artificial pancreas: one of the more
  difficult series of hurdles,'' \emph{Journal of Diabetes Science and
  Technology}, vol.~9, no.~6, pp. 1217--1226, 2015.

\bibitem{bondia2018}
J.~Bondia, S.~Romero-Vivo, B.~Ricarte, and J.~L. Diez, ``Insulin estimation and
  prediction: A review of the estimation and prediction of subcutaneous insulin
  pharmacokinetics in closed-loop glucose control,'' \emph{IEEE Control
  Systems}, vol.~38, no.~1, pp. 47--66, 2018.

\bibitem{moser2018}
O.~Moser, J.~Yardley, and R.~Bracken, ``Interstitial glucose and physical
  exercise in type 1 diabetes: Integrative physiology, technology, and the gap
  in-between,'' \emph{Nutrients}, vol.~10, no.~1, p.~93, 2018.

\bibitem{peters2018dual}
T.~Peters and A.~Haidar, ``Dual-hormone artificial pancreas: benefits and
  limitations compared with single-hormone systems,'' \emph{Diabetic Medicine},
  vol.~35, no.~4, pp. 450--459, 2018.

\bibitem{yavelberg2018}
L.~Yavelberg, D.~Zaharieva, A.~Cinar, M.~C. Riddell, and V.~Jamnik, ``A pilot
  study validating select research-grade and consumer-based wearables
  throughout a range of dynamic exercise intensities in persons with and
  without type 1 diabetes: A novel approach,'' \emph{Journal of Diabetes
  Science and Technology}, vol.~12, no.~3, pp. 569--576, 2018.

\bibitem{buxton2010}
O.~M. Buxton, M.~Pavlova, E.~W. Reid, W.~Wang, D.~C. Simonson, and G.~K. Adler,
  ``Sleep restriction for 1 week reduces insulin sensitivity in healthy men,''
  \emph{Diabetes}, vol.~59, no.~9, pp. 2126--2133, 2010.

\bibitem{donga2010}
E.~Donga, M.~Van~Dijk, J.~G. Van~Dijk, N.~R. Biermasz, G.-J. Lammers,
  K.~Van~Kralingen, R.~P. Hoogma, E.~P. Corssmit, and J.~A. Romijn, ``Partial
  sleep restriction decreases insulin sensitivity in type 1 diabetes,''
  \emph{Diabetes Care}, vol.~33, no.~7, pp. 1573--1577, 2010.

\bibitem{farabi2016}
S.~S. Farabi, ``Type 1 diabetes and sleep,'' \emph{Diabetes Spectrum}, vol.~29,
  no.~1, pp. 10--13, 2016.

\bibitem{qian2018}
J.~Qian, C.~Dalla~Man, C.~J. Morris, C.~Cobelli, and F.~A. Scheer,
  ``Differential effects of the circadian system and circadian misalignment on
  insulin sensitivity and insulin secretion in humans,'' \emph{Diabetes,
  Obesity and Metabolism}, vol.~20, no.~10, pp. 2481--2485, 2018.

\bibitem{bertachi2018}
A.~Bertachi, C.~M. Ramkissoon, J.~Bondia, and J.~Veh{\'\i}, ``Automated blood
  glucose control in type 1 diabetes: A review of progress and challenges,''
  \emph{Endocrinolog{\'\i}a, Diabetes y Nutrici{\'o}n (English ed.)}, vol.~65,
  no.~3, pp. 172--181, 2018.

\bibitem{weinzimer2012}
S.~A. Weinzimer, J.~L. Sherr, E.~Cengiz, G.~Kim, J.~L. Ruiz, L.~Carria,
  G.~Voskanyan, A.~Roy, and W.~V. Tamborlane, ``Effect of pramlintide on
  prandial glycemic excursions during closed-loop control in adolescents and
  young adults with type 1 diabetes,'' \emph{Diabetes care}, vol.~35, no.~10,
  pp. 1994--1999, 2012.

\bibitem{nih2011}
\BIBentryALTinterwordspacing
{National Institutes of Health}, ``Safety auxiliary feedback element for the
  artificial pancreas in type 1 diabetes,'' 2015. [Online]. Available:
  \url{https://projectreporter.nih.gov/project}
\BIBentrySTDinterwordspacing

\bibitem{nih2012}
\BIBentryALTinterwordspacing
------, ``Relative efficacy of single-, bi-, and tri-hormonal closed-loop
  control systems,'' 2015. [Online]. Available:
  \url{https://projectreporter.nih.gov/project}
\BIBentrySTDinterwordspacing

\bibitem{steil2013}
G.~M. Steil, ``Algorithms for a closed-loop artificial pancreas: the case for
  proportional-integral-derivative control,'' \emph{Journal of Diabetes Science
  and Technology}, vol.~7, no.~6, pp. 1621--1631, 2013.

\bibitem{steil2004}
G.~M. Steil, A.~E. Panteleon, and K.~Rebrin, ``Closed-loop insulin delivery-the
  path to physiological glucose control,'' \emph{Advanced Drug Delivery
  Reviews}, vol.~56, no.~2, pp. 125--144, 2004.

\bibitem{huyett2015}
L.~M. Huyett, E.~Dassau, H.~C. Zisser, and F.~J. Doyle~III, ``Design and
  evaluation of a robust pid controller for a fully implantable artificial
  pancreas,'' \emph{Industrial \& Engineering Chemistry Research}, vol.~54,
  no.~42, pp. 10\,311--10\,321, 2015.

\bibitem{mohammadridha2018}
T.~MohammadRidha, M.~Ait-Ahmed, L.~Chaillous, M.~Krempf, I.~Guilhem, J.-Y.
  Poirier, and C.~H. Moog, ``Model free ipid control for glycemia regulation of
  type-1 diabetes,'' \emph{IEEE Transactions on Biomedical Engineering},
  vol.~65, no.~1, pp. 199--206, 2018.

\bibitem{steil2011}
G.~M. Steil, C.~C. Palerm, N.~Kurtz, G.~Voskanyan, A.~Roy, S.~Paz, and F.~R.
  Kandeel, ``The effect of insulin feedback on closed loop glucose control,''
  \emph{The Journal of Clinical Endocrinology \& Metabolism}, vol.~96, no.~5,
  pp. 1402--1408, 2011.

\bibitem{ruiz2012}
J.~L. Ruiz, J.~L. Sherr, E.~Cengiz, L.~Carria, A.~Roy, G.~Voskanyan, W.~V.
  Tamborlane, and S.~A. Weinzimer, ``Effect of insulin feedback on closed-loop
  glucose control: a crossover study,'' \emph{Journal of Diabetes Science and
  Technology}, vol.~6, no.~5, pp. 1123--1130, 2012.

\bibitem{debock2017}
M.~de~Bock, J.~Dart, A.~Roy, R.~Davey, W.~Soon, C.~Berthold, A.~Retterath,
  B.~Grosman, N.~Kurtz, and E.~Davis, ``Exploration of the performance of a
  hybrid closed loop insulin delivery algorithm that includes insulin delivery
  limits designed to protect against hypoglycemia,'' \emph{Journal of Diabetes
  Science and Technology}, vol.~11, no.~1, pp. 68--73, 2017.

\bibitem{revert2013}
A.~Revert, F.~Garelli, J.~Pic{\'o}, H.~De~Battista, P.~Rossetti, J.~Veh{\'\i},
  and J.~Bondia, ``Safety auxiliary feedback element for the artificial
  pancreas in type 1 diabetes,'' \emph{IEEE Transactions on Biomedical
  Engineering}, vol.~60, no.~8, pp. 2113--2122, 2013.

\bibitem{fushimi2018}
E.~Fushimi, N.~Rosales, H.~De~Battista, and F.~Garelli, ``Artificial pancreas
  clinical trials: Moving towards closed-loop control using insulin-on-board
  constraints,'' \emph{Biomedical Signal Processing and Control}, vol.~45, pp.
  1--9, 2018.

\bibitem{cinar2018turksoy-dual}
A.~Cinar and K.~Turksoy, ``Dual-hormone (insulin and glucagon) ap systems,'' in
  \emph{Advances in Artificial Pancreas Systems}.\hskip 1em plus 0.5em minus
  0.4em\relax Springer, 2018, pp. 83--87.

\bibitem{cameron2011}
F.~Cameron, B.~W. Bequette, D.~M. Wilson, B.~A. Buckingham, H.~Lee, and
  G.~Niemeyer, ``A closed-loop artificial pancreas based on risk management,''
  \emph{Journal of Diabetes Science and Technology}, vol.~5, no.~2, pp.
  368--379, 2011.

\bibitem{forlenza2018}
G.~P. Forlenza, F.~M. Cameron, T.~T. Ly, D.~Lam, D.~P. Howsmon, N.~Baysal,
  G.~Kulina, L.~Messer, P.~Clinton, and C.~Levister, ``Fully closed-loop
  multiple model probabilistic predictive controller artificial pancreas
  performance in adolescents and adults in a supervised hotel setting,''
  \emph{Diabetes Technology \& Therapeutics}, vol.~20, no.~5, pp. 335--343,
  2018.

\bibitem{magni2009run}
L.~Magni, M.~Forgione, C.~Toffanin, C.~D. Man, B.~Kovatchev, G.~D. Nicolao, and
  C.~Cobelli, ``Run-to-run tuning of model predictive control for type 1
  diabetes subjects: In silico trial,'' \emph{Journal of Diabetes Science and
  Technology}, vol.~3, no.~5, pp. 1091--1098, 2009.

\bibitem{bequette2013}
B.~W. Bequette, ``Algorithms for a closed-loop artificial pancreas: the case
  for model predictive control,'' \emph{Journal of Diabetes Science and
  Technology}, vol.~7, no.~6, pp. 1632--1643, 2013.

\bibitem{grosman2010}
B.~Grosman, E.~Dassau, H.~C. Zisser, L.~Jovanovi{\v{c}}, and F.~J. Doyle~III,
  ``Zone model predictive control: a strategy to minimize hyper-and
  hypoglycemic events,'' \emph{Journal of Diabetes Science and Technology},
  vol.~4, no.~4, pp. 961--975, 2010.

\bibitem{gondhalekar2016}
R.~Gondhalekar, E.~Dassau, and F.~J. Doyle~III, ``Periodic zone-mpc with
  asymmetric costs for outpatient-ready safety of an artificial pancreas to
  treat type 1 diabetes,'' \emph{Automatica}, vol.~71, pp. 237--246, 2016.

\bibitem{chakrabarty2018}
A.~Chakrabarty, S.~Zavitsanou, F.~J. Doyle, and E.~Dassau, ``Event-triggered
  model predictive control for embedded artificial pancreas systems,''
  \emph{IEEE Transactions on Biomedical Engineering}, vol.~65, no.~3, pp.
  575--586, 2018.

\bibitem{gondhalekar2018}
R.~Gondhalekar, E.~Dassau, and F.~J. Doyle, ``Velocity-weighting \&
  velocity-penalty mpc of an artificial pancreas: Improved safety \&
  performance,'' \emph{Automatica}, vol.~91, pp. 105--117, 2018.

\bibitem{toffanin2013artificial}
C.~Toffanin, M.~Messori, F.~D. Palma, G.~D. Nicolao, C.~Cobelli, and L.~Magni,
  ``Artificial pancreas: Model predictive control design from clinical
  experience,'' \emph{Journal of Diabetes Science and Technology}, vol.~7,
  no.~6, pp. 1470--1483, 2013.

\bibitem{buckingham2009}
B.~Buckingham, E.~Cobry, P.~Clinton, V.~Gage, K.~Caswell, E.~Kunselman,
  F.~Cameron, and H.~P. Chase, ``Preventing hypoglycemia using predictive alarm
  algorithms and insulin pump suspension,'' \emph{Diabetes Technology \&
  Therapeutics}, vol.~11, no.~2, pp. 93--97, 2009.

\bibitem{ellingsen2009}
C.~Ellingsen, E.~Dassau, H.~Zisser, B.~Grosman, M.~W. Percival,
  L.~Jovanovi{\v{c}}, and F.~J. Doyle~III, ``Safety constraints in an
  artificial pancreatic $\beta$ cell: an implementation of model predictive
  control with insulin on board,'' \emph{Journal of Diabetes Science and
  Technology}, vol.~3, no.~3, pp. 536--544, 2009.

\bibitem{hajizadeh2018}
I.~Hajizadeh, M.~Rashid, and A.~Cinar, ``Considering plasma insulin
  concentrations in adaptive model predictive control for artificial pancreas
  systems,'' in \emph{2018 40th Annual International Conference of the IEEE
  Engineering in Medicine and Biology Society (EMBC)}.\hskip 1em plus 0.5em
  minus 0.4em\relax IEEE, 2018, Conference Proceedings, pp. 4452--4455.

\bibitem{shi2018}
D.~Shi, E.~Dassau, and F.~J. Doyle, ``Adaptive zone model predictive control of
  artificial pancreas based on glucose-and velocity-dependent control
  penalties,'' \emph{IEEE Transactions on Biomedical Engineering}, vol.~66,
  no.~4, pp. 1045--1054, 2018.

\bibitem{goodwin2015}
G.~C. Goodwin, A.~M. Medioli, D.~S. Carrasco, B.~R. King, and Y.~Fu, ``A
  fundamental control limitation for linear positive systems with application
  to type 1 diabetes treatment,'' \emph{Automatica}, vol.~55, pp. 73--77, 2015.

\bibitem{messori2016}
M.~Messori, C.~Toffanin, S.~Del~Favero, G.~De~Nicolao, C.~Cobelli, and
  L.~Magni, ``Model individualization for artificial pancreas,'' \emph{Computer
  methods and programs in biomedicine}, 2016.

\bibitem{garcia2018}
J.~Garcia-Tirado, C.~Zuluaga-Bedoya, and M.~D. Breton, ``Identifiability
  analysis of three control-oriented models for use in artificial pancreas
  systems,'' \emph{Journal of Diabetes Science and Technology}, vol.~12, no.~5,
  pp. 937--952, 2018.

\bibitem{fischer1987}
U.~Fischer, W.~Schenk, E.~Salzsieder, G.~Albrecht, P.~Abel, and E.-J. Freyse,
  ``Does physiological blood glucose control require an adaptive control
  strategy?'' \emph{IEEE Transactions on Biomedical Engineering}, no.~8, pp.
  575--582, 1987.

\bibitem{messori2018}
M.~Messori, G.~P. Incremona, C.~Cobelli, and L.~Magni, ``Individualized model
  predictive control for the artificial pancreas: In silico evaluation of
  closed-loop glucose control,'' \emph{IEEE Control Systems}, vol.~38, no.~1,
  pp. 86--104, 2018.

\bibitem{turksoy2014}
K.~Turksoy and A.~Cinar, ``Adaptive control of artificial pancreas systems-a
  review,'' \emph{Journal of Healthcare Engineering}, vol.~5, no.~1, pp. 1--22,
  2014.

\bibitem{cinar2018turksoy-various}
A.~Cinar and K.~Turksoy, ``Various control philosophies for ap systems,'' in
  \emph{Advances in Artificial Pancreas Systems}.\hskip 1em plus 0.5em minus
  0.4em\relax Springer, 2018, pp. 55--63.

\bibitem{turksoy2018}
K.~Turksoy, E.~Littlejohn, and A.~Cinar, ``Multimodule, multivariable
  artificial pancreas for patients with type 1 diabetes: Regulating glucose
  concentration under challenging conditions,'' \emph{IEEE Control Systems},
  vol.~38, no.~1, pp. 105--124, 2018.

\bibitem{kovatchev2009control}
B.~Kovatchev, S.~Patek, E.~Dassau, I.~Francis J.~Doyle, L.~Magni, G.~D.
  Nicolao, C.~Cobelli, and J.~D. R. F. A.~P. Consortium, ``Control to range for
  diabetes: Functionality and modular architecture,'' \emph{Journal of Diabetes
  Science and Technology}, vol.~3, no.~5, pp. 1058--1065, 2009.

\bibitem{patek2012}
S.~D. Patek, L.~Magni, E.~Dassau, C.~Karvetski, C.~Toffanin, G.~De~Nicolao,
  S.~Del~Favero, M.~Breton, C.~Dalla~Man, and E.~Renard, ``Modular closed-loop
  control of diabetes,'' \emph{IEEE Transactions on Biomedical Engineering},
  vol.~59, no.~11, pp. 2986--2999, 2012.

\bibitem{atlas2010}
E.~Atlas, R.~Nimri, S.~Miller, E.~A. Gurmberg, and M.~Phillip, ``Md-logic
  artificial pancreas system: A pilot study in adults with type 1 diabetes
  mellitus,'' \emph{Diabetes Care}, vol.~33, no.~5, pp. 1072--1076, 2010.

\bibitem{mauseth2013}
R.~Mauseth, I.~B. Hirsch, J.~Bollyky, R.~Kircher, D.~Matheson, S.~Sanda, and
  C.~Greenbaum, ``Use of a "fuzzy logic" controller in a closed-loop artificial
  pancreas,'' \emph{Diabetes Technology \& Therapeutics}, vol.~15, no.~8, pp.
  628--633, 2013.

\bibitem{grant2007}
P.~Grant, ``A new approach to diabetic control: fuzzy logic and insulin pump
  technology,'' \emph{Medical Engineering \& Physics}, vol.~29, no.~7, pp.
  824--827, 2007.

\bibitem{nimri2017}
R.~Nimri, N.~Bratina, O.~Kordonouri, M.~Avbelj~Stefanija, M.~Fath, T.~Biester,
  I.~Muller, E.~Atlas, S.~Miller, and A.~Fogel, ``Md-logic overnight type 1
  diabetes control in home settings: A multicentre, multinational, single blind
  randomized trial,'' \emph{Diabetes, Obesity and Metabolism}, vol.~19, no.~4,
  pp. 553--561, 2017.

\bibitem{Dose2016}
\BIBentryALTinterwordspacing
{Dose Safety}, 2016. [Online]. Available: \url{http://www.dosesafety.com/}
\BIBentrySTDinterwordspacing

\bibitem{doyle2001}
F.~Doyle, B.~Srinivasan, and D.~Bonvin, ``Run-to-run control strategy for
  diabetes management,'' in \emph{Engineering in Medicine and Biology Society,
  2001. Proceedings of the 23rd Annual International Conference of the IEEE},
  vol.~4.\hskip 1em plus 0.5em minus 0.4em\relax IEEE, 2001, Conference
  Proceedings, pp. 3159--3162.

\bibitem{zisser2005}
H.~Zisser, L.~Jovanovic, F.~Doyle~III, P.~Ospina, and C.~Owens, ``Run-to-run
  control of meal-related insulin dosing,'' \emph{Diabetes Technology \&
  Therapeutics}, vol.~7, no.~1, pp. 48--57, 2005.

\bibitem{owens2006}
C.~Owens, H.~Zisser, L.~Jovanovic, B.~Srinivasan, D.~Bonvin, and F.~J.
  Doyle~III, ``Run-to-run control of blood glucose concentrations for people
  with type 1 diabetes mellitus,'' \emph{IEEE Transactions on Biomedical
  Engineering}, vol.~53, no.~6, pp. 996--1005, 2006.

\bibitem{palerm2008run}
C.~C. Palerm, H.~Zisser, L.~Jovanovi{\v{c}}, and F.~J. Doyle~III, ``A
  run-to-run control strategy to adjust basal insulin infusion rates in type 1
  diabetes,'' \emph{Journal of process control}, vol.~18, no. 3-4, pp.
  258--265, 2008.

\bibitem{lee2015}
J.~B. Lee, E.~Dassau, and F.~J. Doyle~III, ``A run-to-run approach to enhance
  continuous glucose monitor accuracy based on continuous wear,''
  \emph{IFAC-PapersOnLine}, vol.~48, no.~20, pp. 237--242, 2015.

\bibitem{contreras2018}
I.~Contreras and J.~Vehi, ``Artificial intelligence for diabetes management and
  decision support: Literature review,'' \emph{Journal of Medical Internet
  Research}, vol.~20, no.~5, 2018.

\bibitem{ahmad2017}
S.~Ahmad, N.~Ahmed, M.~Ilyas, and W.~Khan, ``Super twisting sliding mode
  control algorithm for developing artificial pancreas in type 1 diabetes
  patients,'' \emph{Biomedical Signal Processing and Control}, vol.~38, pp.
  200--211, 2017.

\bibitem{thedoylegroup2015}
\BIBentryALTinterwordspacing
{The Doyle Group}, ``Artificial clinical trial database,'' 2015. [Online].
  Available: \url{https://www.thedoylegroup.org/apdatabase}
\BIBentrySTDinterwordspacing

\bibitem{iturralde2017}
E.~Iturralde, M.~L. Tanenbaum, S.~J. Hanes, S.~C. Suttiratana, J.~M. Ambrosino,
  T.~T. Ly, D.~M. Maahs, D.~Naranjo, N.~Walders-Abramson, and S.~A. Weinzimer,
  ``Expectations and attitudes of individuals with type 1 diabetes after using
  a hybrid closed loop system,'' \emph{The Diabetes Educator}, vol.~43, no.~2,
  pp. 223--232, 2017.

\bibitem{foerster2017}
V.~Foerster and M.~Severn, ``A hybrid closed-loop insulin delivery system for
  the treatment of type 1 diabetes,'' in \emph{CADTH Issues in Emerging Health
  Technologies}.\hskip 1em plus 0.5em minus 0.4em\relax Canadian Agency for
  Drugs and Technologies in Health, 2017.

\bibitem{boughton2018}
C.~Boughton and R.~Hovorka, ``Is an artificial pancreas (closed-loop system)
  for type 1 diabetes effective?'' \emph{Diabetic Medicine}, vol.~36, no.~3,
  pp. 279--286, 2019.

\bibitem{trevitt2016}
S.~Trevitt, S.~Simpson, and A.~Wood, ``Artificial pancreas device systems for
  the closed-loop control of type 1 diabetes: what systems are in
  development?'' \emph{Journal of Diabetes Science and Technology}, vol.~10,
  no.~3, pp. 714--723, 2016.

\bibitem{ramkissoon2017}
C.~M. Ramkissoon, B.~Aufderheide, B.~W. Bequette, and J.~Veh{\'\i}, ``A review
  of safety and hazards associated with the artificial pancreas,'' \emph{IEEE
  Reviews in Biomedical Engineering}, vol.~10, pp. 44--62, 2017.

\bibitem{food2014infusion}
\BIBentryALTinterwordspacing
{Food and Drug Administration Center for Devices and Radiological Health},
  ``Infusion pumps total product life cycle: Guidance for industry and fda
  staff,'' 2014. [Online]. Available:
  \url{https://www.fda.gov/media/78369/download}
\BIBentrySTDinterwordspacing

\bibitem{food2012}
\BIBentryALTinterwordspacing
------, ``Guidance for industry and food and drug administration staff: The
  content of investigational device exemption (ide) and premarket approval
  (pma) applications for artificial pancreas device systems,'' 2012. [Online].
  Available: \url{https://www.fda.gov/media/80644/download}
\BIBentrySTDinterwordspacing

\bibitem{lewis2018what}
\BIBentryALTinterwordspacing
D.~Lewis, ``What is \#openaps?'' 2018. [Online]. Available:
  \url{https://openaps.org/frequently-asked-questions/.}
\BIBentrySTDinterwordspacing

\bibitem{lewis2016real}
D.~Lewis, S.~Leibrand, and .~O. Community, ``Real-world use of open source
  artificial pancreas systems,'' \emph{Journal of Diabetes Science and
  Technology}, vol.~10, no.~6, pp. 1411--1411, 2016.

\bibitem{lewis2016poster}
\BIBentryALTinterwordspacing
D.~Lewis, ``Real-world use of open source artificial pancreas systems - poster
  presented at american diabetes association scientific sessions,'' 2016.
  [Online]. Available: \url{https://openaps.org/2016/06/11/real- world- use-
  of- open- source- artificial- pancreas- systems- poster- \\ presented- at-
  american- diabetes- association- scientific- sessions/}
\BIBentrySTDinterwordspacing

\bibitem{lewis2018frequently}
\BIBentryALTinterwordspacing
------, ``Frequently asked questions,'' 2018. [Online]. Available:
  \url{https://openaps.org/what-is-openaps/}
\BIBentrySTDinterwordspacing

\bibitem{bibian2006}
S.~Bibian, ``Automation in clinical anesthesia,'' Ph.D. dissertation,
  University of British Columbia, 2006.

\bibitem{nascu2017explicit}
I.~Na{\c{s}}cu, R.~Oberdieck, and E.~N. Pistikopoulos, ``Explicit hybrid model
  predictive control strategies for intravenous anaesthesia,'' \emph{Computers
  \& Chemical Engineering}, vol. 106, pp. 814--825, 2017.

\bibitem{borgeat1992}
A.~Borgeat, O.~Wilder-Smith, M.~Saiah, and K.~Rifat, ``Does propofol have an
  anti-emetic effect?'' \emph{Anaesthesia and Intensive Care}, vol.~20, no.~2,
  p. 260, 1992.

\bibitem{mcmanus2001neuromuscular}
M.~C. McManus, ``Neuromuscular blockers in surgery and intensive care, part
  2,'' \emph{American Journal of Health-System Pharmacy}, vol.~58, no.~24, pp.
  2381--2395, 2001.

\bibitem{zhusubaliyev2015}
Z.~T. Zhusubaliyev, A.~Medvedev, and M.~M. Silva, ``Bifurcation analysis of
  pid-controlled neuromuscular blockade in closed-loop anesthesia,''
  \emph{Journal of Process Control}, vol.~25, pp. 152--163, 2015.

\bibitem{abdulla2012thes}
S.~A. Abdulla, ``Depth of anaesthesia control techniques and human body
  models,'' Ph.D. dissertation, University of Southern Queensland, 2012.

\bibitem{sinha2007}
P.~K. Sinha and T.~Koshy, ``Monitoring devices for measuring the depth of
  anaesthesia-an overview,'' \emph{Indian Journal of Anaesthesia}, vol.~51,
  no.~5, p. 365, 2007.

\bibitem{lan2012}
J.-Y. Lan, M.~F. Abbod, R.-G. Yeh, S.-Z. Fan, and J.-S. Shieh, ``Intelligent
  modeling and control in anesthesia,'' \emph{Journal of Medical and Biological
  Engineering}, vol.~32, no.~5, pp. 293--308, 2012.

\bibitem{chang2015}
J.~J. Chang, S.~Syafiie, R.~Kamil, and T.~A. Lim, ``Automation of anaesthesia:
  a review on multivariable control,'' \emph{Journal of Clinical Monitoring and
  Computing}, vol.~29, no.~2, pp. 231--239, 2015.

\bibitem{mantzaridis1997}
H.~Mantzaridis and G.~Kenny, ``Auditory evoked potential index: a quantitative
  measure of changes in auditory evoked potentials during general
  anaesthesia,'' \emph{Anaesthesia}, vol.~52, no.~11, pp. 1030--1036, 1997.

\bibitem{musizza2010}
B.~Musizza and S.~Ribaric, ``Monitoring the depth of anaesthesia,''
  \emph{Sensors}, vol.~10, no.~12, pp. 10\,896--10\,935, 2010.

\bibitem{bibian2003}
S.~Bibian, C.~R. Ries, M.~Huzmezan, and G.~A. Dumont, ``Clinical anesthesia and
  control engineering: terminology, concepts and issues,'' in \emph{European
  Control Conference (ECC), 2003}.\hskip 1em plus 0.5em minus 0.4em\relax IEEE,
  2003, Conference Proceedings, pp. 2430--2440.

\bibitem{bailey2005}
J.~M. Bailey and W.~M. Haddad, ``Drug dosing control in clinical
  pharmacology,'' \emph{IEEE Control Systems}, vol.~25, no.~2, pp. 35--51,
  2005.

\bibitem{struys2004}
M.~M. Struys, T.~De~Smet, S.~Greenwald, A.~R. Absalom, S.~Bing{\'e}, and E.~P.
  Mortier, ``Performance evaluation of two published closed-loop control
  systems using bispectral index monitoringa simulation study,''
  \emph{Anesthesiology: The Journal of the American Society of
  Anesthesiologists}, vol. 100, no.~3, pp. 640--647, 2004.

\bibitem{struys2003}
M.~M. Struys, H.~Vereecke, A.~Moerman, E.~W. Jensen, D.~Verhaeghen, N.~De~Neve,
  F.~J. Dumortier, and E.~P. Mortier, ``Ability of the bispectral index,
  autoregressive modelling with exogenous input-derived auditory evoked
  potentials, and predicted propofol concentrations to measure patient
  responsiveness during anesthesia with propofol and remifentanil,''
  \emph{Anesthesiology: The Journal of the American Society of
  Anesthesiologists}, vol.~99, no.~4, pp. 802--812, 2003.

\bibitem{zikov2002}
T.~Zikov, ``Monitoring the anesthetic-induced unconsciousness (hypnosis) using
  wavelet analysis of the electroencephalogram,'' Thesis, 2002.

\bibitem{bibian2005}
S.~Bibian, C.~R. Ries, M.~Huzmezan, and G.~Dumont, ``Introduction to automated
  drug delivery in clinical anesthesia,'' \emph{European Journal of Control},
  vol.~11, no.~6, pp. 535--557, 2005.

\bibitem{schuller2015response}
P.~Schuller, S.~Newell, P.~Strickland, and J.~Barry, ``Response of bispectral
  index to neuromuscular block in awake volunteers,'' \emph{British Journal of
  Anaesthesia}, vol. 115, no. suppl\_1, pp. i95--i103, 2015.

\bibitem{malviya2007effect}
S.~Malviya, T.~Voepel-Lewis, A.~R. Tait, M.~F. Watcha, S.~Sadhasivam, and R.~H.
  Friesen, ``Effect of age and sedative agent on the accuracy of bispectral
  index in detecting depth of sedation in children,'' \emph{Pediatrics}, vol.
  120, no.~3, pp. e461--e470, 2007.

\bibitem{yamashita2008age}
K.~Yamashita, Y.~Terao, C.~Inadomi, M.~Takada, M.~Fukusaki, and K.~Sumikawa,
  ``Age-dependent relationship between bispectral index and sedation level,''
  \emph{Journal of Clinical Anesthesia}, vol.~20, no.~7, pp. 492--495, 2008.

\bibitem{cascella2016mechanisms}
M.~Cascella, ``Mechanisms underlying brain monitoring during anesthesia:
  limitations, possible improvements, and perspectives,'' \emph{Korean Journal
  of Anesthesiology}, vol.~69, no.~2, p. 113, 2016.

\bibitem{prichep2004patient}
L.~Prichep, L.~Gugino, E.~John, R.~Chabot, B.~Howard, H.~Merkin, M.~Tom,
  S.~Wolter, L.~Rausch, and W.~Kox, ``The patient state index as an indicator
  of the level of hypnosis under general anaesthesia,'' \emph{British Journal
  of Anaesthesia}, vol.~92, no.~3, pp. 393--399, 2004.

\bibitem{chen2002comparison}
X.~Chen, J.~Tang, P.~F. White, R.~H. Wender, H.~Ma, A.~Sloninsky, and
  R.~Kariger, ``A comparison of patient state index and bispectral index values
  during the perioperative period,'' \emph{Anesthesia \& Analgesia}, vol.~95,
  no.~6, pp. 1669--1674, 2002.

\bibitem{lee2015patient}
K.~Lee, Y.~Kim, Y.~Sung, and M.~Oh, ``The patient state index is well balanced
  for propofol sedation,'' \emph{Hippokratia}, vol.~19, no.~3, p. 235, 2015.

\bibitem{derendorf1999}
H.~Derendorf and B.~Meibohm, ``Modeling of pharmacokinetic/pharmacodynamic
  (pk/pd) relationships: concepts and perspectives,'' \emph{Pharmaceutical
  Research}, vol.~16, no.~2, pp. 176--185, 1999.

\bibitem{schnider1998}
T.~W. Schnider, C.~F. Minto, P.~L. Gambus, C.~Andresen, D.~B. Goodale, S.~L.
  Shafer, and E.~J. Youngs, ``The influence of method of administration and
  covariates on the pharmacokinetics of propofol in adult volunteers,''
  \emph{Anesthesiology: The Journal of the American Society of
  Anesthesiologists}, vol.~88, no.~5, pp. 1170--1182, 1998.

\bibitem{schuttler2000population}
J.~Sch{\"u}ttler and H.~Ihmsen, ``Population pharmacokinetics of propofola
  multicenter study,'' \emph{Anesthesiology: The Journal of the American
  Society of Anesthesiologists}, vol.~92, no.~3, pp. 727--738, 2000.

\bibitem{minto1997influence}
C.~F. Minto, T.~W. Schnider, T.~D. Egan, E.~Youngs, H.~J. Lemmens, P.~L.
  Gambus, V.~Billard, J.~F. Hoke, K.~H. Moore, and D.~J. Hermann, ``Influence
  of age and gender on the pharmacokinetics and pharmacodynamics of
  remifentanili. model development,'' \emph{Anesthesiology: The Journal of the
  American Society of Anesthesiologists}, vol.~86, no.~1, pp. 10--23, 1997.

\bibitem{minto1997anesthesiology}
C.~F. Minto, T.~W. Schnider, and S.~L. Shafer, ``Pharmacokinetics and
  pharmacodynamics of remifentanilii. model application,''
  \emph{Anesthesiology: The Journal of the American Society of
  Anesthesiologists}, vol.~86, no.~1, pp. 24--33, 1997.

\bibitem{hallynack1981}
T.~Hallynck, H.~Soep, J.~Thomis, J.~Boelaert, R.~Daneels, and L.~Dettli,
  ``Should clearance be normalised to body surface or to lean body mass?''
  \emph{British Journal of Clinical pharmacology}, vol.~11, no.~5, pp.
  523--526, 1981.

\bibitem{sheiner1979}
L.~B. Sheiner, D.~R. Stanski, S.~Vozeh, R.~D. Miller, and J.~Ham,
  ``Simultaneous modeling of pharmacokinetics and pharmacodynamics: application
  to d-tubocurarine,'' \emph{Clinical Pharmacology \& Therapeutics}, vol.~25,
  no.~3, pp. 358--371, 1979.

\bibitem{shafer1998}
S.~Shafer, ``Principles of pharmacokinetics and pharmacodynamics,''
  \emph{Principles and Practice of Anesthesiology}, pp. 1159--1210, 1998.

\bibitem{nino2009}
J.~Nino, R.~De~Keyser, S.~Syafiie, C.~Ionescu, and M.~Struys,
  ``Epsac-controlled anesthesia with online gain adaptation,''
  \emph{International Journal of Adaptive Control and Signal Processing},
  vol.~23, no.~5, pp. 455--471, 2009.

\bibitem{constant2010}
I.~Constant and A.~Rigouzzo, ``Which model for propofol tci in children,''
  \emph{Pediatric Anesthesia}, vol.~20, no.~3, pp. 233--239, 2010.

\bibitem{dahmani2010}
S.~Dahmani, I.~Stany, C.~Brasher, C.~Lejeune, B.~Bruneau, C.~Wood, Y.~Nivoche,
  I.~Constant, and I.~Murat, ``Pharmacological prevention of sevoflurane-and
  desflurane-related emergence agitation in children: a meta-analysis of
  published studies,'' \emph{British Journal of Anaesthesia}, vol. 104, no.~2,
  pp. 216--223, 2010.

\bibitem{mani2010}
V.~Mani and N.~S. Morton, ``Overview of total intravenous anesthesia in
  children,'' \emph{Pediatric Anesthesia}, vol.~20, no.~3, pp. 211--222, 2010.

\bibitem{rigouzzo2010}
A.~Rigouzzo, F.~Servin, and I.~Constant, ``Pharmacokinetic-pharmacodynamic
  modeling of propofol in children,'' \emph{Anesthesiology: The Journal of the
  American Society of Anesthesiologists}, vol. 113, no.~2, pp. 343--352, 2010.

\bibitem{zhusubaliyev2013bifurcation}
Z.~T. Zhusubaliyev, A.~Medvedev, and M.~M. Silva, ``Bifurcation analysis for
  pid-controller tuning based on a minimal neuromuscular blockade model in
  closed-loop anesthesia,'' in \emph{52nd IEEE Conference on Decision and
  Control}.\hskip 1em plus 0.5em minus 0.4em\relax IEEE, 2013, pp. 115--120.

\bibitem{dasilva2012nonlinear}
M.~M. Da~Silva, T.~Wigren, and T.~Mendon{\c{c}}a, ``Nonlinear identification of
  a minimal neuromuscular blockade model in anesthesia,'' \emph{IEEE
  Transactions on Control Systems Technology}, vol.~20, no.~1, pp. 181--188,
  2012.

\bibitem{ward1983}
S.~Ward, E.~Neill, B.~Weatherley, and I.~Corall, ``Pharmacokinetics of
  atracurium besylate in healthy patients (after a single iv bolus dose),''
  \emph{BJA: British Journal of Anaesthesia}, vol.~55, no.~2, pp. 113--118,
  1983.

\bibitem{weatherley1983}
B.~Weatherley, S.~Williams, and E.~Neill, ``Pharmacokinetics, pharmacodynamics
  and dose-response relationships of atracurium administered iv,'' \emph{BJA:
  British Journal of Anaesthesia}, vol.~55, pp. 39S--45S, 1983.

\bibitem{hodrea2013}
R.~Hodrea, R.~Morar, I.~Nascu, and H.~Vasian, ``Modeling of neuromuscular
  blockade in general anesthesia,'' in \emph{Advanced Topics in Electrical
  Engineering (ATEE), 2013 8th International Symposium on}.\hskip 1em plus
  0.5em minus 0.4em\relax IEEE, 2013, Conference Proceedings, pp. 1--4.

\bibitem{soltesz2012simulated}
K.~Soltesz, G.~A. Dumont, K.~van Heusden, T.~H{\"a}gglund, and J.~M. Ansermino,
  ``Simulated mid-ranging control of propofol and remifentanil using
  eeg-measured hypnotic depth of anesthesia,'' in \emph{2012 IEEE 51st IEEE
  Conference on Decision and Control (CDC)}.\hskip 1em plus 0.5em minus
  0.4em\relax IEEE, 2012, pp. 356--361.

\bibitem{zikov2006}
T.~Zikov, S.~Bibian, G.~A. Dumont, M.~Huzmezan, and C.~R. Ries, ``Quantifying
  cortical activity during general anesthesia using wavelet analysis,''
  \emph{IEEE Transactions on Biomedical Engineering}, vol.~53, no.~4, pp.
  617--632, 2006.

\bibitem{kern2004}
S.~E. Kern, G.~Xie, J.~L. White, and T.~D. Egan, ``A response surface analysis
  of propofol-remifentanil pharmacodynamic interaction in volunteers,''
  \emph{The Journal of the American Society of Anesthesiologists}, vol. 100,
  no.~6, pp. 1373--1381, 2004.

\bibitem{Dasilva2010}
M.~M. Da~Silva, T.~Mendon, and T.~Wigren, ``Online nonlinear identification of
  the effect of drugs in anaesthesia using a minimal parameterization and bis
  measurements,'' in \emph{American Control Conference (ACC), 2010}.\hskip 1em
  plus 0.5em minus 0.4em\relax IEEE, 2010, Conference Proceedings, pp.
  4379--4384.

\bibitem{silva2011prediction}
M.~Silva, ``Prediction error identification of minimally parameterized wiener
  models in anesthesia,'' in \emph{Proc. 18th IFAC World Congress}, 2011, pp.
  5615--5620.

\bibitem{nogueira2016}
F.~N. Nogueira, T.~Mendon{\c{c}}a, and P.~Rocha, ``Positive state observer for
  the automatic control of the depth of anesthesia-clinical results,''
  \emph{Computer Methods and Programs in Biomedicine}, 2016.

\bibitem{mendoncca2012comparing}
T.~Mendon{\c{c}}a, H.~Alonso, M.~M. da~Silva, S.~Esteves, and M.~Seabra,
  ``Comparing different identification approaches for the depth of anesthesia
  using bis measurements,'' \emph{IFAC Proceedings Volumes}, vol.~45, no.~16,
  pp. 781--785, 2012.

\bibitem{tan2010}
Z.~Tan, R.~Kaddoum, and H.~W. Le~Yi~Wang, ``Decision-oriented multi-outcome
  modeling for anesthesia patients,'' \emph{The Open Biomedical Engineering
  journal}, vol.~4, p. 113, 2010.

\bibitem{nakaya2002}
M.~Nakayama, H.~Ichinose, S.~Yamamoto, N.~Kanaya, and A.~Namiki, ``The effect
  of fentanyl on hemodynamic and bispectral index changes during anesthesia
  induction with propofol,'' \emph{Journal of clinical anesthesia}, vol.~14,
  no.~2, pp. 146--149, 2002.

\bibitem{tan2011}
Z.~Tan, ``Monitoring, diagnosis, and control for advanced anesthesia
  management,'' Ph.D. dissertation, Wayne State University, 2011.

\bibitem{silva2014a}
M.~M. Silva, T.~Wigren, and T.~Mendon{\c{c}}a, ``A reduced mimo wiener model
  for recursive identification of the depth of anesthesia,''
  \emph{International Journal of Adaptive Control and Signal Processing},
  vol.~28, no.~12, pp. 1357--1371, 2014.

\bibitem{hall2017}
M.~J. Hall, A.~Schwartzman, J.~Zhang, and X.~Liu, ``Ambulatory surgery data
  from hospitals and ambulatory surgery centers: United states, 2010,''
  \emph{National Health Statistics Reports}, no. 102, pp. 1--15, 2017.

\bibitem{abdulla2012Depth}
S.~Abdulla and P.~Wen, ``Depth of anaesthesia control investigation using
  robust deadbeat control technique,'' in \emph{2012 ICME International
  Conference on Complex Medical Engineering (CME)}.\hskip 1em plus 0.5em minus
  0.4em\relax IEEE, 2012, Conference Proceedings, pp. 107--111.

\bibitem{bruhn2006}
J.~Bruhn, P.~Myles, R.~Sneyd, and M.~Struys, ``Depth of anaesthesia monitoring:
  what's available, what's validated and what's next?'' \emph{BJA: British
  Journal of Anaesthesia}, vol.~97, no.~1, pp. 85--94, 2006.

\bibitem{padula2017}
F.~Padula, C.~Ionescu, N.~Latronico, M.~Paltenghi, A.~Visioli, and G.~Vivacqua,
  ``Optimized pid control of depth of hypnosis in anesthesia,'' \emph{Computer
  Methods and Programs in Biomedicine}, vol. 144, pp. 21--35, 2017.

\bibitem{neckebroek2018}
M.~Neckebroek, C.~Ionescu, K.~van Amsterdam, T.~De~Smet, P.~De~Baets,
  J.~Decruyenaere, R.~De~Keyser, and M.~Struys, ``A comparison of
  propofol-to-bis post-operative intensive care sedation by means of target
  controlled infusion, bayesian-based and predictive control methods: an
  observational, open-label pilot study,'' \emph{Journal of Clinical Monitoring
  and Computing}, pp. 1--12, 2018.

\bibitem{biebuyck1992}
J.~F. Biebuyck, D.~A. O'hara, D.~K. Bogen, and A.~Noordergraaf, ``The use of
  computers for controlling the delivery of anesthesia,'' \emph{Anesthesiology:
  The Journal of the American Society of Anesthesiologists}, vol.~77, no.~3,
  pp. 563--581, 1992.

\bibitem{Asbury1997}
A.~J. Asbury, ``Feedback control in anaesthesia,'' \emph{International journal
  of clinical monitoring and computing}, vol.~14, no.~1, pp. 1--10, 1997.

\bibitem{soltesz2013automation}
K.~Solt{\'e}sz, \emph{On automation in anesthesia}.\hskip 1em plus 0.5em minus
  0.4em\relax Department of Automatic Control, Lund University, 2013.

\bibitem{absalom2003}
A.~R. Absalom and G.~Kenny, ``Closed-loop control of propofol anaesthesia using
  bispectral index$^{TM}$: performance assessment in patients receiving
  computer-controlled propofol and manually controlled remifentanil infusions
  for minor surgery,'' \emph{British Journal of Anaesthesia}, vol.~90, no.~6,
  pp. 737--741, 2003.

\bibitem{glen1998}
J.~Glen, ``The development of 'diprifusor': a {TCI} system for propofol,''
  \emph{Anaesthesia}, vol.~53, pp. 13--21, 1998.

\bibitem{merigo2017}
L.~Merigo, M.~Beschi, F.~Padula, N.~Latronico, M.~Paltenghi, and A.~Visioli,
  ``Event-based control of depth of hypnosis in anesthesia,'' \emph{Computer
  Methods and Programs in Biomedicine}, vol. 147, pp. 63--83, 2017.

\bibitem{liu2006}
N.~Liu, T.~Chazot, A.~Genty, A.~Landais, A.~Restoux, K.~McGee, P.-A. Laloë,
  B.~Trillat, L.~Barvais, and M.~Fischler, ``Titration of propofol for
  anesthetic induction and maintenance guided by the bispectral index:
  Closed-loop versus manual controla prospective, randomized, multicenter
  study,'' \emph{Anesthesiology: The Journal of the American Society of
  Anesthesiologists}, vol. 104, no.~4, pp. 686--695, 2006.

\bibitem{liu2011}
N.~Liu, T.~Chazot, S.~Hamada, A.~Landais, N.~Boichut, C.~Dussaussoy,
  B.~Trillat, L.~Beydon, E.~Samain, and D.~I. Sessler, ``Closed-loop
  coadministration of propofol and remifentanil guided by bispectral index: a
  randomized multicenter study,'' \emph{Anesthesia \& Analgesia}, vol. 112,
  no.~3, pp. 546--557, 2011.

\bibitem{ionescu2008}
C.~M. Ionescu, R.~De~Keyser, B.~C. Torrico, T.~De~Smet, M.~M. Struys, and J.~E.
  Normey-Rico, ``Robust predictive control strategy applied for propofol dosing
  using bis as a controlled variable during anesthesia,'' \emph{IEEE
  Transactions on Biomedical Engineering}, vol.~55, no.~9, pp. 2161--2170,
  2008.

\bibitem{desmet2008}
T.~De~Smet, M.~M. Struys, M.~M. Neckebroek, K.~Van~den Hauwe, S.~Bonte, and
  E.~P. Mortier, ``The accuracy and clinical feasibility of a new
  bayesian-based closed-loop control system for propofol administration using
  the bispectral index as a controlled variable,'' \emph{Anesthesia \&
  Analgesia}, vol. 107, no.~4, pp. 1200--1210, 2008.

\bibitem{struys2001}
M.~M. Struys, T.~De~Smet, L.~F. Versichelen, S.~Van~de Velde, R.~Van~den
  Broecke, and E.~P. Mortier, ``Comparison of closed-loop controlled
  administration of propofol using bispectral index as the controlled variable
  versus “standard practice” controlled administration,''
  \emph{Anesthesiology: The Journal of the American Society of
  Anesthesiologists}, vol.~95, no.~1, pp. 6--17, 2001.

\bibitem{dumont2012}
G.~A. Dumont, ``Closed-loop control of anesthesia-a review,'' \emph{IFAC
  Proceedings Volumes}, vol.~45, no.~18, pp. 373--378, 2012.

\bibitem{dumont2013}
G.~A. Dumont and J.~M. Ansermino, ``Closed-loop control of anesthesia: a primer
  for anesthesiologists,'' \emph{Anesthesia \& Analgesia}, vol. 117, no.~5, pp.
  1130--1138, 2013.

\bibitem{ilyas2017}
M.~Ilyas, M.~F.~U. Butt, M.~Bilal, K.~Mahmood, A.~Khaqan, and R.~Ali~Riaz, ``A
  review of modern control strategies for clinical evaluation of propofol
  anesthesia administration employing hypnosis level regulation,'' \emph{BioMed
  Research International}, vol. 2017, 2017.

\bibitem{chilcoat1980}
R.~Chilcoat, ``A review of the control of depth of anaesthesia,''
  \emph{Transactions of the Institute of Measurement and Control}, vol.~2,
  no.~1, pp. 38--45, 1980.

\bibitem{sieber2000model}
T.~Sieber, C.~Frei, M.~Derighetti, P.~Feigenwinter, D.~Leibundgut, and A.~M.
  Zbinden, ``Model-based automatic feedback control versus human control of
  end-tidal isoflurane concentration using low-flow anaesthesia,''
  \emph{British Journal of Anaesthesia}, vol.~85, no.~6, pp. 818--825, 2000.

\bibitem{struys2004time}
M.~Struys, A.~Kalmar, L.~De~Baerdemaeker, E.~Mortier, G.~Rolly, J.~Manigel, and
  W.~Buschke, ``Time course of inhaled anaesthetic drug delivery using a new
  multifunctional closed-circuit anaesthesia ventilator. in vitro comparison
  with a classical anaesthesia machine,'' \emph{British Journal of Anaesthesi},
  vol.~94, no.~3, pp. 306--317, 2004.

\bibitem{lortat2009assessing}
B.~Lortat-Jacob, V.~Billard, W.~Buschke, and F.~Servin, ``Assessing the
  clinical or pharmaco-economical benefit of target controlled desflurane
  delivery in surgical patients using the zeus{\textregistered} anaesthesia
  machine,'' \emph{Anaesthesia}, vol.~64, no.~11, pp. 1229--1235, 2009.

\bibitem{connor2019optimizing}
C.~W. Connor, ``Optimizing target control of the vessel rich group with
  volatile anesthetics,'' \emph{Journal of Clinical Monitoring and Computing},
  vol.~33, no.~3, pp. 445--454, 2019.

\bibitem{mahfouf2003unconstrained}
M.~Mahfouf, A.~Asbury, and D.~Linkens, ``Unconstrained and constrained
  generalised predictive control of depth of anaesthesia during surgery,''
  \emph{Control Engineering Practice}, vol.~11, no.~12, pp. 1501--1515, 2003.

\bibitem{krieger2013modeling}
A.~Krieger, N.~Panoskaltsis, A.~Mantalaris, M.~C. Georgiadis, and E.~N.
  Pistikopoulos, ``Modeling and analysis of individualized pharmacokinetics and
  pharmacodynamics for volatile anesthesia,'' \emph{IEEE Transactions on
  Biomedical Engineering}, vol.~61, no.~1, pp. 25--34, 2013.

\bibitem{krieger2014}
A.~Krieger and E.~N. Pistikopoulos, ``Model predictive control of anesthesia
  under uncertainty,'' \emph{Computers \& Chemical Engineering}, vol.~71, pp.
  699--707, 2014.

\bibitem{krieger2013modellingb}
A.~Krieger, ``Modelling, optimisation and explicit model predictive control of
  anaesthesia drug delivery systems,'' Ph.D. dissertation, Imperial College
  London, 2013.

\bibitem{das2014}
S.~Das, S.~Das, and K.~Maharatna, ``Control strategy for anaesthetic drug
  dosage with interaction among human physiological organs using optimal
  fractional order pid controller,'' in \emph{Proceedings of The 2014
  International Conference on Control, Instrumentation, Energy and
  Communication (CIEC)}.\hskip 1em plus 0.5em minus 0.4em\relax IEEE, 2014,
  Conference Proceedings, pp. 66--70.

\bibitem{higgen2018mid}
F.~L. Higgen, G.~G. Supp, J.~F. Hipp, A.~K. Engel, and M.~Siegel, ``Mid-latency
  auditory evoked potentials differentially predict sedation and drug level
  under opioid and hypnotic agents,'' \emph{Frontiers in Pharmacology}, vol.~9,
  p. 1427, 2018.

\bibitem{kenny1999}
G.~Kenny and H.~Mantzaridis, ``Closed-loop control of propofol anaesthesia,''
  \emph{British Journal of Anaesthesia}, vol.~83, no.~2, pp. 223--228, 1999.

\bibitem{moreley2000}
A.~Morley, J.~Derrick, P.~Mainland, B.~Lee, and T.~Short, ``Closed loop control
  of anaesthesia: an assessment of the bispectral index as the target of
  control,'' \emph{Anaesthesia}, vol.~55, no.~10, pp. 953--959, 2000.

\bibitem{hemmerling2010}
T.~M. Hemmerling, S.~Charabati, C.~Zaouter, C.~Minardi, and P.~A. Mathieu, ``A
  randomized controlled trial demonstrates that a novel closed-loop propofol
  system performs better hypnosis control than manual administration,''
  \emph{Canadian Journal of Anesthesia/Journal canadien d'anesth{\'e}sie},
  vol.~57, no.~8, pp. 725--735, 2010.

\bibitem{liu2012}
N.~Liu, M.~Le~Guen, F.~Benabbes-Lambert, T.~Chazot, B.~Trillat, D.~I. Sessler,
  and M.~Fischler, ``Feasibility of closed-loop titration of propofol and
  remifentanil guided by the spectral m-entropy monitor,''
  \emph{Anesthesiology: The Journal of the American Society of
  Anesthesiologists}, vol. 116, no.~2, pp. 286--295, 2012.

\bibitem{west2013}
N.~West, G.~A. Dumont, K.~van Heusden, C.~L. Petersen, S.~Khosravi, K.~Soltesz,
  A.~Umedaly, E.~Reimer, and J.~M. Ansermino, ``Robust closed-loop control of
  induction and maintenance of propofol anesthesia in children,''
  \emph{Pediatric Anesthesia}, vol.~23, no.~8, pp. 712--719, 2013.

\bibitem{vanheusden2014}
K.~Van~Heusden, G.~A. Dumont, K.~Soltesz, C.~L. Petersen, A.~Umedaly, N.~West,
  and J.~M. Ansermino, ``Design and clinical evaluation of robust pid control
  of propofol anesthesia in children,'' \emph{IEEE Transactions on Control
  Systems Technology}, vol.~22, no.~2, pp. 491--501, 2014.

\bibitem{simalatsar2016cascaded}
A.~Simalatsar, M.~Guidi, and T.~Buclin, ``Cascaded pid controller for
  anaesthesia delivery,'' in \emph{2016 38th Annual International Conference of
  the IEEE Engineering in Medicine and Biology Society (EMBC)}.\hskip 1em plus
  0.5em minus 0.4em\relax IEEE, 2016, pp. 533--536.

\bibitem{soltesz2013individualized}
K.~Soltesz, J.-O. Hahn, T.~H{\"a}gglund, G.~A. Dumont, and J.~M. Ansermino,
  ``Individualized closed-loop control of propofol anesthesia: A preliminary
  study,'' \emph{Biomedical Signal Processing and Control}, vol.~8, no.~6, pp.
  500--508, 2013.

\bibitem{astrom2006}
K.~J. {\AA}str{\"o}m, T.~H{\"a}gglund, and K.~J. Astrom, \emph{Advanced PID
  control}.\hskip 1em plus 0.5em minus 0.4em\relax ISA-The Instrumentation,
  Systems, and Automation Society Research Triangle, 2006, vol. 461.

\bibitem{ejaz2004}
K.~Ejaz and J.-S. Yang, ``Controlling depth of anesthesia using pid tuning: a
  comparative model-based study,'' in \emph{Proceedings of the 2004 IEEE
  International Conference on Control Applications}, vol.~1.\hskip 1em plus
  0.5em minus 0.4em\relax IEEE, 2004, Conference Proceedings, pp. 580--585.

\bibitem{padula2016}
F.~Padula, C.~Ionescu, N.~Latronico, M.~Paltenghi, A.~Visioli, and G.~Vivacqua,
  ``Inversion-based propofol dosing for intravenous induction of hypnosis,''
  \emph{Communications in Nonlinear Science and Numerical Simulation}, vol.~39,
  pp. 481--494, 2016.

\bibitem{anna2010}
S.~Anna and P.~Wen, ``Depth of anesthesia control using internal model control
  techniques,'' in \emph{IEEE/ICME International Conference on Complex Medical
  Engineering}.\hskip 1em plus 0.5em minus 0.4em\relax IEEE, 2010, Conference
  Proceedings, pp. 294--300.

\bibitem{hahn2012}
J.-O. Hahn, G.~A. Dumont, and J.~M. Ansermino, ``Robust closed-loop control of
  hypnosis with propofol using wavcns index as the controlled variable,''
  \emph{Biomedical Signal Processing and Control}, vol.~7, no.~5, pp. 517--524,
  2012.

\bibitem{merigo2018a}
L.~Merigo, F.~Padula, A.~Pawlowski, S.~Dormido, J.~L.~G. Sánchez,
  N.~Latronico, M.~Paltenghi, and A.~Visioli, ``A model-based control scheme
  for depth of hypnosis in anesthesia,'' \emph{Biomedical Signal Processing and
  Control}, vol.~42, pp. 216--229, 2018.

\bibitem{dumont2009}
G.~A. Dumont, A.~Martinez, and J.~M. Ansermino, ``Robust control of depth of
  anesthesia,'' \emph{International Journal of Adaptive Control and Signal
  Processing}, vol.~23, no.~5, pp. 435--454, 2009.

\bibitem{bhole2015}
K.~Bhole and S.~Agashe, ``Automating intravenous anesthesia with a fuzzy
  inference system coupled with a proportional integral derivative (pid)
  controller,'' \emph{American Journal of Biomedical Science and Engineering},
  vol.~1, no.~6, pp. 93--99, 2015.

\bibitem{merigo2018optimized}
L.~Merigo, F.~Padula, N.~Latronico, T.~Mendon{\c{c}}a, M.~Paltenghi, P.~Rocha,
  and A.~Visioli, ``Optimized pid tuning for the automatic control of
  neuromuscular blockade,'' \emph{IFAC-PapersOnLine}, vol.~51, no.~4, pp.
  66--71, 2018.

\bibitem{panagopoulos2002}
H.~Panagopoulos, K.~Astrom, and T.~Hagglund, ``Design of pid controllers based
  on constrained optimisation,'' \emph{IEE Proceedings-Control Theory and
  Applications}, vol. 149, no.~1, pp. 32--40, 2002.

\bibitem{castillo2005}
O.~Castillo, G.~Huesca, and F.~Valdez, ``Evolutionary computing for optimizing
  type-2 fuzzy systems in intelligent control of non-linear dynamic plants,''
  in \emph{Fuzzy Information Processing Society, 2005. NAFIPS 2005. Annual
  Meeting of the North American}.\hskip 1em plus 0.5em minus 0.4em\relax IEEE,
  2005, Conference Proceedings, pp. 247--251.

\bibitem{soltesz2011individualized}
K.~Soltesz, J.-O. Hahn, G.~A. Dumont, and J.~M. Ansermino, ``Individualized pid
  control of depth of anesthesia based on patient model identification during
  the induction phase of anesthesia,'' in \emph{2011 50th IEEE Conference on
  Decision and Control and European Control Conference}.\hskip 1em plus 0.5em
  minus 0.4em\relax IEEE, 2011, pp. 855--860.

\bibitem{padula2015}
F.~Padula, C.~Ionescu, N.~Latronico, M.~Paltenghi, A.~Visioli, and G.~Vivacqua,
  ``A gain-scheduled pid controller for propofol dosing in anesthesia,''
  \emph{IFAC-PapersOnLine}, vol.~48, no.~20, pp. 545--550, 2015.

\bibitem{sawaguchi2003model}
Y.~Sawaguchi, E.~Furutani, G.~Shirakami, M.~Araki, and K.~Fukuda, ``A model
  predictive sedation control system under total intravenous anesthesia,'' in
  \emph{IEEE EMBS Asian-Pacific Conference on Biomedical Engineering,
  2003.}\hskip 1em plus 0.5em minus 0.4em\relax IEEE, 2003, pp. 358--359.

\bibitem{furutani2005}
E.~Furutani, Y.~Sawaguchi, G.~Shirakami, M.~Araki, and K.~Fukuda, ``A hypnosis
  control system using a model predictive controller with online identification
  of individual parameters,'' in \emph{Control Applications, 2005. CCA 2005.
  Proceedings of 2005 IEEE Conference on}.\hskip 1em plus 0.5em minus
  0.4em\relax IEEE, 2005, Conference Proceedings, pp. 154--159.

\bibitem{bamdadian2008generalized}
A.~Bamdadian, F.~Towhidkhah, and M.~H. Moradi, ``Generalized predictive control
  of depth of anesthesia by using a pharmocokinetic-pharmacodynamic model of
  the patient,'' in \emph{2008 2nd International Conference on Bioinformatics
  and Biomedical Engineering}.\hskip 1em plus 0.5em minus 0.4em\relax IEEE,
  2008, Conference Proceedings, pp. 1276--1279.

\bibitem{goodwin2019}
G.~Goodwin, A.~Medioli, K.~Murray, R.~Sykes, and C.~Stephen, \emph{Applications
  of MPC in the Area of Health Care}.\hskip 1em plus 0.5em minus 0.4em\relax
  Springer, 2019, pp. 529--550.

\bibitem{ingole2015}
D.~Ingole, J.~Holaza, B.~Takács, and M.~Kvasnica, ``Fpga-based explicit model
  predictive control for closed-loop control of intravenous anesthesia,'' in
  \emph{2015 20th International Conference on Process Control (PC)}.\hskip 1em
  plus 0.5em minus 0.4em\relax IEEE, 2015, Conference Proceedings, pp. 42--47.

\bibitem{nacscu2016a}
I.~Na{\c{s}}cu and E.~N. Pistikopoulos, ``Multiparametric model predictive
  control and state estimation of the hypnotic component in anesthesia,'' in
  \emph{2016 IEEE International Conference on Automation, Quality and Testing,
  Robotics (AQTR)}.\hskip 1em plus 0.5em minus 0.4em\relax IEEE, 2016, pp.
  1--6.

\bibitem{darby2007}
M.~L. Darby and M.~Nikolaou, ``A parametric programming approach to
  moving-horizon state estimation,'' \emph{Automatica}, vol.~43, no.~5, pp.
  885--891, 2007.

\bibitem{nacscu2016b}
I.~Na{\c{s}}cu and E.~N. Pistikopoulos, ``Multiparametric model predictive
  control strategies of the hypnotic component in intravenous anesthesia,'' in
  \emph{2016 IEEE International Conference on Systems, Man, and Cybernetics
  (SMC)}.\hskip 1em plus 0.5em minus 0.4em\relax IEEE, 2016, pp.
  002\,828--002\,833.

\bibitem{nascu2017modeling}
------, ``Modeling, estimation and control of the anaesthesia process,''
  \emph{Computers \& Chemical Engineering}, vol. 107, pp. 318--332, 2017.

\bibitem{pawlowski2017event}
A.~Paw{\l}owski, L.~Merigo, J.~Guzman, A.~Visioli, and S.~Dormido,
  ``Event-based gpc for depth of hypnosis in anesthesia for efficient use of
  propofol,'' in \emph{2017 3rd International Conference on Event-Based
  Control, Communication and Signal Processing (EBCCSP)}.\hskip 1em plus 0.5em
  minus 0.4em\relax IEEE, 2017, pp. 1--7.

\bibitem{sawaguchi2008}
Y.~Sawaguchi, E.~Furutani, G.~Shirakami, M.~Araki, and K.~Fukuda, ``A
  model-predictive hypnosis control system under total intravenous
  anesthesia,'' \emph{IEEE Transactions on Biomedical Engineering}, vol.~55,
  no.~3, pp. 874--887, 2008.

\bibitem{robayo2010estimating}
F.~Robayo, D.~Sendoya, R.~Hodrea, C.~Ionescu, and R.~de~Keyser, ``Estimating
  the time-delay for predictive control in general anesthesia,'' in
  \emph{Control and Decision Conference (CCDC), 2010 Chinese}.\hskip 1em plus
  0.5em minus 0.4em\relax IEEE, 2010, pp. 3719--3724.

\bibitem{sartori2006line}
V.~Sartori, P.~M. Schumacher, T.~Bouillon, M.~Luginbuehl, and M.~Morari,
  ``On-line estimation of propofol pharmacodynamic parameters,'' in \emph{2005
  IEEE Engineering in Medicine and Biology 27th Annual Conference}.\hskip 1em
  plus 0.5em minus 0.4em\relax IEEE, 2006, pp. 74--77.

\bibitem{ingole2017}
D.~Ingole, J.~Drgo{\v{n}}a, and M.~Kvasnica, ``Offset-free hybrid model
  predictive control of bispectral index in anesthesia,'' in \emph{2017 21st
  International Conference on Process Control (PC)}.\hskip 1em plus 0.5em minus
  0.4em\relax IEEE, 2017, pp. 422--427.

\bibitem{nascu2016advanced}
I.~Na{\c{s}}cu, ``Advanced multiparametric optimization and control studies for
  anaesthesia,'' Ph.D. dissertation, Imperial College London, 2016.

\bibitem{sopasakis2014}
P.~Sopasakis, P.~Patrinos, and H.~Sarimveis, ``Robust model predictive control
  for optimal continuous drug administration,'' \emph{Computer Methods and
  Programs in Biomedicine}, vol. 116, no.~3, pp. 193--204, 2014.

\bibitem{haddad2003adaptive}
W.~M. Haddad, T.~Hayakawa, and J.~M. Bailey, ``Adaptive control for
  non-negative and compartmental dynamical systems with applications to general
  anesthesia,'' \emph{International Journal of Adaptive Control and Signal
  Processing}, vol.~17, no.~3, pp. 209--235, 2003.

\bibitem{haddad2003nonlinear}
------, ``Nonlinear adaptive control for intensive care unit sedation and
  operating room hypnosis,'' in \emph{American Control Conference, 2003.
  Proceedings of the 2003}, vol.~2.\hskip 1em plus 0.5em minus 0.4em\relax
  IEEE, 2003, Conference Proceedings, pp. 1808--1813.

\bibitem{absalom2002}
A.~R. Absalom, N.~Sutcliffe, and G.~N. Kenny, ``Closed-loop control of
  anesthesia using bispectral indexperformance assessment in patients
  undergoing major orthopedic surgery under combined general and regional
  anesthesia,'' \emph{Anesthesiology: The Journal of the American Society of
  Anesthesiologists}, vol.~96, no.~1, pp. 67--73, 2002.

\bibitem{haddad2006}
W.~M. Haddad, T.~Hayakawa, and J.~M. Bailey, ``Adaptive control for nonlinear
  compartmental dynamical systems with applications to clinical pharmacology,''
  \emph{Systems \& Control Letters}, vol.~55, no.~1, pp. 62--70, 2006.

\bibitem{hayakawa2004}
T.~Hayakawa, W.~M. Haddad, N.~Hovakimyan, and J.~M. Bailey, ``Neural network
  adaptive dynamic output feedback control for nonlinear nonnegative systems
  using tapped delay memory units,'' in \emph{American Control Conference,
  2004. Proceedings of the 2004}, vol.~5.\hskip 1em plus 0.5em minus
  0.4em\relax IEEE, 2004, Conference Proceedings, pp. 4505--4510.

\bibitem{haddad2011}
W.~M. Haddad, K.~Y. Volyanskyy, J.~M. Bailey, and J.~J. Im, ``Neuroadaptive
  output feedback control for automated anesthesia with noisy eeg
  measurements,'' \emph{IEEE Transactions on Control Systems Technology},
  vol.~19, no.~2, pp. 311--326, 2011.

\bibitem{bailey2006}
J.~M. Bailey, W.~M. Haddad, J.~J. Im, T.~Hayakawa, and P.~A. Nagel, ``Adaptive
  and neural network adaptive control of depth of anesthesia during surgery,''
  in \emph{American Control Conference, 2006}.\hskip 1em plus 0.5em minus
  0.4em\relax IEEE, 2006, Conference Proceedings, p. 6 pp.

\bibitem{schwilden1987}
H.~Schwilden, J.~Sch{\"u}ttler, and H.~Stoeckel, ``Closed-loop feedback control
  of methohexital anesthesia by quantitative eeg analysis in humans.''
  \emph{Anesthesiology}, vol.~67, no.~3, pp. 341--347, 1987.

\bibitem{schwilden1989}
H.~Schwilden, H.~Stoeckel, and J.~Sch{\"u}ttler, ``Closed-loop feedback control
  of propofol anaesthesia by quantitative eeg analysis in humans,'' \emph{BJA:
  British Journal of Anaesthesia}, vol.~62, no.~3, pp. 290--296, 1989.

\bibitem{navarro2015}
G.~Navarro-Guerrero and Y.~Tang, ``Adaptive control for anesthesia based on a
  simple fractional-order model,'' in \emph{2015 54th IEEE Conference on
  Decision and Control (CDC)}.\hskip 1em plus 0.5em minus 0.4em\relax IEEE,
  2015, pp. 5623--5628.

\bibitem{navarro2017}
------, ``Fractional order model reference adaptive control for anesthesia,''
  \emph{International Journal of Adaptive Control and Signal Processing},
  vol.~31, no.~9, pp. 1350--1360, 2017.

\bibitem{ralph2011}
M.~Ralph, C.~Beck, and M.~Bloom, ``L1-adaptive methods for control of patient
  response to anesthesia,'' in \emph{American Control Conference (ACC),
  2011}.\hskip 1em plus 0.5em minus 0.4em\relax IEEE, 2011, pp. 1729--1735.

\bibitem{kharisov2012}
E.~Kharisov, C.~L. Beck, and M.~Bloom, ``Control of patient response to
  anesthesia using $l$1 adaptive methods,'' \emph{IFAC Proceedings Volumes},
  vol.~45, no.~18, pp. 391--396, 2012.

\bibitem{kharisov2013}
------, ``Regulation of anesthesia delivery via l1-adaptive control,'' in
  \emph{2013 Proceedings of the Conference on Control and its
  Applications}.\hskip 1em plus 0.5em minus 0.4em\relax SIAM, 2013, Conference
  Proceedings, pp. 191--198.

\bibitem{kharisov2015}
------, ``Design of l1 adaptive controllers for human patient anesthesia,''
  \emph{Control Engineering Practice}, vol.~44, pp. 65--77, 2015.

\bibitem{simanski2009}
O.~Simanski, M.~Janda, A.~Schubert, J.~Bajorat, R.~Hofmockel, and B.~Lampe,
  ``Progress of automatic drug delivery in anaesthesia-the 'rostock assistant
  system for anaesthesia control (ran)','' \emph{International Journal of
  Adaptive Control and Signal Processing}, vol.~23, no.~5, pp. 504--521, 2009.

\bibitem{chou2010}
Y.-C. Chou, M.~F. Abbod, J.-S. Shieh, and C.-Y. Hsu, ``Multivariable fuzzy
  logic/self-organizing for anesthesia control,'' \emph{Journal of Medical and
  Biological Engineering}, vol.~30, no.~5, pp. 297--306, 2010.

\bibitem{abdulla2011}
S.~A. Abdulla and P.~Wen, ``Robust internal model control for depth of
  anaesthesia,'' \emph{International Journal of Mechatronics and Automation},
  vol.~1, no.~1, pp. 1--8, 2011.

\bibitem{mendez2018}
J.~A. Mendez, A.~Leon, A.~Marrero, J.~M. Gonzalez-Cava, J.~A. Reboso, J.~I.
  Estevez, and J.~F. Gomez-Gonzalez, ``Improving the anesthetic process by a
  fuzzy rule based medical decision system,'' \emph{Artificial Intelligence in
  Medicine}, vol.~84, pp. 159--170, 2018.

\bibitem{shieh2007an}
J.-S. Shieh, L.-W. Chang, T.-C. Yang, and C.-C. Liu, ``An enhanced patient
  controlled analgesia (epca) for the extracorporeal shock wave lithotripsy
  (eswl),'' \emph{Biomedical Engineering: Applications, Basis and
  Communications}, vol.~19, no.~01, pp. 7--17, 2007.

\bibitem{shieh2007a}
J.-S. Shieh, C.-Y. Dai, Y.-R. Wen, and W.-Z. Sun, ``A novel fuzzy pain demand
  index derived from patient-controlled analgesia for postoperative pain,''
  \emph{IEEE Transactions on Biomedical Engineering}, vol.~54, no.~12, pp.
  2123--2132, 2007.

\bibitem{schubert2008fuzzy}
A.~U. Schubert, M.~Janda, O.~Simanski, J.~Bajorat, B.~Pohl, R.~Hofmockel, and
  B.~Lampe, ``A fuzzy system for regulation of the analgesic remifentanil
  during general anaesthesia,'' in \emph{2008 16th Mediterranean Conference on
  Control and Automation}.\hskip 1em plus 0.5em minus 0.4em\relax IEEE, 2008,
  pp. 1634--1639.

\bibitem{linkens1991}
D.~A. Linkens and S.~Hasnain, ``Self-organising fuzzy logic control and
  application to muscle relaxant anaesthesia,'' in \emph{IEE Proceedings D
  (Control Theory and Applications)}, vol. 138.\hskip 1em plus 0.5em minus
  0.4em\relax IET, 1991, Conference Proceedings, pp. 274--284.

\bibitem{fazendeiro2002}
P.~A. Fazendeiro and J.~V. de~Oliveira, ``A survey of fuzzy control strategies
  for neuromuscular blockade using continuous infusion of atracurium,'' in
  \emph{2002 IEEE World Congress on Computational Intelligence. 2002 IEEE
  International Conference on Fuzzy Systems.}, vol.~1.\hskip 1em plus 0.5em
  minus 0.4em\relax IEEE, 2002, Conference Proceedings, pp. 547--552.

\bibitem{abbod1998}
M.~Abbod, M.~Mahfouf, and D.~Linkens, ``Multi-objective genetic optimisation
  for self-organising fuzzy logic control,'' 1998.

\bibitem{liu2015}
Y.-X. Liu, J.-S. Shieh, S.-Z. Fan, F.~Doctor, and K.-K. Jen, ``Genetic type-2
  self-organising fuzzy logic controller applied to anaesthesia,'' in
  \emph{2015 Conference on Technologies and Applications of Artificial
  Intelligence (TAAI)}.\hskip 1em plus 0.5em minus 0.4em\relax IEEE, 2015,
  Conference Proceedings, pp. 83--88.

\bibitem{shieh2006}
J.-S. Shieh, M.-H. Kao, and C.-C. Liu, ``Genetic fuzzy modelling and control of
  bispectral index (bis) for general intravenous anaesthesia,'' \emph{Medical
  Engineering \& Physics}, vol.~28, no.~2, pp. 134--148, 2006.

\bibitem{muthuswamy1999}
J.~Muthuswamy and R.~J. Roy, ``The use of fuzzy integrals and bispectral
  analysis of the electroencephalogram to predict movement under anesthesia,''
  \emph{IEEE Transactions on Biomedical Engineering}, vol.~46, no.~3, pp.
  291--299, 1999.

\bibitem{shieh2002}
J.-S. Shieh, S.-Z. Fan, and W.-L. Shi, ``The intelligent model of a patient
  using artificial neural networks for inhalational anaesthesia,''
  \emph{Journal of the Chinese Institute of Chemical Engineers}, vol.~33,
  no.~6, pp. 609--620, 2002.

\bibitem{taheriyan2018applying}
F.~Taheriyan, M.~sadat Ghafourian, and A.~Noori, ``Applying ga optimization
  algorithm for interval type-2 fuzzy logic controller parameters of
  multivariable anesthesia system,'' in \emph{Iranian Conference on Electrical
  Engineering (ICEE)}.\hskip 1em plus 0.5em minus 0.4em\relax IEEE, 2018, pp.
  1613--1618.

\bibitem{zhang2002}
X.-S. Zhang, J.~W. Huang, and R.~J. Roy, \emph{Depth of anesthesia control with
  fuzzy logic}.\hskip 1em plus 0.5em minus 0.4em\relax Springer, 2002, pp.
  83--112.

\bibitem{marrero2017}
A.~Marrero, J.~M{\'e}ndez, J.~Reboso, I.~Mart{\'\i}n, and J.~Calvo, ``Adaptive
  fuzzy modeling of the hypnotic process in anesthesia,'' \emph{Journal of
  Clinical Monitoring and Computing}, vol.~31, no.~2, pp. 319--330, 2017.

\bibitem{shieh2009}
J.-S. Shieh, M.~F. Abbod, C.~Hsu, S.-J. Huang, Y.-Y. Han, and S.-Z. Fan,
  \emph{Monitoring and control of anesthesia using multivariable
  self-organizing fuzzy logic structure}.\hskip 1em plus 0.5em minus
  0.4em\relax Springer, 2009, pp. 273--295.

\bibitem{elbardini2011}
M.~El-Bardini and A.~M. El-Nagar, ``Direct adaptive interval type-2 fuzzy logic
  controller for the multivariable anaesthesia system,'' \emph{Ain Shams
  Engineering Journal}, vol.~2, no. 3-4, pp. 149--160, 2011.

\bibitem{allen2001}
R.~Allen and D.~Smith, ``Neuro-fuzzy closed-loop control of depth of
  anaesthesia,'' \emph{Artificial Intelligence in Medicine}, vol.~21, no. 1-3,
  pp. 185--191, 2001.

\bibitem{nunes2005}
C.~S. Nunes, M.~Mahfouf, D.~A. Linkens, and J.~E. Peacock, ``Modelling and
  multivariable control in anaesthesia using neural-fuzzy paradigms: Part i.
  classification of depth of anaesthesia and development of a patient model,''
  \emph{Artificial Intelligence in Medicine}, vol.~35, no.~3, pp. 195--206,
  2005.

\bibitem{mahfouf2005}
M.~Mahfouf, C.~S. Nunes, D.~A. Linkens, and J.~E. Peacock, ``Modelling and
  multivariable control in anaesthesia using neural-fuzzy paradigms: Part ii.
  closed-loop control of simultaneous administration of propofol and
  remifentanil,'' \emph{Artificial Intelligence in Medicine}, vol.~35, no.~3,
  pp. 207--213, 2005.

\bibitem{tosun2010anesthetic}
M.~Tosun and R.~G{\"u}nt{\"u}rk{\"u}n, ``Anesthetic gas control with
  neuro-fuzzy system in anesthesia,'' \emph{Expert Systems with Applications},
  vol.~37, no.~3, pp. 2690--2695, 2010.

\bibitem{elnegar2014}
A.~M. El-Nagar and M.~El-Bardini, ``Interval type-2 fuzzy neural network
  controller for a multivariable anesthesia system based on a
  hardware-in-the-loop simulation,'' \emph{Artificial Intelligence in
  Medicine}, vol.~61, no.~1, pp. 1--10, 2014.

\bibitem{yu2018an}
Y.-N. Yu, F.~Doctor, S.-Z. Fan, and J.-S. Shieh, ``An adaptive monitoring
  scheme for automatic control of anaesthesia in dynamic surgical environments
  based on bispectral index and blood pressure,'' \emph{Journal of Medical
  Systems}, vol.~42, no.~5, p.~95, 2018.

\bibitem{mendez2016}
J.~A. M{\'e}ndez, A.~Marrero, J.~A. Reboso, and A.~Le{\'o}n, ``Adaptive fuzzy
  predictive controller for anesthesia delivery,'' \emph{Control Engineering
  Practice}, vol.~46, pp. 1--9, 2016.

\bibitem{wang2017}
Y.~Wang, Q.~Jin, and R.~Zhang, ``Improved fuzzy pid controller design using
  predictive functional control structure,'' \emph{ISA Transactions}, vol.~71,
  pp. 354--363, 2017.

\bibitem{sadati2018multi}
N.~Sadati, M.~Hosseinzadeh, and G.~A. Dumont, ``Multi-model robust control of
  depth of hypnosis,'' \emph{Biomedical Signal Processing and Control},
  vol.~40, pp. 443--453, 2018.

\bibitem{rigatos2016nonlinear}
G.~Rigatos, E.~Rigatou, and N.~Zervos, ``A nonlinear h-infinity approach to
  optimal control of the depth of anaesthesia,'' in \emph{AIP Conference
  Proceedings}, vol. 1790, no.~1.\hskip 1em plus 0.5em minus 0.4em\relax AIP
  Publishing, 2016, p. 060006.

\bibitem{padmanabhan2019nonovershooting}
R.~Padmanabhan, N.~Meskin, C.~M. Ionescu, and W.~M. Haddad, ``A nonovershooting
  tracking controller for simultaneous infusion of anesthetics and
  analgesics,'' \emph{Biomedical Signal Processing and Control}, vol.~49, pp.
  375--387, 2019.

\bibitem{iskrenovic2017}
O.~Iskrenovic-Momcilovic and A.~Momcilovic, ``Sliding mode control for depth of
  anesthesia,'' \emph{Annals of the Faculty of Engineering Hunedoara}, vol.~15,
  no.~4, pp. 41--46, 2017.

\bibitem{abood2019design}
L.~H. Abood, E.~H. Karam, and A.~H. Issa, ``Design of adaptive neuro sliding
  mode controller for anesthesia drug delivery based on biogeography based
  optimization,'' \emph{International Journal of Advanced Computer Research},
  vol.~9, no.~42, pp. 146--155, 2019.

\bibitem{khaqan2016}
A.~Khaqan, M.~Bilal, M.~Ilyas, B.~Ijaz, and R.~Ali~Riaz, ``Control law design
  for propofol infusion to regulate depth of hypnosis: a nonlinear control
  strategy,'' \emph{Computational and Mathematical Methods in Medicine}, vol.
  2016, 2016.

\bibitem{khaqan2017}
A.~Khaqan, Q.~ul~Hasan, S.~A. Malik, M.~Bilal, M.~F.~U. Butt, and R.~A. Riaz,
  ``Comparison of two nonlinear control strategies for hypnosis regulation,''
  \emph{Arabian Journal for Science and Engineering}, vol.~42, no.~12, pp.
  5165--5178, 2017.

\bibitem{hodrea2014}
R.~Hodrea, I.~Nascu, I.~Nascu, R.~De~Keyser, and H.~Vasian, ``Epsac versus pid
  control of neuromuscular blockade,'' in \emph{2014 IEEE International
  Conference on Automation, Quality and Testing, Robotics}.\hskip 1em plus
  0.5em minus 0.4em\relax IEEE, 2014, Conference Proceedings, pp. 1--6.

\bibitem{yelneedi2009}
S.~Yelneedi, L.~Samavedham, and G.~Rangaiah, ``Advanced control strategies for
  the regulation of hypnosis with propofol,'' \emph{Industrial \& Engineering
  Chemistry Research}, vol.~48, no.~8, pp. 3880--3897, 2009.

\bibitem{yousefi2017formally}
M.~Yousefi, K.~van Heusden, I.~M. Mitchell, J.~M. Ansermino, and G.~A. Dumont,
  ``A formally-verified safety system for closed-loop anesthesia,''
  \emph{IFAC-PapersOnLine}, vol.~50, no.~1, pp. 4424--4429, 2017.

\bibitem{yousefi2019formalized}
M.~Yousefi, K.~van Heusden, N.~West, I.~M. Mitchell, J.~M. Ansermino, and G.~A.
  Dumont, ``A formalized safety system for closed-loop anesthesia with
  pharmacokinetic and pharmacodynamic constraints,'' \emph{Control Engineering
  Practice}, vol.~84, pp. 23--31, 2019.

\bibitem{khodaei2018}
M.~J. Khodaei, A.~Mehrvarz, N.~Jalili \emph{et~al.}, ``An adaptive multi-critic
  neuro-fuzzy control framework for intravenous anesthesia administration,''
  \emph{IFAC-PapersOnLine}, vol.~51, no.~34, pp. 202--207, 2019.

\bibitem{senseonics2018}
\BIBentryALTinterwordspacing
{Senseonics, Inc}, ``{FDA} executive summary - senseonics eversense continuous
  glucose monitoring system,'' 2018. [Online]. Available:
  \url{https://www.fda.gov/downloads}
\BIBentrySTDinterwordspacing

\bibitem{colquitt2011}
R.~B. Colquitt, D.~A. Colquhoun, and R.~H. Thiele, ``In silico modelling of
  physiologic systems,'' \emph{Best Practice \& Research Clinical
  Anaesthesiology}, vol.~25, no.~4, pp. 499--510, 2011.

\bibitem{tucker2017}
G.~T. Tucker, ``Personalized drug dosage-closing the loop,''
  \emph{Pharmaceutical Research}, vol.~34, no.~8, pp. 1539--1543, 2017.

\bibitem{arroyocurras2018}
N.~Arroyo-Curr\'{a}s, G.~Ortega, D.~A. Copp, K.~L. Ploense, Z.~A. Plaxco, T.~E.
  Kippin, J.~P. Hespanha, and K.~W. Plaxco, ``High-precision control of plasma
  drug levels using feedback-controlled dosing,'' \emph{ACS Pharmacology \&
  Translational Science}, vol.~1, no.~2, pp. 110--118, 2018.

\bibitem{yu2018advance}
J.~Yu, Y.~Zhang, J.~Yan, A.~R. Kahkoska, and Z.~Gu, ``Advances in bioresponsive
  closed-loop drug delivery systems,'' \emph{International Journal of
  Pharmaceutics}, vol. 544, no.~2, pp. 350--357, 2018.

\bibitem{vandergraaf2016}
P.~H. van~der Graaf, N.~Benson, and L.~A. Peletier, ``Topics in mathematical
  pharmacology,'' \emph{Journal of Dynamics and Differential Equations},
  vol.~28, no. 3-4, pp. 1337--1356, 2016.

\bibitem{sopasakis2018}
P.~Sopasakis, H.~Sarimveis, P.~Macheras, and A.~Dokoumetzidis, ``Fractional
  calculus in pharmacokinetics,'' \emph{Journal of Pharmacokinetics and
  Pharmacodynamics}, vol.~45, no.~1, pp. 107--125, 2018.

\bibitem{parker2001control}
R.~S. Parker and F.~J. Doyle~III, ``Control-relevant modeling in drug
  delivery,'' \emph{Advanced Drug Delivery Reviews}, vol.~48, no. 2-3, pp.
  211--228, 2001.

\bibitem{parvinian2019credibility}
B.~Parvinian, P.~Pathmanathan, C.~Daluwatte, F.~Yaghouby, R.~A. Gray,
  S.~Weininger, T.~M. Morrison, and C.~G. Scully, ``Credibility evidence for
  computational patient models used in the development of physiological
  closed-loop controlled devices for critical care medicine,'' \emph{Frontiers
  in Physiology}, vol.~10, 2019.

\bibitem{mulugeta2018credibility}
L.~Mulugeta, A.~Drach, A.~Erdemir, C.~A. Hunt, M.~Horner, J.~P. Ku, J.~G.
  Myers~Jr, R.~Vadigepalli, and W.~W. Lytton, ``Credibility, replicability, and
  reproducibility in simulation for biomedicine and clinical applications in
  neuroscience,'' \emph{Frontiers in Neuroinformatics}, vol.~12, 2018.

\bibitem{asme2018}
``Assessing credibility of computational modeling through verification and
  validation: Application to medical devices,'' 2018.

\bibitem{virk2018}
G.~S. Virk, K.~Chinzei, and M.~Brossoit, ``Emergence of medical robots and
  autonomy in medical electrical equipment,'' \emph{Biomedical Instrumentation
  \& Technology}, vol.~52, no.~2, pp. 156--159, 2018.

\bibitem{yang2017}
G.-Z. Yang, J.~Cambias, K.~Cleary, E.~Daimler, J.~Drake, P.~E. Dupont, N.~Hata,
  P.~Kazanzides, S.~Martel, and R.~V. Patel, ``Medical robotics-regulatory,
  ethical, and legal considerations for increasing levels of autonomy,''
  \emph{Science Robotics}, vol.~2, no.~4, p. 8638, 2017.

\bibitem{dey2018}
N.~Dey, A.~S. Ashour, F.~Shi, S.~J. Fong, and J.~M.~R. Tavares, ``Medical
  cyber-physical systems: A survey,'' \emph{Journal of Medical Systems},
  vol.~42, no.~4, p.~74, 2018.

\bibitem{crawford2016}
\BIBentryALTinterwordspacing
M.~Crawford, ``Top 6 robotic applications in medicine,'' 2016. [Online].
  Available:
  \url{https://www.asme.org/engineering-topics/articles/bioengineering/top-6-robotic-applications-in-medicine}
\BIBentrySTDinterwordspacing

\bibitem{beasley2012}
R.~A. Beasley, ``Medical robots: current systems and research directions,''
  \emph{Journal of Robotics}, vol. 2012, 2012.

\end{thebibliography}

\end{document}